\documentclass[a4paper, 12pt]{article}
\usepackage{jheppub}

\usepackage{amsmath,amssymb,amsfonts}
\usepackage{mathrsfs}
\usepackage{xcolor}
\usepackage{graphicx} 
\usepackage{slashed}

\usepackage{lineno}

\allowdisplaybreaks

\title{\boldmath Exploring Invisible New Physics with Exotic Pion Decays}

\def\ucsc{Department of Physics, University of California, Santa Cruz, and Santa Cruz Institute for Particle Physics, Santa Cruz, CA 95064, USA}
\def\uw{Department of Physics, University of Washington, Seattle, WA 98195 USA}
\def\UF{Physics Department, University of Florida, Gainesville, FL 32611, USA}
\def\sjsu{Department of Physics and Astronomy, San Jose State University, San Jose, CA 95192, USA}

\author[a]{Wolfgang~Altmannshofer,}
\author[b]{Jeff~A.~Dror,}
\author[a]{Pierce~Giffin,}
\author[a]{Stefania~Gori,}
\author[a]{Ollie~Jackson,}
\author[a]{Khai~Luong,}
\author[c]{Patrick~Schwendimann,}
\author[d]{and Se~Rang~Seo}

\affiliation[a]{\ucsc}
\affiliation[b]{\UF}
\affiliation[c]{\uw}
\affiliation[d]{\sjsu}

\emailAdd{waltmann@ucsc.edu}
\emailAdd{jeffdror@ufl.edu}
\emailAdd{pgiffin@ucsc.edu}
\emailAdd{sgori@ucsc.edu}
\emailAdd{oajackso@ucsc.edu}
\emailAdd{khhluong@ucsc.edu}
\emailAdd{schwenpa@uw.edu}
\emailAdd{seo.simon.se.rang@gmail.com}

\abstract{We study the sensitivity of past and future stopped-pion experiments to light, invisible dark sector particles produced in exotic pion decays. We consider two-body decays involving sterile neutrinos, $\pi^+ \to \ell^+ N$, as well as three-body decays $\pi^+ \to \ell^+ \nu_\ell X$, with $X$ an invisible scalar, axion-like particle, or dark vector. We recast existing limits from the PIENU experiment and project the reach of the planned PIONEER experiment using detailed simulations based on the current detector design. We find that PIONEER can improve bounds on exotic pion branching ratios by at least one order of magnitude below current limits. We compare the projected sensitivities with complementary constraints from lepton anomalous magnetic moments, mono-photon searches, and beam-dump experiments, identifying weak-violating axion-like particles as a particularly well-motivated benchmark. Our results establish PIONEER as a powerful and complementary probe of light, invisible dark sectors.}

\begin{document}
\maketitle
\flushbottom

\section{Introduction} \label{sec:intro}

Precision measurements of rare meson decays offer a powerful avenue to test the Standard Model (SM) and search for new physics. Among these, charged pion decays play a special role~\cite{Bryman:2025pet}. They are theoretically exceptionally clean and experimentally accessible. Moreover, the pion's relatively small decay width makes them sensitive to a broad class of new light particles that may couple feebly to the Standard Model and that can be produced from pion decays. The proposed PIONEER experiment at PSI~\cite{PIONEER:2022yag, PIONEER:2022alm} will collect an unprecedented sample of pions decaying at rest and presents a unique opportunity to test light new particles.

In this work, we investigate the potential of the PIONEER experiment to discover or constrain exotic decays of charged pions to invisible particles. We consider both two-body decays such as $\pi^+\to \ell^+ N$, where $N$ is a sterile neutrino, and three-body decays involving other dark sector particles in the final state, $\pi^+ \to \ell^+ \nu_\ell X$. Such three-body decays are a general prediction of well-motivated extensions of the SM, including models with axion-like particles (ALPs), light scalars, and dark vector bosons. The considered decay topologies capture a wide class of new physics scenarios in which light, weakly coupled states are produced in pion decays and escape detection, see e.g.~\cite{Aditya:2012ay, Batell:2017cmf, Alves:2017avw, Altmannshofer:2019yji, Dror:2020fbh, Hostert:2020xku, Bauer:2020jbp, Bauer:2021wjo, Bauer:2021mvw, Gallo:2021ame, Dutta:2021cip, Bickendorf:2022buy, Altmannshofer:2022ckw, Guerrera:2022ykl,  DiLuzio:2023ndz, Hostert:2023tkg, Eberhart:2025lyu, DiLuzio:2025ojt, Okawa:2025sak}. We compare the reach of the PIONEER experiment with that of previous pion experiments, including the PIENU experiment~\cite{PIENU:2011aa, PiENu:2015seu, PIENU:2017wbj, PIENU:2019usb, PIENU:2020loi, PIENU:2021clt}, as well as the one of kaon factories, such as NA62 \cite{NA62:2020mcv, NA62:2021bji, NA62:2025csa}. We demonstrate that PIONEER will have access to significant new regions of parameter space of invisible dark particles. 

The paper is structured as follows: in Sec. \ref{sec:exp}, we give an overview of the PIENU and PIONEER experiments, their main physics goals, and the possible searches for new light particles. In Sec. \ref{sec:2body}, we discuss exotic two-body decays of charged pions into sterile neutrinos. In Sec. \ref{sec:3body}, we analyze exotic three-body decays of charged pions using a simplified model framework to describe decays into a scalar, an axion-like particle, or a dark vector. In all cases, we compare the projected PIONEER sensitivity with existing experimental constraints on these simplified models. Finally, Sec. \ref{sec:conclusions} summarizes our conclusions. Details on the recast of PIENU searches for exotic two-body and three-body decays are given in Appendices \ref{app:recast} and \ref{app:recast2}, respectively. 

\section{Exotic Pion Decays at PIENU and PIONEER} \label{sec:exp} 

The main goal of Phase~I of the proposed PIONEER experiment is the precision measurement of the ratio of the charged pion branching ratios to electrons and muons 
\begin{equation}
R_\pi\equiv \frac{{\rm{BR}}(\pi^+\to e^+\nu_e (\gamma))}{{\rm{BR}}(\pi^+\to \mu^+\nu_\mu (\gamma))} ~.
\end{equation}
This observable is extremely sensitive to a wide variety of new physics effects and is one of the most accurate probes of lepton flavor universality.
The ratio can be predicted with remarkable precision in the SM, as uncertainties from the CKM matrix element $V_{ud}$ and the pion decay constant $f_\pi$ cancel, and radiative corrections (indicated by $\Delta_\text{rad}$ in the equation below) are well understood. The main uncertainty in the SM prediction is due to structure dependent radiative corrections that are estimated at next-to-leading order in chiral perturbation theory. The SM prediction reads ~\cite{Marciano:1993sh, Knecht:1999ag, Cirigliano:2007xi, Cirigliano:2007ga, PIONEER:2022yag}
\begin{equation} \label{eq:Rpi_SM}
R_\pi^\text{SM} = \frac{m_e^2}{m_\mu^2} \frac{(m_\pi^2 - m_e^2)^2}{(m_\pi^2 - m_\mu^2)^2} \big( 1 + \Delta_\text{rad} \big)= 1.23524(15) \times 10^{-4} ~.
\end{equation}
On the experimental side, the most precise measurement of $R_\pi$ comes from the PIENU experiment~\cite{PiENu:2015seu}. Combining the PIENU result with previous measurements~\cite{Bryman:1985bv, Britton:1992pg, Czapek:1993kc}, the PDG gives the following world average~\cite{ParticleDataGroup:2024cfk} 
\begin{equation} \label{eq:Rpi_exp}
R_\pi^\text{exp} = 1.2327(23) \times 10^{-4} ~.
\end{equation}
Combining with the SM prediction from above, this gives the following 90\% C.L. limits on new physics contributions to $R_\pi$ 
\begin{equation}\label{eq:Rpi_current_constraint}
-5.1 \times 10^{-3} < R_\pi / R_\pi^\text{SM} - 1 < 1.0 \times 10^{-3} ~.
\end{equation}

The goal of PIONEER Phase I is to measure $R_\pi$ with a precision of $0.01\%$, a factor of $\sim 20$ more precisely than the PIENU measurement. Assuming the central value will coincide with the SM prediction, we find the following sensitivity to new physics contributions at 90\% C.L.
\begin{equation} \label{eq:Rpi_future_constraint}
-2.6 \times 10^{-4} < R_\pi / R_\pi^\text{SM} - 1 < 2.6 \times 10^{-4} ~.
\end{equation}
The $R_\pi$ measurement by itself can be used to constrain regions of parameter space of models that predict exotic decays of charged pions to light new physics particles. In many scenarios however, dedicated searches for the distinct signatures of exotic pion decays can lead to a even better sensitivity. Stopped pion experiments combine very large pion yields with a simple and known initial state (the stopped pion), giving access to potentially very small exotic branching ratios.
 
As emphasized for example in \cite{Altmannshofer:2019yji}, searching for exotic decays of charged pions is very well motivated, considering the small charged pion decay width, $\Gamma_{\pi^+}\simeq 2.5\times 10^{-17}$\,GeV, and the fact that the prominent SM decay modes $\pi^+ \to e^+ \nu_e$~\cite{PiENu:2015seu} and $\pi^+ \to \pi^0 e^+ \nu_e$~\cite{Pocanic:2003pf} are helicity or phase space suppressed and therefore have very small branching ratios. Even very feeble couplings of a new physics particle to the pion or leptons could lead to a detectable exotic branching ratio. 

Exotic two-body decays into a sterile neutrino $\pi^+ \to \ell^+ N$ are particularly clean, as they yield a mono-energetic charged lepton whose energy shifts if $N$ has a mass. This makes peak searches in the lepton energy spectrum extremely powerful. As we will discuss in section~\ref{sec:2body}, the best limits on certain sterile neutrinos in the mass range below the pion mass come indeed from charged pion decay experiments. Also the production of new light particles in three-body decays, $\pi^+ \to \ell^+ \nu_e X$, modifies the charged lepton energy spectrum. However, in contrast to the SM decays $\pi^+ \to \ell^+ \nu_\ell$ and the decays into sterile neutrinos $\pi^+ \to \ell^+ N$, the three-body decays give a continuous spectrum of charged lepton energies. Precise measurements of the charged lepton energy spectra can set stringent constraints on the parameter space of new physics models. 

In addition to exotic pion decays, pion factories can also look for exotic decays of secondary muons given the timing separation between the pion stop and the muon decay. Thanks to the very large statistics of muons, as well as to the small muon decay width, $\Gamma_\mu\sim 3\times 10^{-19}$ GeV, pion decay experiments looking for $\mu^+ \to e^+ X$ can have sensitivity to tiny flavor changing couplings of the muon.

For all these reasons, the search for light new  particles has been broadly pursued by the PIENU experiment, which conducted searches for sterile neutrinos $\pi^+ \to e^+ N$~\cite{PIENU:2017wbj} and $\pi^+ \to \mu^+ N$~\cite{PIENU:2019usb} as well as for exotic three-body decay modes of the charged pion $\pi^+ \to \ell^+ \nu_\ell X$~\cite{PIENU:2021clt}, and the two-body decay of the muon $\mu^+ \to e^+ X$~ \cite{PIENU:2020loi}, with a light new particle $X$ that escapes detection.
The PIENU search for $\mu^+ \to e^+ X$ achieved a sensitivity that rivals the ones of past dedicated experiments~\cite{Jodidio:1986mz, TWIST:2014ymv} (but cannot compete with future experiments searching for $\mu \to e$ transitions~\cite{Calibbi:2020jvd, Jho:2022snj, Hill:2023dym, Knapen:2023zgi}).

Many searches for exotic pion decays can be performed in the future at the PIONEER experiment. One can classify the new exotic decay modes according to their multiplicity. We consider final states that contain two particles, as in $\pi^+\to \ell^+ N$, 
where the new state $N$ has to be electrically neutral and a fermion (to be specific, we will consider an electrically neutral lepton); 
or three particles, as in $\pi^+\to \ell^+ \nu_\ell X$, with SM leptons $\ell=e,\mu$ 
and where the new state $X$ has to be an electrically neutral boson.
In the interesting region of parameter space, the sterile neutrinos are either long-lived and decay far outside the detector, or they decay into other invisible states.\footnote{As discussed in section \ref{sec:sterile_theory}, regions of parameter space where sterile neutrinos produced from pion decays decay promptly or slightly displaced to visible SM particles are already completely probed.} 
In either case, they do not give any visible signature in the detector. 
For the boson $X$ there are several options: 
\begin{itemize}
\item[(i)] $X$ does not give any visible signatures because it is either long lived on detector scales or it decays into invisible particles; 
\item[(ii)] $X$ decays into fully visible resonant final states, in particular the two possible two-body modes $X \to \gamma \gamma$ or $X \to e^+e^-$; 
\item[(iii)] $X$ decays into semi-visible final states, e.g. three-body $X \to \gamma \gamma Y$ or $X \to e^+e^- Y$ with another invisible state $Y$.
\end{itemize}
The visible and semi-visible decays can be either prompt or displaced. Displacements up to a few mm might be accessible to the PIONEER experiment whose active target is $6$\,mm thick~\cite{PIONEER:2022yag}. 

Systematically studying PIONEER's sensitivity to the full set of signatures is an important task. In this article, we will focus on two-body and three-body pion decays into invisible new physics particles. We leave the study of visible and semi-visible exotic decays for future work~\cite{Altmannshofer_inprep}. 

The main features of the PIENU and PIONEER experiments that are relevant for our study are reported in Table~\ref{tab:Exp}. 
The PIENU experiment relied on a target built from a 8\,mm thick scintillator where the incoming pion beam was stopped. Muons from pion decays at rest have a
kinetic energy of approximately 4\,MeV, thus mostly remain inside the target. A cylindrical
NaI crystal calorimeter with a length of 19 radiation lengths was used to measure the energy of the emitted
positrons. CsI crystals surrounded the NaI crystal to better contain electromagnetic showers.
The setup is described in more details in \cite{PiENu:2015seu}.

Based on the lessons learned from PIENU, the PIONEER experiment will use a highly segmented
active target (ATAR) with 4800 LGAD channels read out instead of a single 8\,mm scintillator.
This enables tracking inside the stopping target
and thus it will be possible to better match incoming pions to outgoing positrons which in turn
allows for higher pion beam rates that result in larger statics. Additionally, it becomes possible
to observe the detailed stopping behavior of pions and muons, thus providing additional handles
to distinguish between \(\pi^+ \rightarrow e^+ \nu (\gamma)\) events and \(\pi\rightarrow \mu \rightarrow e\)
decay chains.

Additionally, the PIONEER experiment aims for a spherical state of the art calorimeter.
As for PIENU, it was designed to be 19 radiation lengths deep. The spherical shape
is a major improvement as it covers a larger solid angle and provides a more uniform detector
response as a function of the polar angle. This reduces the variation in the \(\pi^+ \rightarrow e^+ \nu (\gamma)\)
line shape. Moreover, the PIONEER calorimeter will feature a much faster decay time compared
to the PIENU detector, which is crucial to support the higher anticipated beam rate. Combined with
the improved ATAR capabilities, the PIONEER experiment aims to achieve a \(\pi^+ \rightarrow e^+ \nu (\gamma)\)
event selection that provides a sample pure enough for a direct line shape measurement across the
detector.

\setlength{\tabcolsep}{8pt}
\renewcommand{\arraystretch}{1.6}
\begin{table}[]
    \centering
    \begin{tabular}{p{5 cm}cc}
        \hline\hline
         & PIENU & PIONEER Phase I \\
        \hline
        \(N_{\pi\rightarrow e}\) reconstructed & $10^7$& \(2\times10^{8}\)  \\
        \(N_{\pi\rightarrow e}\) selected      & \( \approx 1.4 \times 10^6 \)& \(0.5\times10^{8}\) \\
        Signal to Background \newline below \(56\)\,MeV & \( \approx 0.3\) & \( \approx 10\) \\
        Resolution Function (\(e^+\)) & \(\frac{\sigma_E}{E} \approx 3\% ~@~ 40\,\text{MeV}\) & \(\frac{\sigma_E}{E} \approx\left (0.43 + \frac{4.95}{\sqrt{E}} + \frac{41.3}{E}\right) \%\) \\
        \hline
        \(N_{\pi\rightarrow \mu \rightarrow e}\) selected & \(9\times10^{6}\) & \(\mathcal{O}\left(10^{10}\right)\) \\
        Resolution (\(\mu^+\) at \(4\)\,MeV) & \( \approx 4 \% \)  & \( 5 \% \) \\
        \hline\hline
    \end{tabular}
    \caption{Summary of key experimental parameters relevant for exotic pion decay searches at the PIENU experiment and the PIONEER Phase I setup, including event yields, background levels, and energy resolutions for positrons and muons. The PIENU parameters are estimated from \cite{PIENU:2017wbj, PIENU:2019usb}. (In the case of the muon decays, they correspond to the analysis with muon energies above 1.2\,MeV.) }
    \label{tab:Exp}
\end{table}
\renewcommand{\arraystretch}{1.0}

\section{Exotic Two-Body Decays of Pions to Sterile Neutrinos} \label{sec:2body} 

In this section, we investigate the two-body decays $\pi^+ \to \ell^+ N$ as probes of light sterile neutrinos that mix with the active neutrino flavors. We begin by outlining the theoretical framework and the relevant parameters governing sterile neutrino production in pion decays. We then reproduce existing constraints from the PIENU experiment and derive projected sensitivities for the PIONEER experiment.

\subsection{Sterile neutrinos} \label{sec:sterile_theory} 

Sterile neutrinos are a well-motivated extension of the Standard Model, introduced in many frameworks to account, e.g., for neutrino masses, dark matter, or anomalies in neutrino oscillation experiments. Pion decay experiments can search for sterile neutrinos in the mass range below the charged pion mass~\cite{Shrock:1980vy, Shrock:1980ct}. See e.g.~\cite{deGouvea:2015euy, Bryman:2019bjg, Abdullahi:2022jlv} for reviews.

Sterile neutrinos can mix with the SM neutrinos through the renormalizable neutrino portal. This mixing determines the strength of the weak interactions of the sterile neutrinos. The interactions with the $W$ boson can be written as
\begin{equation}
\mathcal L\supset- \frac{g}{\sqrt 2} U_{jN} W_\mu^-\bar\ell_j \gamma^\mu N ~+~ \text{h.c.} ~,
\end{equation}
where $g$ is the $SU(2)_L$ gauge coupling, $N$ is the sterile neutrino, and $\ell_j$, with $j = 1,2,3$, are the three generations of left-handed charged leptons of the SM. In this simplified model, the sterile neutrino parameter space consists of the three mixing angles $U_{jN}$ and the mass of the sterile neutrino $m_N$. The sterile neutrino can be a Dirac fermion or a Majorana fermion. 

In this minimal setup, sterile neutrinos decay only via their mixing with active neutrinos, with a decay width scaling as $\Gamma_N \propto |U_{jN}|^2m_N^5/v^4$, with $v$ the vacuum expectation value of the SM Higgs. For sterile neutrino masses below the pion mass, this scaling implies that even sizable mixing angles completely probed by past experiments, $|U_{jN}|\sim 0.1$, correspond to macroscopic decay lengths of the order of $\gtrsim 10~{\rm m}$. Sterile neutrinos in this regime therefore escape the detector and manifest as missing energy in pion decay experiments. As we will mention in more detail in section~\ref{sec:sterile_sensitivity_PIONEER}, Big Bang Nucleosynthesis (BBN) places stringent bounds on such long-lived states, essentially excluding the minimal scenario in this mass range for interesting mixing angles. A way to evade the BBN bound is to allow additional decay modes of the sterile neutrino into dark-sector states (see e.g.,~\cite{Deppisch:2024izn}). 
These channels can reduce the lifetime to the level required by BBN ($\tau_N \lesssim 0.1$ s), while still ensuring that sterile neutrinos appear as missing energy in laboratory experiments.

A sterile neutrino of Majorana nature could plausibly be involved in the mechanism that provides masses to the SM neutrinos. Parametrically, the see-saw mechanism predicts that the SM neutrinos acquire a mass of the order of $m_\nu \sim |U_{j N}|^2 m_N$. From neutrino oscillations we know that at least two neutrinos are massive with masses $m_\nu \gtrsim 0.008$\,eV and $m_\nu \gtrsim 0.05$\,eV~\cite{Esteban:2024eli}. The sum of neutrino masses is constrained by cosmology to be below $\sum m_\nu \lesssim \mathcal O(0.1$\,eV)~\cite{Planck:2019nip, ACT:2023kun, DESI:2024mwx}. For these reasons, as we will adopt in the next section, a good representative choice for the see-saw region is $0.01\,{\rm eV} < |U_{j N}|^2 m_N < 0.1\,{\rm eV}$. We emphasize, however, that these boundaries are not sharply defined and should be regarded as indicative rather than exact. 

\subsection{Experimental signatures at pion decay experiments} \label{sec:sterile_signature} 

The mixing of the active neutrinos with the sterile neutrino has two effects. First, it reduces the standard leptonic decay rates of the pion into a positron and electron-neutrino, or muon and muon-neutrino
\begin{equation} \label{eq:sterile_mix}
\frac{\Gamma(\pi^+ \to e^+ \nu_e)}{\Gamma(\pi^+ \to e^+ \nu_e)_\text{SM}} = 1 - |U_{eN}|^2~,\quad \frac{\Gamma(\pi^+ \to \mu^+ \nu_\mu)}{\Gamma(\pi^+ \to \mu^+ \nu_\mu)_\text{SM}} = 1 - |U_{\mu N}|^2~.
\end{equation}
Second, the mixing introduces new two-body decay modes of the pion into the sterile neutrino, $\pi^+ \to e^+ N$ and $\pi^+ \to \mu^+ N$. If those decays are kinematically allowed, i.e. if $m_N < m_\pi- m_e$ or $m_N < m_\pi - m_\mu$, the corresponding branching ratios are to a very good approximation
\begin{eqnarray} \label{eq:sterile_exact1}
\frac{\text{BR}(\pi^+ \to e^+ N)}{\text{BR}(\pi^+ \to e^+ \nu_e)} &=& \frac{|U_{eN}|^2}{1 - |U_{eN}|^2} \frac{\rho(\delta_e^\pi, \delta_N^\pi)}{\delta_e^\pi (1 - \delta_e^\pi)^2} \\ \label{eq:sterile_exact2}
&=& \frac{|U_{e N}|^2}{1-|U_{e N}|^2} \frac{m_N^2}{m_e^2} \left( 1 - \frac{m_e^2}{m_\pi^2} \right)^{-2} \sqrt{\lambda\left(1, \frac{m_N^2}{m_\pi^2}, \frac{m_e^2}{m_\pi^2} \right)} \nonumber \\ 
&& \qquad\qquad\qquad \times \left( 1 - \frac{m_N^2}{m_\pi^2} + \frac{m_e^2}{m_N^2} + \frac{2 m_e^2}{m_\pi^2} - \frac{m_e^4}{m_N^2 m_\pi^2} \right)  \\ \label{eq:sterile_approx}
&\simeq& \frac{|U_{e N}|^2}{1-|U_{e N}|^2} \frac{m_N^2}{m_e^2} \left( 1 - \frac{m_N^2}{m_\pi^2} \right)^2~,
\end{eqnarray}
where $\lambda(a,b,c) = a^2 + b^2 + c^2 - 2(ab + ac + bc)$. 
The first line is a compact form introduced in~\cite{Shrock:1980vy, Shrock:1980ct} that uses the dimensionless variables $\delta_e^\pi = m_e^2/m_\pi^2$ and $\delta_N^\pi = m_N^2/ m_\pi^2$ and the function $\rho(x,y) = [x + y - (x-y)^2]~ [\lambda(1,x,y)]^{1/2}$.
The last line holds under the assumption that the electron is much lighter than the sterile neutrino, $m_e \ll m_N$.
For the muonic decay $\pi^+ \to \mu^+ N$, expressions analogous to equations~\eqref{eq:sterile_exact1} and~\eqref{eq:sterile_exact2} hold, but the analog to equation~\eqref{eq:sterile_approx} is never a good approximation because of the large muon mass. 

The experimental signature of the new decay modes is a peak in the positron or muon energy spectrum at
\begin{equation} \label{eq:E_peak}
\bar E_e = \frac{m_\pi}{2} \left(1 - \frac{m_N^2}{m_\pi^2} +\frac{m_e^2}{m_\pi^2} \right) ~,\quad  \bar E_\mu = \frac{m_\pi}{2} \left(1 - \frac{m_N^2}{m_\pi^2} +\frac{m_\mu^2}{m_\pi^2} \right) ~.
\end{equation}
If the positron or muon energy spectra of the SM processes $\pi^+ \to e^+ \nu_e$ and $\pi^+ \to \mu^+ \nu_\mu$ (including bremsstrahlung and detector effects) and all relevant backgrounds can be modeled reliably, one can perform a bump hunt for the sterile neutrino signal.

Alternatively, one can simply consider the precisely measured ratio $R_\pi$ and compare it to the SM prediction. One has to distinguish three mass regimes: 

\underline{(1) $m_N < m_\pi - m_\mu$}: in this regime both the $\pi^+ \to e^+ N$ and $\pi^+ \to \mu^+ N$ decays are kinematically open, and one has
\begin{multline}
   R_\pi = \frac{\text{BR}(\pi^+ \to e^+ + \text{invisible}) }{\text{BR}(\pi^+ \to \mu^+ + \text{invisible})} = \frac{\text{BR}(\pi^+ \to e^+ \nu_e) + r_e ~\text{BR}(\pi^+ \to e^+ N)}{\text{BR}(\pi^+ \to \mu^+ \nu_\mu) + r_\mu ~ \text{BR}(\pi^+ \to \mu^+ N)} \\[16pt]
   = R_\pi^\text{SM} \times \Bigg(1 - |U_{e N}|^2 \Bigg[ 1 -  \frac{m_N^2}{m_e^2} \left( 1 - \frac{m_e^2}{m_\pi^2} \right)^{-2} \sqrt{\lambda\left(1, \frac{m_N^2}{m_\pi^2}, \frac{m_e^2}{m_\pi^2} \right)} \\ \times \left( 1 - \frac{m_N^2}{m_\pi^2} + \frac{m_e^2}{m_N^2} + \frac{2 m_e^2}{m_\pi^2} - \frac{m_e^4}{m_N^2 m_\pi^2} \right)\Bigg] \Bigg)\\
   \times \Bigg(1 - |U_{\mu N}|^2 \Bigg[ 1 - \frac{m_N^2}{m_\mu^2} \left( 1 - \frac{m_\mu^2}{m_\pi^2} \right)^{-2} \sqrt{\lambda\left(1, \frac{m_N^2}{m_\pi^2}, \frac{m_\mu^2}{m_\pi^2} \right)} \\ \times \left( 1 - \frac{m_N^2}{m_\pi^2} + \frac{m_\mu^2}{m_N^2} + \frac{2 m_\mu^2}{m_\pi^2} - \frac{m_\mu^4}{m_N^2 m_\pi^2} \right) \Bigg] \Bigg)^{-1} ~.
\end{multline}
The SM prediction $R_\pi^\text{SM}$ is given in equation~\eqref{eq:Rpi_SM}. The coefficients $r_e$ and $r_\mu$ are analysis acceptance ratios of the decays into sterile neutrinos versus the decays into SM neutrinos. We have set them to $r_e = r_\mu = 1$ in the second line. Deviations from $1$ may arise if strong selection cuts are used in the analysis. 

\underline{(2) $m_\pi - m_\mu < m_N < m_\pi - m_e$}: for such sterile neutrino masses one has instead
\begin{multline}
   R_\pi = R_\pi^\text{SM} \times \Bigg(1 - |U_{e N}|^2 \Bigg[ 1 -  \frac{m_N^2}{m_e^2} \left( 1 - \frac{m_e^2}{m_\pi^2} \right)^{-2} \sqrt{\lambda\left(1, \frac{m_N^2}{m_\pi^2}, \frac{m_e^2}{m_\pi^2} \right)} \\ \times \left( 1 - \frac{m_N^2}{m_\pi^2} + \frac{m_e^2}{m_N^2} + \frac{2 m_e^2}{m_\pi^2} - \frac{m_e^4}{m_N^2 m_\pi^2} \right)\Bigg] \Bigg) \frac{1}{1 - |U_{\mu N}|^2} ~,
\end{multline}
where we again set the acceptance ratio $r_e = 1$. The ratio $r_\mu$ is not relevant as the $\pi^+ \to \mu^+ N$ decay is not open. 

\underline{(3) $m_\pi - m_e < m_N$}: in this mass regime,
the only observable effect is the modification of the $\pi^+ \to e^+ \nu_e$ and $\pi^+ \to \mu^+ \nu_\mu$ rates due to the mixing, cf. equation~\eqref{eq:sterile_mix}, 
\begin{equation}
   R_\pi = R_\pi^\text{SM} \times \frac{1 - |U_{e N}|^2}{1 - |U_{\mu N}|^2} ~.
\end{equation}
Measurements of $R_\pi$ can be used to constrain the sterile neutrino parameters. In this approach, one does not need to precisely know the positron and muon energy spectra of $\pi^+ \to e^+ \nu_e$ and $\pi^+ \to \mu^+ \nu_\mu$, and the derived results are arguably more robust. However, without spectral information, one does not disentangle the effects in the electron and muon channels and one cannot distinguish a sterile neutrino effect from the effect of other new physics that modifies $R_\pi$. 

\subsection{Reproducing the existing constraints on \texorpdfstring{$|U_{eN}|$}{Ue} and \texorpdfstring{$|U_{\mu N}|$}{Umu}} \label{sec:sterile_constraints_UeUmu} 

Bump searches for $\pi^+ \to e^+ N$ and $\pi^+ \to \mu^+ N$ have been performed at PIENU~\cite{PIENU:2017wbj, PIENU:2019usb} (See~\cite{Bryman:1983cja, Azuelos:1986eg, DeLeener-Rosier:1991luz, Britton:1992xv, Abela:1981nf, Minehart:1981fv, Daum:1987bg, Daum:1995hs, Bryman:1996xd, Assamagan:1998vy, Daum:2000ac} for previous sterile neutrino searches in pion decays). Upper limits on the mixing parameter $|U_{e N}|^2$ were set in the ballpark of $10^{-7}$ to $10^{-8}$, for sterile neutrino masses in the range of $62 - 134$~MeV, corresponding to a positron energy of $E_e \lesssim 56$\,MeV. For smaller masses, the sterile neutrino peak moves to larger positron energies and it becomes increasingly challenging to precisely model the SM spectrum (see the discussion in section~\ref{sec:sterile_sensitivity_PIONEER} below).
Similar constraints on $|U_{e N}|^2$ were recently obtained by the NA62 experiment in the mass range $95 - 126$~MeV~\cite{NA62:2025csa} (See~\cite{Asano:1981he, Hayano:1982wu, E949:2014gsn, NA62:2017qcd} for earlier sterile neutrino searches in kaon decays at higher masses.). The mixing parameter $|U_{\mu N}|^2$ is constrained by PIENU at the level of $\sim 10^{-5}$ for sterile neutrino masses between $16$~MeV and $34$~MeV.

Before producing sensitivity estimates for PIONEER, we first reproduced the PIENU results from the bump searches, both in $\pi^+ \to e^+ N$ and $\pi^+ \to \mu^+ N$ to validate our approach. Details about the PIENU recast that closely mirrors our PIONEER sensitivity study are given in appendix~\ref{app:recast}. 

\begin{figure}[tb]
\centering
\includegraphics[width=0.9\textwidth]{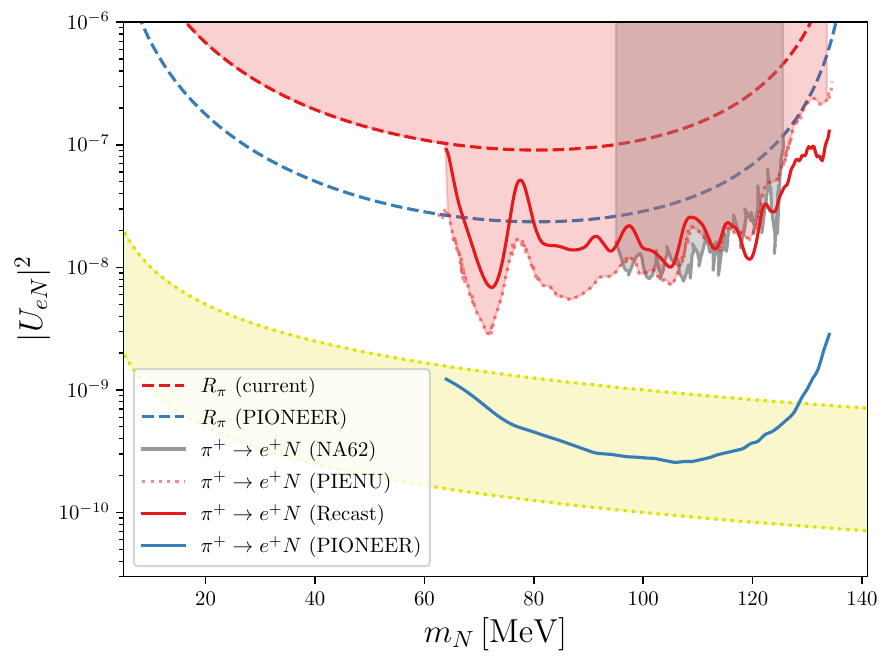} 
\caption{Constraints from pion decays $\pi^+ \to e^+ N$ on the sterile neutrino mixing angle $|U_{e N}|^2$ at 90\% C.L. as a function of the sterile neutrino mass $m_N$. The PIENU result from~\cite{PIENU:2017wbj} is shown by the dotted red line. Our reproduction of the PIENU bound is shown in solid red. The corresponding limit from NA62~\cite{NA62:2025csa} is shown in gray. The bound from the $R_\pi$ measurement is shown in dashed red. The red and gray shaded regions are excluded. The solid blue and the dashed blue curves correspond to our sensitivity estimates of a PIONEER search for sterile neutrinos and the PIONEER $R_\pi$ measurement, respectively. The yellow shaded region indicates a see-saw target (see text for more details). This region can be reached by PIONEER for $64~{\rm{MeV}}\lesssim m_N\lesssim 130$ MeV.}
\label{fig:sterile_bounds}
\end{figure}

From the $\pi^+ \to e^+ N$ analysis, we obtain the 90\% C.L. bound on the sterile neutrino mixing angle with electron neutrinos, $|U_{e N}|^2$, shown in figure~\ref{fig:sterile_bounds}. Our reproduction of the bound is given by the solid red curve as a function of the sterile neutrino mass in the range between 62~MeV and 134~MeV. Our reproduction agrees reasonably well with the bound obtained by the PIENU analysis~\cite{PIENU:2017wbj} shown in dotted red. The bound we find is slightly weaker for low masses and slightly stronger for high masses. 
For comparison, the plot also shows the limit from NA62~\cite{NA62:2025csa} in gray. We also show, as a dashed red curve, the bound we derived from existing measurements of $R_\pi$~\cite{ParticleDataGroup:2022pth} which is in good agreement with the one obtained in~\cite{Bryman:2019bjg}. To find this bound, we assume that the sterile neutrino mixes only with the electron neutrino, and that no additional new physics affects $R_\pi$. For sterile neutrino masses between 64~MeV and 135~MeV, the bump hunt provides constraints that are more stringent by approximately one order of magnitude. 

\begin{figure}[tb]
\centering
\includegraphics[width=0.9\textwidth]{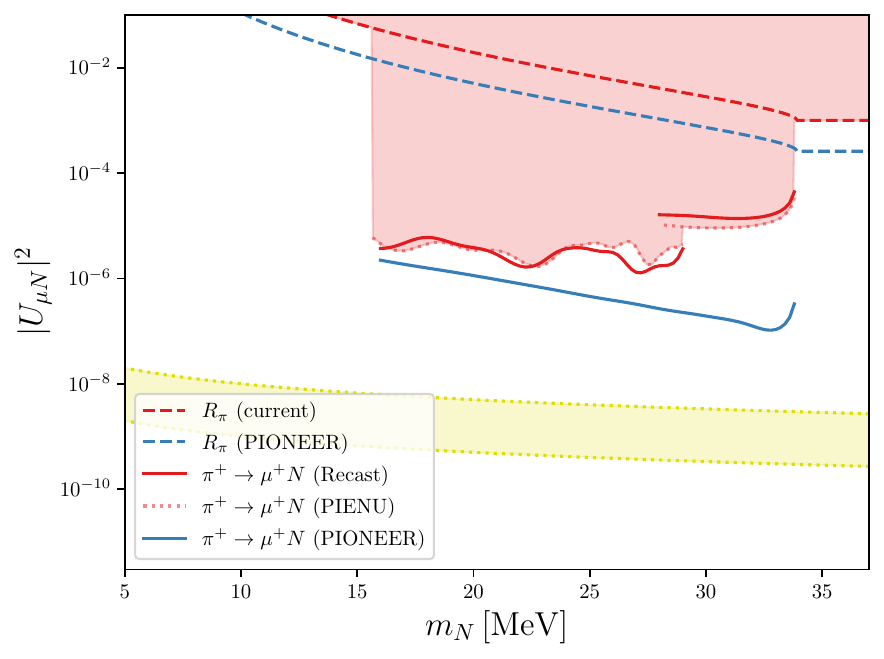} 
\caption{Constraints on the sterile neutrino mixing angle $|U_{\mu N}|^2$ at 90\% C.L. as a function of the sterile neutrino mass $m_N$. The PIENU result from~\cite{PIENU:2019usb} is shown by the dotted red lines. Our reproduction of the PIENU bound is shown in solid red. The bound from the $R_\pi$ measurement is shown in dashed red. The shaded regions are excluded. The solid blue and the dashed blue curves correspond to our sensitivity estimates of a PIONEER search for sterile neutrinos and the PIONEER $R_\pi$ measurement, respectively. The see-saw target (indicated with the yellow band) is out of the PIONEER reach.} 
\label{fig:sterile_bounds_mu}
\end{figure}

Similarly, the limits on $|U_{\mu N}|^2$ that we reproduce from PIENU's $\pi^+ \to \mu^+ N$ analysis are shown in figure~\ref{fig:sterile_bounds_mu}. The solid red curves correspond to our reproduction of the 90\% C.L. bounds from PIENU shown in dotted red~\cite{PIENU:2019usb}. 
Overall, we find remarkable good agreement. The constraint from the $R_\pi$ measurement (assuming no additional new physics than a sterile neutrino mixing with a muon neutrino) is shown by the red dashed curve.

\subsection{Expected sensitivities at PIONEER} \label{sec:sterile_sensitivity_PIONEER} 

After successfully reproducing the limits on the sterile neutrino mixing angles from PIENU, we next estimate the sensitivity that can be expected at the PIONEER experiment.

To assess the sensitivity of PIONEER to sterile neutrinos, we follow the strategy of bump hunts in the charged lepton energy spectrum, similar to the searches performed at PIENU. In particular, we simulate bump searches in the channels $\pi^+ \to e^+ N$ and $\pi^+ \to \mu^+ N$.

\paragraph{\boldmath The $\pi^+ \to e^+ N$ search:} The search for sterile neutrinos in the decay $\pi^+ \to e^+ N$ at PIONEER profits from the increased statistics, the much better suppression of the muon decay backgrounds at small positron energy, and the better energy resolution (see Table \ref{tab:Exp}).

We used simulated data including both inclusive \(\pi^+\rightarrow e^+ \nu_e (\gamma)\) events as well as muon decay in flight events (\(\pi^+\rightarrow \mu^+ \nu_\mu\) where the \(\mu^+\rightarrow e^+\nu_e \bar{\nu}_\mu\) follows before the muon comes to rest, within less than \(\approx 13\)~ps). Preliminary studies suggest that other backgrounds can be suppressed well enough using time differences and the very characteristic \(4\)~MeV energy deposited by a stopping muon along with other selection criteria involving the segmentation of the active target. Therefore, those backgrounds are considered negligible for this study. This assumption will need to be tested by the PIONEER experiment.

\begin{figure}[tb]
\centering
\includegraphics[width=0.6\textwidth]{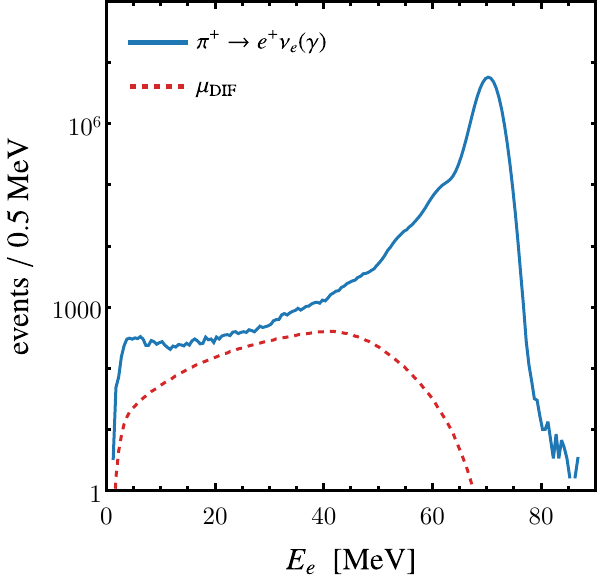} 
\caption{The expected positron energy spectrum at phase~I of PIONEER after background suppression cuts. Shown in blue are $0.5 \times 10^8$ $\pi^+ \to e^+ \nu_e (\gamma)$ events and in red dashed an $8.8\%$ contamination from muon decays in flight, $\mu_{\rm{DIF}}$.} 
\label{fig:PIONEER_espec}
\end{figure}

While the PIONEER experiment requires at least \(2\times10^8\) reconstructed \(\pi^+\rightarrow e^+ \nu_e (\gamma)\) events in an unbiased fashion to measure the branching ratio \(R_{e/\mu}\) to the desired precision, a set of background rejections will be used to measure the \(\pi\rightarrow e \nu \left(\gamma\right)\) line shape. Until more detailed exotic search strategies are developed by the collaboration, it is reasonable to assume that those searches will follow a similar set of background rejections to find features without a corresponding SM explanation in this line shape. Early reconstruction and background suppression studies suggest that a sample with only \(8.8\%\) muon decay in flight contamination in the search region can be achieved while preserving \(0.5\times10^8\) of all \(\pi^+\rightarrow e^+ \nu (\gamma)\) events~\cite{patrick1, patrick2}.
The corresponding positron energy spectra are shown in figure~\ref{fig:PIONEER_espec}. The shapes of both the $\pi^+ \to e^+ \nu_e$ decay and the muon decay in flight background are taken directly from Monte Carlo simulations~\cite{patrick1, patrick2}. 

For the hypothetical sterile neutrino signal, we model the bump by a Gaussian peak whose width is determined by the expected energy resolution of the calorimeter. The energy dependence of the resolution is implemented using a parameterized curve and varies from $\sim 1.5\%$ for a positron energy of $E_e = 70$\,MeV to $\sim 6\%$ for a positron energy of $E_e = 10$\,MeV (see Table \ref{tab:Exp}).
The parameters are obtained using the full simulation chain which convolutes experimental calorimeter crystal responses taken from beam measurements~\cite{Beesley:2024mts}, proper dead material models and early calibration strategies.\footnote{As a robustness check, we have tested an alternative modeling of the signal shape by using the shape of the SM $\pi^+ \to e^+ \nu_e$ peak and rescaling its width according to the relative energy resolution. This approach yields results consistent with the Gaussian modeling, confirming that our sensitivity projections are not very sensitive to the precise choice of signal parameterization.}

The sensitivity of the sterile neutrino search is evaluated through a binned $\chi^2$ fit of the simulated positron energy spectra. We use $0.5$\,MeV wide bins and include all bins in the range $(4-56)$\,MeV, following the approach of the PIENU analysis. The $\chi^2$ function accounts for statistical uncertainties only, based on an assumed total of $0.5 \times 10^8$ reconstructed $\pi^+ \to e^+ \nu_e$ events. In the fit, the overall normalizations of the hypothetical sterile neutrino signal, the $\pi^+ \to e^+ \nu_e$ component, and the muon decay in flight background are allowed to float while their shapes are kept fixed. 

\paragraph{\boldmath The $\pi^+ \to \mu^+ N$ search:} 
We follow a similar approach for the sterile neutrino search in the decay $\pi^+ \to \mu^+ N$ at PIONEER. Our analysis is based on a simulated data set consisting of $\pi^+ \to \mu^+ \nu_\mu$ events that produce muons with a kinetic energy of approximately $4$\,MeV with a tail to lower energies. Additional background components are expected to be negligible as such low-energy muons are highly ionizing and stop within a few strips of the active target. This results in a characteristic Bragg peak in the energy loss rate per unit distance that can be used as selection criteria. 
The $4$\,MeV peak has a width of about $0.2$\,MeV, and we model the sterile neutrino signal as a Gaussian peak of the same fixed width, independent of its position. The $\chi^2$ fit is performed using $0.06$\,MeV wide bins over the energy range $0$\,MeV - $3.2$\,MeV, similar to the PIENU analysis. Only statistical uncertainties on the event counts in each bin are included, assuming a total of $10^{10}$ reconstructed $\pi^+ \to \mu^+ \nu_\mu$ events. 
This is an early order-of-magnitude estimate based on few \(10^{12}\) pion decays required to obtain \(2\times 10^8\) \(\pi\rightarrow e \nu (\gamma)\) events, combined with a prescale factor of \(50\) and the fact that about \(20\%\) of pion decays will satisfy the \(\pi\rightarrow \mu \rightarrow e\) trigger conditions on their own. A so far unknown amount of \(\pi^+\rightarrow \mu^+ \nu_\mu\) events will have their trigger completed by accidental background, adding to the number. In the fit, the normalizations of the $\pi^+ \to \mu^+ N$ signal and the $\pi^+ \to \mu^+ \nu_\mu$ component are allowed to float while their shapes are kept fixed. 

\bigskip
The sensitivity projections that we obtain from the described bump searches for $\pi^+ \to e^+ N$ and $\pi^+ \to \mu^+ N$ are shown by the solid blue curves in figures~\ref{fig:sterile_bounds} and~\ref{fig:sterile_bounds_mu}. Under the assumption of no signal, we show the $90\%$ C.L. upper limits on the mixing angle of the sterile neutrino with the electron neutrino $|U_{e N}|^2$ and the muon neutrino $|U_{\mu N}|^2$.  
The sensitivities of the expected PIONEER $R_\pi$ measurement, see equation~\eqref{eq:Rpi_future_constraint}, are shown in the same plots by the dashed blue curves. We observe that the bump hunt at PIONEER will be able to improve the sensitivity to the squared mixing angles by more than an order of magnitude compared to PIENU. The improvement when using the expected $R_\pi$ measurement is approximately a factor of four, see the upper limits quoted in equations~\eqref{eq:Rpi_current_constraint} and~\eqref{eq:Rpi_future_constraint}. The see-saw target regions are also shown in the plots of figures~\ref{fig:sterile_bounds} and~\ref{fig:sterile_bounds_mu}. Interestingly, the PIONEER Phase~I sensitivity of $\pi^+ \to e^+ N$ reaches into the shown see-saw target region.

Assuming systematic uncertainties can be kept under control, the sensitivities of the bump-hunt analyses scale, to good approximation, with the square root of the total number of events. As mentioned above, we consider $0.5 \times 10^8$ $\pi^+ \to e^+ \nu_e$ events, corresponding approximately to the expected sample size during PIONEER Phase~I. Improved sensitivity to $|U_{e N}|^2$ could be achieved in later phases of PIONEER with larger data sets. For the $\pi^+ \to \mu^+ N$ search, we have assumed $10^{10}$ $\pi^+ \to \mu^+ \nu_\mu$ events. However, already during Phase~I, up to $10^{12}$ $\pi^+ \to \mu^+ \nu_\mu$ decays are expected. If a larger fraction of these events can be recorded and reconstructed, the sensitivity to $|U_{\mu N}|^2$ could be significantly improved already at Phase~I.

\begin{figure}[tb]
\centering
\includegraphics[width=0.9\textwidth]{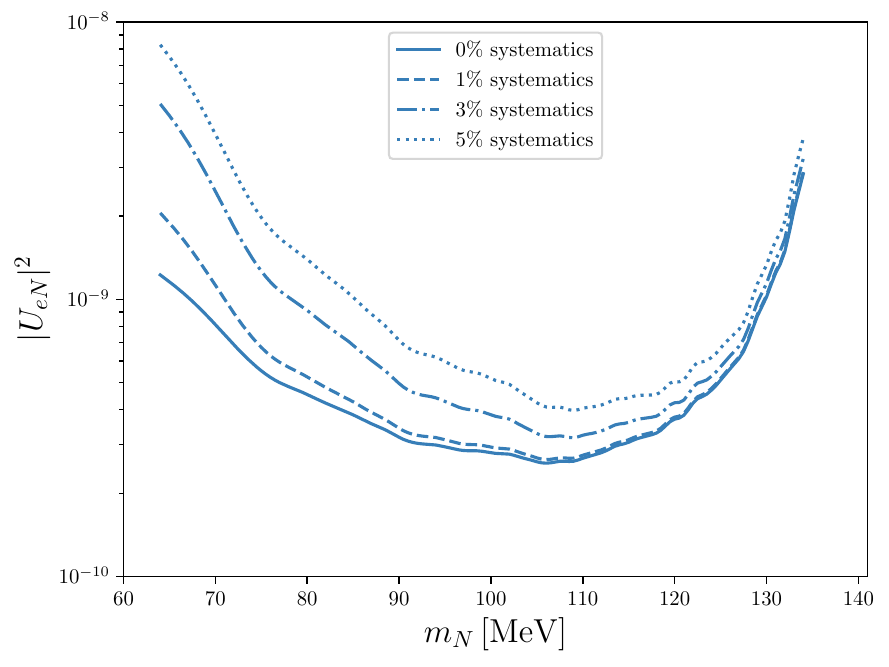} 
\caption{Sensitivity to the sterile neutrino mixing angle $|U_{e N}|^2$ at 90\% C.L. as a function of the sterile neutrino mass $m_N$ from a bump search at PIONEER. The solid line takes into account statistical uncertainties on the event numbers and allows for a floating normalization of background keeping the background shape fixed. The dashed, dot-dashed, and dotted lines include in addition uncorrelated systematic uncertainties of 1\%/3\%/5\% in each positron energy bin.}
\label{fig:sterile_bounds_unc}
\end{figure}

As already observed by PIENU, and expected to be even more relevant for PIONEER, photonuclear effects in the calorimeter leave a characteristic imprint on the tail of the positron energy spectrum in the $\pi^+ \to e^+ \nu_e$ decay, in particular a prominent feature at $E_e \simeq 60$\,MeV, see figure~\ref{fig:PIONEER_espec}. Accurately modeling these effects is challenging. In our default bump-hunt analysis described above we therefore only include data below $E_e \leq 56$\,MeV. The background shape is taken from simulation and kept fixed in the fit, with only its overall normalization allowed to float. To assess the potential impact of imperfect background modeling at lower energies, we introduced additional uncorrelated systematic uncertainties of 1\%, 3\%, and 5\% on the background event counts in each energy bin. This treatment is conservative, as it allows the background yield to fluctuate independently from bin to bin. The corresponding results are shown in figure~\ref{fig:sterile_bounds_unc}, which compares the sensitivity to $|U_{e N}|^2$ from our default analysis with that obtained when including these additional uncertainties. We find that the impact of background shape uncertainties is most significant at lower sterile neutrino masses, corresponding to higher positron energies. In this case, the bound of the analysis that includes a $5\%$ systematics can differ by almost an order of magnitude if compared to the bound with no systematics. This highlights the importance of a precise understanding of the line shape and, consequently, of a careful calibration of the PIONEER calorimeter response, which requires dedicated positron-beam measurements.

\bigskip
Finally, we mention that several other complementary probes can be sensitive to similar regions in sterile neutrino parameter space: 
\begin{itemize}
\item Searches for kaon decays $K \to e N$ and $K \to \mu N$ in~\cite{NA62:2020mcv, NA62:2021bji} by the NA62 collaboration constrain mixing angles at the level of $|U_{e N}|^2 \lesssim 10^{-9}$ and $|U_{\mu N}|^2 \lesssim 10^{-8}$. However, limits are only reported for sterile neutrino masses above the pion mass, outside the region shown in figures~\ref{fig:sterile_bounds} and~\ref{fig:sterile_bounds_mu}.
\item Big bang nucleosynthesis (BBN) gives relevant constraints both on electron and muon mixed sterile neutrinos and can exclude large regions of the parameter space shown in figures~\ref{fig:sterile_bounds} and~\ref{fig:sterile_bounds_mu}~\cite{Boyarsky:2020dzc, Bondarenko:2021cpc}. The BBN constraints can be avoided if the sterile neutrinos have a sufficiently short lifetime $\tau_N \lesssim 0.02\,\text{s} - 0.1$\,s. 
Additional invisible decay modes of the sterile neutrino are required to achieve such a sterile neutrino lifetime in the regions of parameter space that can be covered by PIONEER. Note that not all scenarios with additional invisible decay modes are cosmologically viable, as discussed in~\cite{Deppisch:2024izn, Dev:2025pru}.
\item A single sterile Majorana neutrino that mixes with the electron neutrino is strongly constrained by neutrinoless double beta decay~\cite{Bolton:2019pcu, Dekens:2020ttz, Deppisch:2020ztt, deVries:2024mla}. For masses around 100\,MeV, the constraints are around $|U_{e N}|^2 \lesssim 10^{-9}$. The constraints can be much weaker if one has a pair of Majorana neutrinos with a small mass splitting that form a quasi-Dirac state. If the sterile neutrino is Dirac, there is no constraint from neutrinoless double beta decay.
\item Interesting sensitivity to sterile neutrino parameter space can also be achieved at the DUNE near detector, that can look for the visible decays of sterile neutrinos that are produced from meson decays at the DUNE target station~\cite{Ballett:2019bgd, Berryman:2019dme, Coloma:2020lgy, Breitbach:2021gvv}. In the relevant mass range of minimal models, DUNE can reach a sensitivity that is comparable to that of the PIENU search for $\pi^+ \to e^+ N$, and might be competitive with PIONEER's $\pi^+ \to \mu^+ N$ search. However, in contrast to PIENU and PIONEER, DUNE is not sensitive to short lived sterile neutrinos that avoid the BBN constraints via decays into invisible final states.
\end{itemize}

\section{Exotic Three-Body Decays of Pions} \label{sec:3body} 

In this section we study the three-body pion decays $\pi^+ \to \ell^+ \nu_\ell X$ as probes of light dark sector particles. We first introduce the benchmark models under consideration and describe their expected experimental signatures. We then present a recast of existing PIENU searches and a projection of the sensitivity achievable at PIONEER. Finally, we compare the resulting sensitivities with complementary probes, including anomalous magnetic moments of the electron and muon, monophoton searches at $B$ factories, and beam dump experiments, and discuss the combined implications for the various model scenarios.
In this section, we focus on bounds derived from laboratory-based experiments. We note that observations of SN1987A might also provide complementary constraints on the parameter space, generally at significantly smaller values of the coupling between the new dark particles and leptons (see e.g. \cite{Carenza:2021pcm, Fiorillo:2025sln}).

\subsection{Simplified dark sector models and their pion decay spectra} \label{sec:3body_models} 

One can consider a variety of dark sector states $X$ that can be produced in pion decays. In the following, we will focus on a class of models in which dark spin $0$ or spin $1$ states couple to the charged leptons. More specifically, we consider scalars, axion-like particles, as well as vectors with a variety of couplings. In contrast to the masses of axion-like particles and vectors, masses of scalars are susceptible to large radiative corrections. Light scalars require either a protection mechanism or are fine-tuned.

\paragraph{Scalars:} The first example we consider is a light scalar $s$ interacting with electrons or muons with the following Lagrangian 
\begin{equation} \label{eq:scalar_couplings}
 \mathcal L_\text{scalar} = g_s s (\bar \ell \ell) ~.
\end{equation}
Invariance under the SM gauge group implies that this interaction arises from a dimension 5 operator and is therefore suppressed. A typical expectation would be $g_s \sim m_\ell/\Lambda$ with some high new physics scale $\Lambda$. The coupling could be as large as $g_s \sim v / \Lambda$, with the Higgs vacuum expectation value $v \simeq 246$\,GeV.

The corresponding double differential decay rate of $\pi^+ \to \ell^+ \nu_\ell \,s$ as a function of the lepton energy $E_\ell$ and the scalar energy $E_s$ is given by
\begin{multline} \label{eq:BR_scalar}
\frac{d \text{BR}(\pi^+ \to \ell^+ \nu_\ell \,s )}{\text{BR}(\pi^+ \to \ell^+ \nu_\ell )} = \frac{g_s^2}{4 \pi^2} \frac{dE_\ell dE_s}{m_\pi^2} \\ \times \frac{1}{(1 - x_\ell)^2} \left[ \frac{x_{\ell\nu} x_{\ell s}}{x_\ell (x_{\ell s} - x_\ell)} + \frac{x_{\ell s} ( 3 + x_s - 4 x_{\ell s}) - x_s + x_\ell}{(x_{\ell s} - x_\ell)^2} \right] ~,
\end{multline}
where we used a convenient normalization to the branching ratio of the corresponding two-body decay $\pi^+ \to \ell^+ \nu_\ell$. In the above expression we introduced the following shorthand notation for squared mass ratios
\begin{equation} \label{eq:x_mass}
x_\ell = \frac{m_\ell^2}{m_\pi^2} ~,\quad x_s = \frac{m_s^2}{m_\pi^2} ~, 
\end{equation}
and ratios of squared invariant masses to the squared pion mass
\begin{equation} \label{eq:x_invariant_mass}
x_{\ell \nu} = \frac{1}{m_\pi^2} (p_\ell + p_\nu)^2 = 1 + \frac{m_s^2}{m_\pi^2} - \frac{2 E_s}{m_\pi} ~,\quad x_{\ell s} = \frac{1}{m_\pi^2} (p_\ell + p_s)^2 =\frac{2 E_s}{m_\pi} + \frac{2 E_\ell}{m_\pi} - 1~.
\end{equation}
We note that the branching ratio in equation~\eqref{eq:BR_scalar} contains a term proportional to $1/x_\ell = m_\pi^2/m_\ell^2$, signaling that the three-body decay into a scalar can overcome the helicity suppression of the two-body decay, $\pi^+\to\ell^+\nu_\ell$. 

\paragraph{Axion-like particles:} Next, we consider an axion-like particle (ALP) $a$ with the following derivative interactions
\begin{equation} \label{eq:axion_couplings}
 \mathcal L_\text{ALP} = \frac{\partial_\alpha a}{2 m_\ell} \Big( \bar g_a (\bar \ell \gamma^\alpha \ell) + g_a (\bar \ell \gamma^\alpha \gamma_5 \ell) + g_{a,\nu} (\bar \nu \gamma^\alpha P_L \nu ) \Big)  ~.
\end{equation}
The generic expectation for the ALP couplings is $g_a, \bar g_a, g_{a,\nu} \sim m_\ell/f_a$, with the axion decay constant $f_a$. One also expects a relation due to $SU(2)_L$ invariance, $\bar g_a - g_a - g_{a,\nu} \sim v^2/\Lambda^2 \ll 1$, with a new physics scale $\Lambda$ not necessarily related to $f_a$. We refer to an ALP with $\bar g_a - g_a - g_{a,\nu} \neq 0$ as a ``weak violating ALP'', i.e., an ALP with different couplings to SM neutrinos and left-handed charged leptons (see~\cite{Altmannshofer:2022ckw} for more details and possible UV completions of such an ALP).

The corresponding normalized differential branching ratio for the three-body decay $\pi^+ \to \ell^+ \nu_\ell \,a$ is
\begin{multline} \label{eq:axion_shape}
\frac{d \text{BR}(\pi^+ \to \ell^+ \nu_\ell \,a )}{\text{BR}(\pi^+ \to \ell^+ \nu_\ell )} = \frac{1}{4 \pi^2} \frac{dE_\ell dE_a}{m_\pi^2} \\ \times \frac{1}{x_\ell (1 - x_\ell)^2} \bigg[ g_a^2~ \frac{x_{\ell a} x_a (x_{\ell a} - 1)}{(x_{\ell a} - x_\ell)^2} + g_a (\bar g_a - g_{a,\nu}) \frac{x_{\ell a} (x_{\ell \nu} - x_a) + x_a - x_\ell}{x_{\ell a} - x_\ell} \\
+ (g_a - \bar g_a + g_{a,\nu})^2 \frac{1}{4 x_\ell} \big( x_{\ell a} (x_{\nu a} - x_\ell) - x_a + x_\ell \big) \bigg] ~,
\end{multline}
where $x_a$, $x_{\ell\nu}$, and $x_{\ell a}$ are defined analogously to equations~\eqref{eq:x_mass} and~\eqref{eq:x_invariant_mass}, and we have in addition
\begin{equation} \label{eq:x_invariant_mass_2}
x_{\nu a} = \frac{1}{m_\pi^2} (p_\nu + p_a)^2 = 1 + \frac{m_\ell^2}{m_\pi^2} - \frac{2 E_\ell}{m_\pi}~.
\end{equation}
The $\pi^+ \to \ell^+ \nu_\ell \,a$ branching ratio overcomes the helicity suppression if the ALP is weak violating (see the last term in equation (\ref{eq:axion_shape})).

\paragraph{Vectors:} We also consider a light vector $V$ with the following set of couplings to the SM leptons
\begin{multline} \label{eq:vector_couplings}
 \mathcal L_\text{vector} = V_\alpha \Big( g_V (\bar \ell \gamma^\alpha \ell) + g_A (\bar \ell \gamma^\alpha \gamma_5 \ell) + (g_V - g_A) (\bar \nu \gamma^\alpha P_L \nu ) \Big) \\
 + \frac{V_{\alpha\beta}}{\Lambda} \Big( g_T (\bar \ell \sigma^{\alpha \beta} \ell) + g_{T5} (\bar \ell \sigma^{\alpha\beta} i \gamma_5 \ell)  \Big) ~.
\end{multline}
The vector coupling $g_V$ could for example arise if the vector $V$ is the gauge boson of an abelian gauge symmetry that includes electron or muon number. The axial vector coupling $g_A$ (or combinations of $g_A$ and $g_V$) can for example be engineered in the effective $Z^\prime$ construction from~\cite{Fox:2011qd}.
Note that we have fixed the coupling of neutrinos using $SU(2)_L$ gauge invariance. Exploring ``weak violating vectors'' is beyond the scope of this work.

The second line of the above Langrangian shows dipole interactions with the field strength tensor $V_{\alpha\beta} = \partial_\alpha V_\beta - \partial_\beta V_\alpha$. It is plausible that such dipole interactions are loop suppressed. Due to $SU(2)_L$ invariance, we expect $g_T, g_{T5} \lesssim v/(16 \pi^2 \Lambda)$, with a typical expectation of $g_T, g_{T5} \sim m_\ell/(16 \pi^2 \Lambda)$. The simultaneous presence of both $g_T$ and $g_{T5}$ leads to 1-loop contributions of electric dipole moments, which are very strongly constrained. In our numerical analysis, we will therefore consider only one of them at a time.  

Taking into account only the vector and axial vector couplings, we find the following expression for the $\pi^+ \to \ell^+ \nu_\ell \,V $ differential branching ratio
\begin{multline} \label{eq:VA_shape}
\frac{d \text{BR}(\pi^+ \to \ell^+ \nu_\ell \,V )}{\text{BR}(\pi^+ \to \ell^+ \nu_\ell )} = \frac{1}{2 \pi^2} \frac{dE_\ell dE_V}{m_\pi^2} \frac{1}{(1 - x_\ell)^2} \bigg[ g_V^2 \bigg( 2 + \frac{x_{\ell V}-2 + x_\ell}{x_{\nu V}} \\ 
- \frac{x_V(1-x_\ell)}{x_{\nu V}^2} + \frac{2(1-x_\ell)^2 + 2 x_\ell x_V}{(x_{\ell V} - x_\ell) x_{\nu V}} - \frac{2(1-x_\ell) - x_{\nu V}}{x_{\ell V} - x_\ell} - \frac{(2 x_\ell + x_V)(1-x_\ell)}{(x_{\ell V} - x_\ell)^2} \bigg) \\ + g_A^2 \bigg( \frac{x_{\ell V} + 2 - 3 x_\ell}{x_{\nu V}} - \frac{2(x_{\ell V} - 1 + x_{\nu V})}{x_V} - \frac{(1- x_\ell) x_V}{x_{\nu V}^2} + \frac{2 x_\ell}{x_{\ell V} - x_\ell} \Big( \frac{x_V}{x_{\nu V}} - \frac{x_{\nu V}}{x_V} \Big) \\ - \frac{2(1- x_\ell)^2}{(x_{\ell V} - x_\ell) x_{\nu V}} + \frac{2 + x_{\nu V} - 6 x_\ell}{x_{\ell V} - x_\ell} - \frac{(1-x_\ell)(x_V - 4 x_\ell)}{(x_{\ell V} - x_\ell)^2} \bigg)\\  - 2 g_V g_A \bigg( \frac{x_{\ell V} - x_\ell}{x_{\nu V}} - \frac{x_V(1 - x_\ell)}{x_{\nu V}^2} - \frac{x_{\nu V}+ 2x_\ell}{x_{\ell V}- x_\ell} + \frac{2 x_\ell x_V}{(x_{\ell V}- x_\ell)x_{\nu V}} \\ + \frac{x_V(1 + x_\ell) - 2 x_\ell x_{\nu V}}{(x_{\ell V}- x_\ell)^2}\bigg) \bigg] ~.
\end{multline}
Interestingly, this three-body decay features the same helicity suppression as the two-body decay $\pi^+ \to \ell^+ \nu_\ell$. 

If instead we only consider the dipole interactions, we find
\begin{multline} \label{eq:T_shape}
\frac{d \text{BR}(\pi^+ \to \ell^+ \nu_\ell \,V )}{\text{BR}(\pi^+ \to \ell^+ \nu_\ell )} = \frac{1}{\pi^2} \frac{dE_\ell dE_V}{m_\pi^2} \frac{m_\pi^2}{\Lambda^2} \frac{1}{(1 - x_\ell)^2} \\
\times  \bigg[ g_T^2 \bigg( 2 + 8 x_V -2x_{\ell V}
- \frac{x_V ( 8 - 16 x_\ell - 2x_V + x_{\nu V})}{x_{\ell V} - x_\ell} - \frac{(1-x_\ell)x_V (8 x_\ell + x_V)}{(x_{\ell V} - x_\ell)^2} \bigg) \\ + g_{T5}^2 \bigg(2 - 4 x_V 
-2 x_{\ell V} +\frac{x_V ( 4 - 8 x_\ell + 2x_V - x_{\nu V})}{x_{\ell V} - x_\ell} 
+ \frac{(1 - x_\ell) x_V(4 x_\ell-x_V)}{(x_{\ell V} - x_\ell)^2}  \bigg) \\ + (g_T^2 + g_{T5}^2) \frac{1}{x_\ell} \Big( x_{\ell V} ( 2 x_{\nu V} - x_V ) - x_V ( 1 + x_{\nu V} - x_V  ) \Big) \bigg] ~.
\end{multline}
Both the $g_T$ and the $g_{T5}$ couplings provide a term enhanced by $\sim m_\pi^2/m_\ell^2$. 

\bigskip
In all the cases discussed in this section, the phase space boundaries of the lepton energy $E_\ell$ and the energy of the dark sector particle $E_X$ with $X = s, a , V$, are given by the expressions
\begin{equation}
E_- < E_X < E_+ ~, \quad m_\ell < E_\ell < \frac{m_\pi}{2} \left( 1 - x_X + x_\ell \right) ~, 
\end{equation} 
\begin{equation}
E_\pm = \frac{m_\pi}{2} \Bigg[1 - \frac{E_\ell}{m_\pi} \pm \sqrt{\frac{E_\ell^2}{m_\pi^2} - x_\ell} + \frac{x_X \left(1 - E_\ell/m_\pi \mp \sqrt{E_\ell^2/m_\pi^2 - x_\ell}\right)}{1 + x_\ell - 2 E_\ell/m_\pi} \Bigg] ~.
\end{equation} 

\begin{figure}[!tb]
\centering
\includegraphics[width=0.85\textwidth]{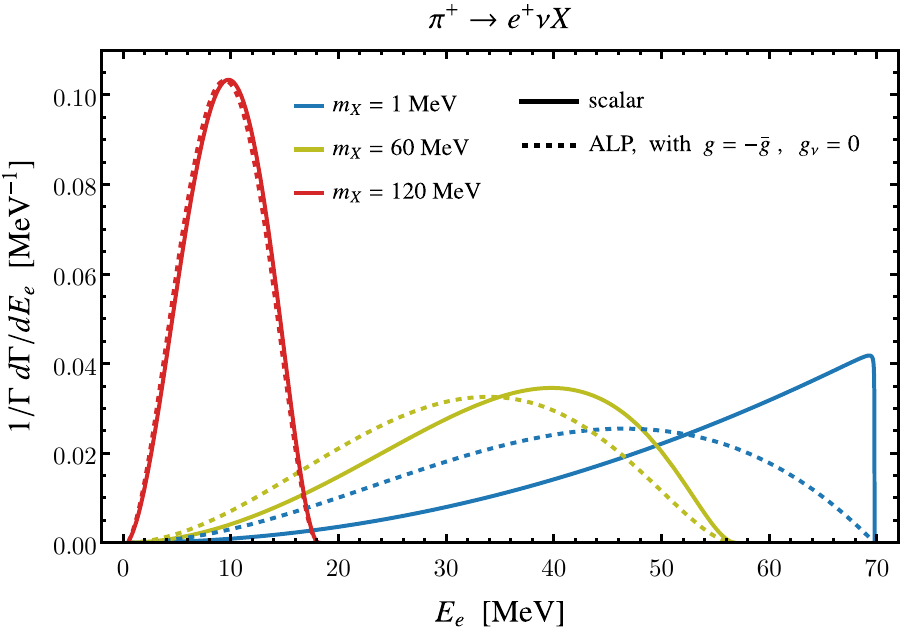} \\[16pt]
\includegraphics[width=0.85\textwidth]{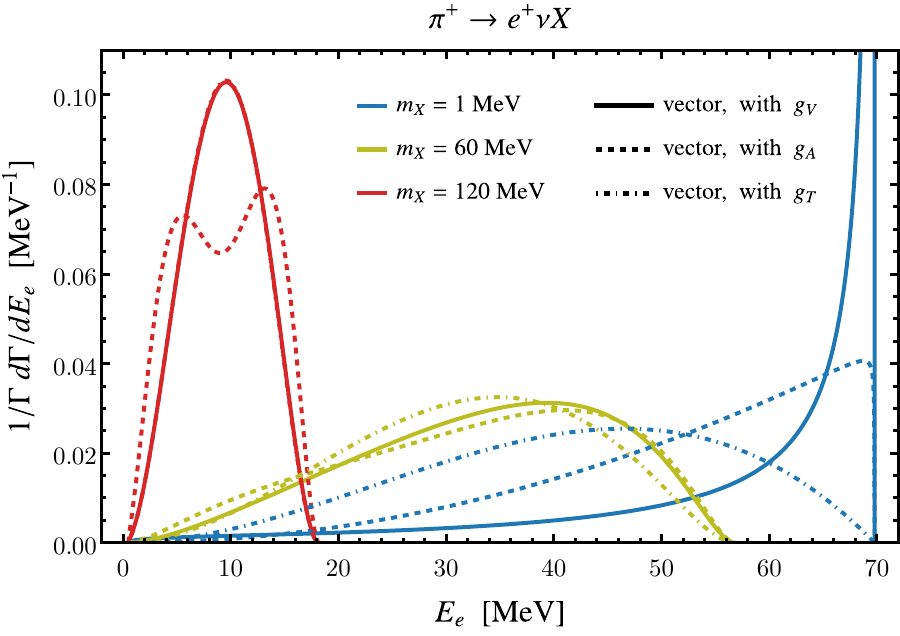} 
\caption{Representative examples of normalized signal shapes of the $\pi^+ \to e^+ \nu_e X$ decay as a function of the positron energy. Top: scalar and ALP models; bottom: various vector models. For a mass of $m_X = 120$\,MeV, the spectra of the vector models with $g_V$ coupling and $g_T$ coupling are almost indistinguishable. The spectra of the vector model with $g_{T5}$ coupling look virtually identical to the one with $g_T$ coupling and are therefore not shown.} 
\label{fig:X_spectrum_e}
\end{figure}

\begin{figure}[!tb]
\centering
\includegraphics[width=0.85\textwidth]{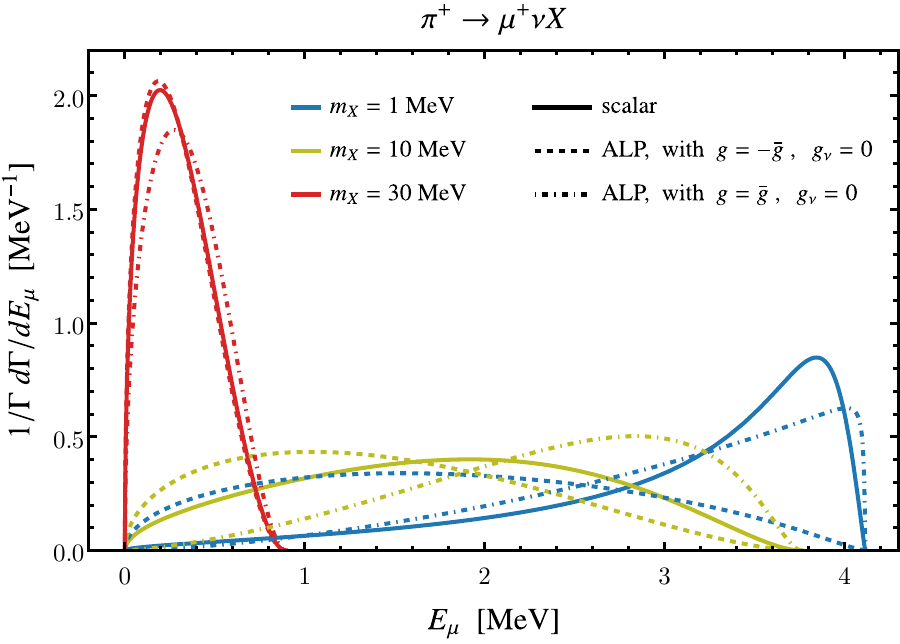} \\[16pt]
\includegraphics[width=0.85\textwidth]{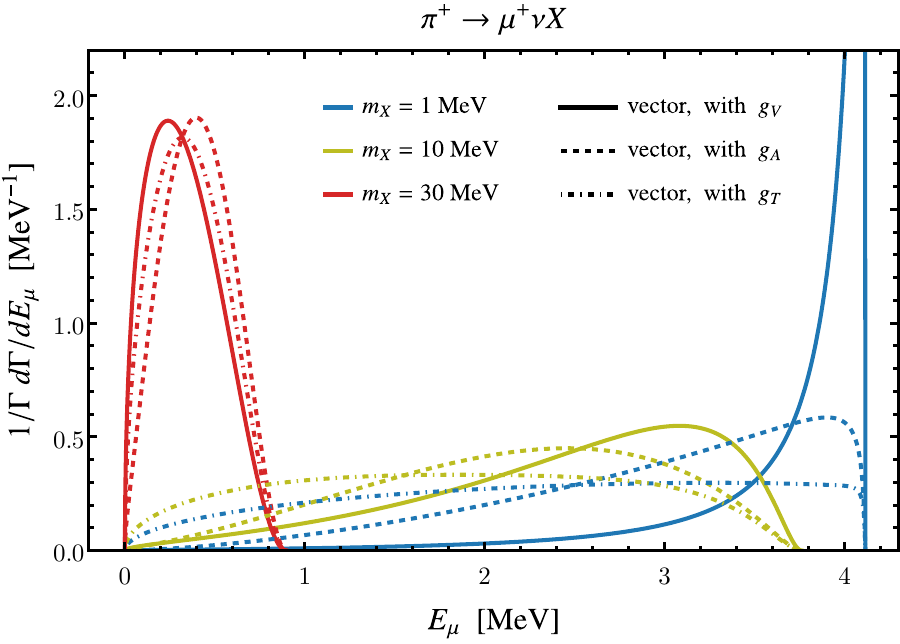} 
\caption{Representative examples of normalized signal shapes of the $\pi^+ \to \mu^+ \nu_\mu X$ decay as a function of the muon kinetic energy. Top: scalar and ALP models; bottom: various vector models.}
\label{fig:X_spectrum_mu}
\end{figure}

In figures~\ref{fig:X_spectrum_e} and~\ref{fig:X_spectrum_mu} we show example spectra for the decays $\pi^+ \to e^+ \nu_e X$ and $\pi^+ \to \mu^+ \nu_\mu X$, for scalars, ALPs, and vectors with representative choices of couplings and various choices of $X$ masses. Each scenario gives its own characteristic signal shape. 


\subsection{Experimental signatures at pion decay experiments} \label{sec:3body_signatures} 

There are several types of experimental signatures of the $\pi^+ \to \ell^+ \nu_\ell X$ decays, depending on the lifetime and the decay modes of $X$. The most relevant are 
\begin{equation}
\pi^+ \to \ell^+ \nu_\ell + \text{invisible} ~,\quad \pi^+ \to \ell^+ \nu_\ell + \gamma\gamma ~,\quad \pi^+ \to \ell^+ \nu_\ell + e^+ e^- ~. 
\end{equation}
The photons and the electron-positron pair could be prompt or displaced. As already mentioned in section~\ref{sec:exp}, in this work we focus on the signature $\pi^+ \to \ell^+ \nu_\ell + \text{invisible}$ which resembles most closely the SM decay $\pi^+ \to \ell^+ \nu_\ell$ but features a non-standard continuous lepton energy spectrum. The shape of the lepton energy spectrum can be predicted in a given model and it depends on the mass of $X$, its spin, and the type of coupling to the SM particles. Searches for a non-standard component in the lepton energy spectrum from $\pi^+ \to \ell^+ \nu_\ell X$ decays are therefore necessarily model-dependent. 

Without directly using any kinematic information, we can still bound the branching ratios into light exotic states $X$, by making use of the $R_\pi$ measurement. Assuming that the only new physics effects are the $\pi^+ \to e^+ \nu_e \,X$ or $\pi^+ \to \mu^+ \nu_\mu \, X$ decays, the leading new physics correction to $R_\pi$ is given by
\begin{equation} 
\frac{R_\pi}{R_\pi^\text{SM}} \simeq 1 + r_e \times \frac{\text{BR}(\pi^+ \to e^+ \nu_e \,X)}{\text{BR}(\pi^+ \to e^+ \nu_e)_\text{SM}} - r_\mu \times \frac{\text{BR}(\pi^+ \to \mu^+ \nu_\mu \,X)}{\text{BR}(\pi^+ \to \mu^+ \nu_\mu)_\text{SM}} ~,
\end{equation}
where $r_e$ and $r_\mu$ are efficiency ratios due to the different kinematics of the three-body decays into the dark sector state as compared to the two-body SM decays. We will assume $r_e = r_\mu = 1$.

\subsection{Recast of existing pion decay constraints} \label{sec:3body_recast} 

The PIENU experiment has performed a search for the three-body decay $\pi^+ \to \ell^+ \nu_\ell X$ with an invisible dark sector particle $X$, as reported in~\cite{PIENU:2021clt}. The signal shape used in their analysis is based on the model from~\cite{Batell:2017cmf}, and closely resembles the contribution from a scalar or a weak-preserving axion-like particle, as shown in equations~\eqref{eq:BR_scalar} and~\eqref{eq:axion_shape}. PIENU set upper limits on the branching ratio $\text{BR}(\pi^+ \to e^+ \nu_e X)$ at the level of $10^{-7}$ to $10^{-6}$ for dark sector masses $m_X \lesssim 120$\,MeV. For decays involving muons, $\pi^+ \to \mu^+ \nu_\mu X$, the corresponding branching ratio limits reach approximately $10^{-5}$ for masses $m_X \lesssim 35$\,MeV.

\begin{figure}[tb]
\centering
\includegraphics[width=0.47\textwidth]{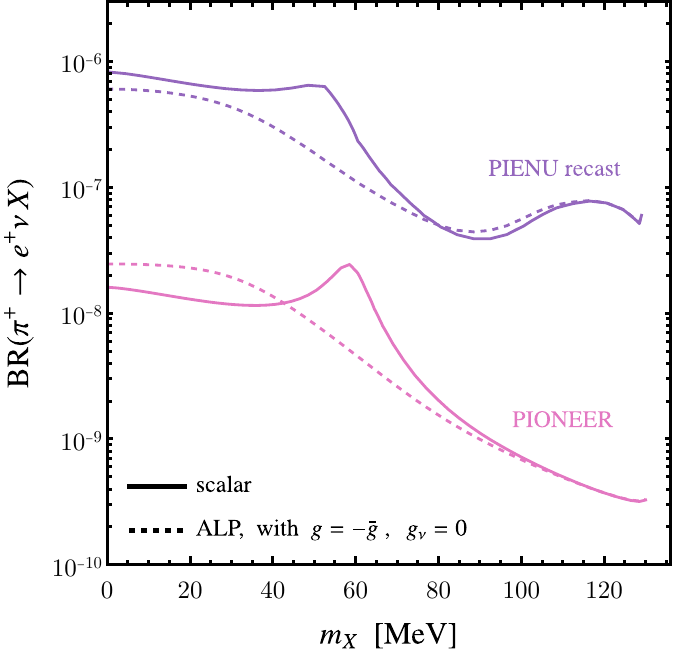} \quad
\includegraphics[width=0.47\textwidth]{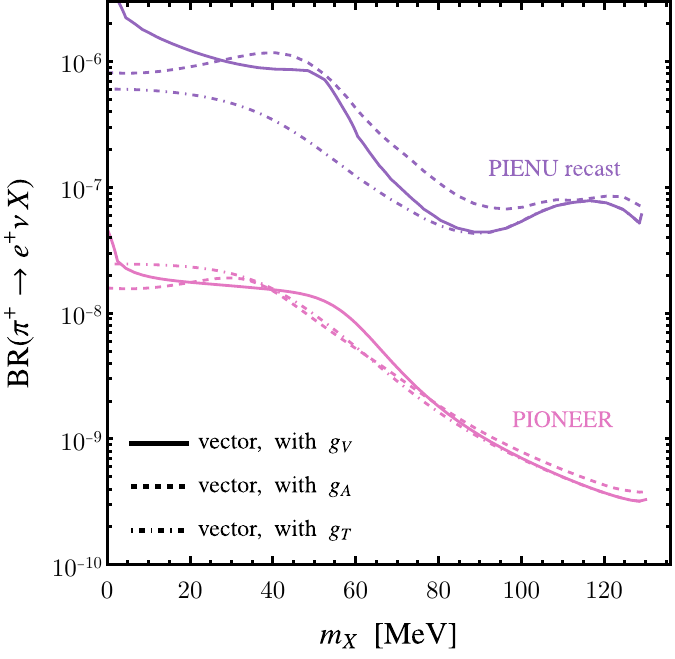} 
\caption{Recasted PIENU limits and expected PIONEER sensitivities to the $\pi^+ \to e^+ \nu_e X$ branching ratio for various signal models as described in section~\ref{sec:3body_models}. The PIONEER sensitivity assumes $0.5 \times 10^8$ recorded $\pi^+ \to e^+ \nu_e$ events.} 
\label{fig:3body_BR_limits_e}
\end{figure}

\begin{figure}[tb]
\centering
\includegraphics[width=0.47\textwidth]{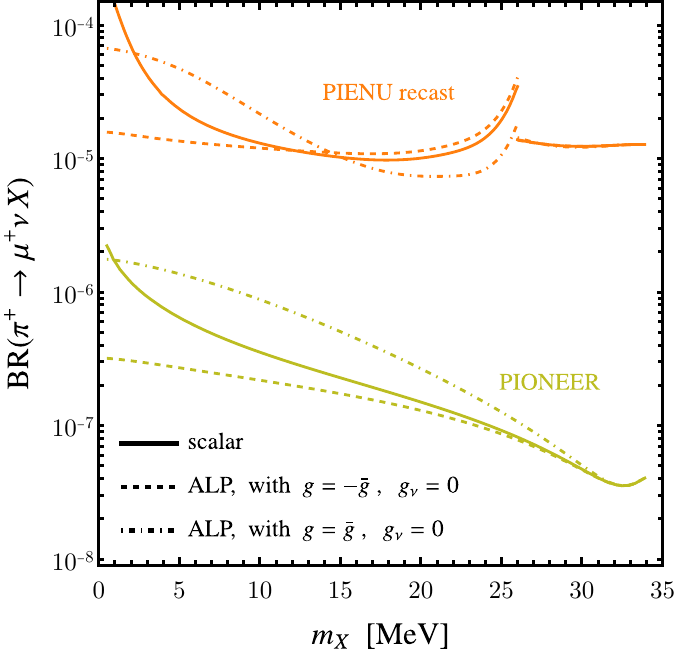} \quad
\includegraphics[width=0.47\textwidth]{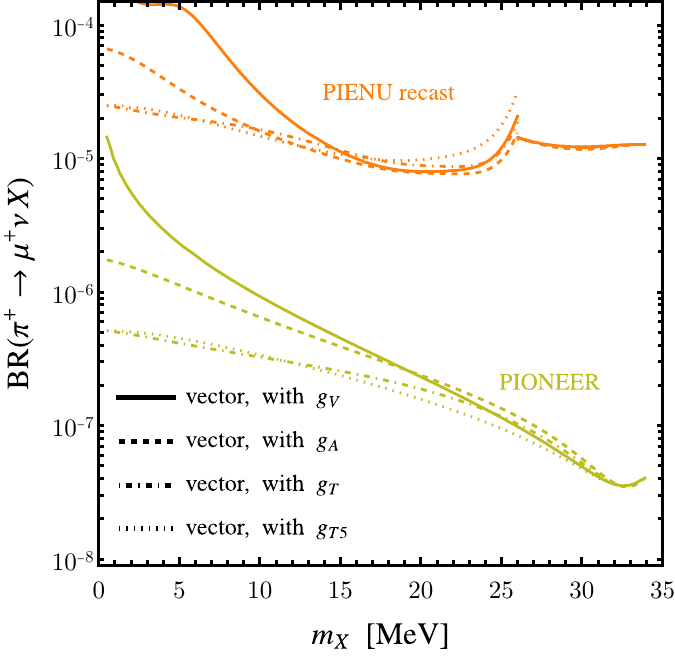}
\caption{Recasted PIENU limits and expected PIONEER sensitivities to the $\pi^+ \to \mu^+ \nu_\mu X$ branching ratio for various signal models as described in section~\ref{sec:3body_models}. The PIONEER sensitivity assumes $10^{10}$ recorded $\pi^+ \to \mu^+ \nu_\mu$ events.} 
\label{fig:3body_BR_limits_mu}
\end{figure}

As described in detail in appendix~\ref{app:recast2} we were able to successfully reproduce the PIENU limits on the $\text{BR}(\pi^+ \to e^+ \nu_e X)$ and $\text{BR}(\pi^+ \to \mu^+ \nu_\mu X)$ branching ratios using the signal model from~\cite{Batell:2017cmf}. 
As a next step, we recast the searches in the context of all the alternative models considered in our study. We substitute the signal shape from~\cite{Batell:2017cmf} with the predicted spectra of other models from equations~\eqref{eq:BR_scalar}, \eqref{eq:axion_shape}, \eqref{eq:VA_shape}, and~\eqref{eq:T_shape}, and apply the same analysis pipeline. We obtain the branching ratio limits as a function of the $X$ mass as shown in figures~\ref{fig:3body_BR_limits_e} and~\ref{fig:3body_BR_limits_mu}. 

The limits on the $\pi^+ \to e^+ \nu_e X$ branching ratio range from approximately $10^{-6}$ at low masses $m_X \ll m_\pi$ to approximately $10^{-7}$ for the largest masses $m_X \simeq m_\pi$. Similarly, we find limits on the $\pi^+ \to \mu^+ \nu_\mu X$ branching ratio that vary from $\sim 10^{-4}$ to $\sim 10^{-5}$ depending on the mass of $X$. 
The precise shape of the exclusion curves are model dependent.
In the case of the decay $\pi^+ \to e^+ \nu_e X$ we do not show the weak preserving axion-like particle and the vector with $g_{T5}$ coupling. The corresponding curves are almost identical to the scalar and vector with $g_T$ coupling, respectively. The relative differences in the respective differential decay rates are indeed proportional to $m_e^2/m_\pi^2$ and therefore tiny. In the case of the $\pi^+ \to \mu^+ \nu_\mu X$ decay, the curves of all models differ visibly.


As we discuss in section~\ref{sec:3body_results} below, these branching ratio limits can be translated into constraints on the underlying coupling constants of the respective models, providing a model-specific interpretation of the experimental sensitivity.

\subsection{Expected sensitivities at PIONEER} \label{sec:3body_sensitivity} 

Having validated our analysis framework through a successful reproduction of the PIENU search results for $\pi^+ \to \ell^+ \nu_\ell X$ decays, we now turn to estimating the future sensitivity of PIONEER. To assess PIONEER's discovery potential, we analyze the same set of benchmark models and use simulated PIONEER data and an analysis pipeline closely mirroring that employed for the PIENU recast. In the following, we discuss the projected sensitivities for the $\pi^+ \to e^+ \nu_e X$ and $\pi^+ \to \mu^+ \nu_\mu X$ channels.

\paragraph{\boldmath The $\pi^+ \to e^+ \nu_e X$ search:} Our sensitivity study of $\pi^+ \to e^+ \nu_e X$ is based on the same simulated dataset and analysis framework as the sterile neutrino search described in section~\ref{sec:sterile_sensitivity_PIONEER}. The simulation includes $0.5 \times 10^8$ reconstructed $\pi^+ \to e^+ \nu_e$ events and a subdominant component of approximately $8.8\%$ from muon decay in flight. Other backgrounds are negligible.  

The main difference from the sterile neutrino analysis lies in the modeling of the signal. Instead of a narrow peak in the positron energy spectrum, the decay $\pi^+ \to e^+ \nu_e X$ leads to a continuous positron energy distribution that depends on the mass, spin, and couplings of the dark particle $X$. For each model under consideration, the theoretical positron spectrum is convolved with a Gaussian resolution function, using the same energy-dependent width as in the sterile neutrino study.

Sensitivities are obtained from a binned $\chi^2$ fit of the simulated positron energy spectrum using 0.5\,MeV bins in the range $(4-56)$\,MeV. The normalizations of the $\pi^+ \to e^+ \nu_e X$ signal, as well as the $\pi^+ \to e^+ \nu_e$ and muon decay in flight components are allowed to float while their shapes are kept fixed. Assuming no signal, we derive projected 90\%~C.L. upper limits on $\text{BR}(\pi^+ \to e^+ \nu_e X)$ for all the benchmark models discussed above.

\paragraph{\boldmath The $\pi^+ \to \mu^+ \nu_\mu X$ search:} The sensitivity to the decay $\pi^+ \to \mu^+ \nu_\mu X$ is evaluated using simulated $\pi^+ \to \mu^+ \nu_\mu$ events corresponding to $10^{10}$ reconstructed decays, with negligible background contributions. 
The dark sector signal is a three-body spectrum convolved with a Gaussian resolution function, analogous to the treatment of $\pi^+ \to e^+ \nu_e X$.
The width of the Gaussian is fixed to $0.2$\,MeV. 
A binned $\chi^2$ fit is performed over the muon kinetic energy range $0$–$3.2$\,MeV with 0.06\,MeV bins. The normalizations of the signal and the standard $\pi^+ \to \mu^+ \nu_\mu$ component are allowed to float, while their shapes are fixed. Under the assumption of no signal, we derive projected 90\% C.L. upper limits on $\text{BR}(\pi^+ \to \mu^+ \nu_\mu X)$ for the benchmark models introduced above.

\bigskip
The PIONEER sensitivities to the $\pi^+ \to e^+ \nu_e X$ and $\pi^+ \to \mu^+ \nu_\mu X$ branching ratios are shown in figures~\ref{fig:3body_BR_limits_e} and~\ref{fig:3body_BR_limits_mu} and compared to our recast of the existing PIENU limits. We find that PIONEER will be able to improve over PIENU by one to two orders of magnitude. For $X$ masses around $100$\,MeV, $\pi^+ \to e^+ \nu_e X$ branching ratios below $10^{-9}$ are in reach. In the case of  $\pi^+ \to \mu^+ \nu_\mu X$, PIONEER will be able to probe branching ratios as low as few $\times ~10^{-8}$.

To a good approximation, the sensitivity to the $\pi^+ \to \ell^+ \nu_\ell X$ branching ratios scales with the square root of events. 
Therefore, further improved sensitivity to the branching ratios can be expected in later phases of PIONEER with larger data sets. Already during Phase~I, a better sensitivity to $\pi^+ \to \mu^+ \nu_\mu X$ could be achieved if more than $10^{10}$ $\pi^+ \to \mu^+ \nu$ decays can be recorded and reconstructed.

\subsection{Comparison to other dark sector searches} \label{sec:comparison} 

The dark sector scenarios introduced in section~\ref{sec:3body_models} are subject to various additional constraints beyond those that can be found from pion decays. The most relevant constraints in the mass range that we are considering are the anomalous magnetic moments of the electron and the muon, mono-photon searches at the $B$ factories, and searches at electron beam dumps. Also the rare kaon decays $K \to \pi + \text{invisible}$ give constraints, which are, however, to some extent model dependent. 

\subsubsection{Anomalous magnetic moments} \label{sec:g-2}

On the experimental side, the anomalous magnetic moments of the electron and the muon are measured with extremely high precision. 
In the case of the electron the most precise determination is given in~\cite{Fan:2022eto}. For the muon we use the world average provided by the Fermilab muon g-2 collaboration~\cite{Muong-2:2025xyk}.
Comparing with SM predictions~\cite{Aliberti:2025beg}, one finds
\begin{eqnarray}
\Delta a_e = a_e^\text{exp} - a_e^\text{SM} &=& \begin{cases} (-10.0 \pm 2.6) \times 10^{-13}~, \quad & \text{for} ~\alpha_\text{em} ~\text{from Cs~\cite{Parker:2018vye}~,} \\ (+3.5 \pm 1.6) \times 10^{-13}~,
\quad & \text{for} ~\alpha_\text{em} ~\text{from Rb~\cite{Morel:2020dww}~,}\end{cases} \\
\Delta a_\mu = a_\mu^\text{exp} - a_\mu^\text{SM} &=& (38.5 \pm 63.7)\times 10^{-11}~\text{\cite{Aliberti:2025beg, Muong-2:2025xyk}} ~.
\end{eqnarray}
For the Standard Model prediction of the electron $g-2$, the latest updates on the 5-loop QED contribution are included~\cite{Aoyama:2019ryr, Volkov:2024yzc, Aoyama:2024aly} and we quote two values corresponding to the two most precise determination of the fine structure constant from Rubidium atom recoil measurements~\cite{Morel:2020dww} and Cesium atom recoil measurements~\cite{Parker:2018vye}, respectively. For the muon, we take the latest SM prediction by the Muon g-2 Theory
Initiative~\cite{Aliberti:2025beg} which is based on lattice determinations of the hadronic vacuum polarization contribution and which updates the result in their first white paper~\cite{Aoyama:2020ynm}.

In the case of the electron, in view of the large discrepancy of the fine structure constant determinations using Rubidium and Cesium, we impose the $2\sigma$ upper bound from Rb and the $2\sigma$ lower bound from Cs. For the muon we use the $2\sigma$ bound. In our numerical analysis, possible new physics contributions are thus bounded by
\begin{equation}
-15.2 \times 10^{-13} < \Delta a_e < 6.7 \times 10^{-13} ~, \quad  -0.89 \times 10^{-9} < \Delta a_\mu < 1.66 \times 10^{-9} ~.
\end{equation}

Next, we discuss the new physics contributions to the anomalous magnetic moments in the scenarios that we consider.
We calculated the contributions of the scalar, the ALP, and the vector and find the following expressions
\begin{eqnarray}
\Delta a_\ell^s &=& \frac{g_s^2}{16\pi^2} \Big( 3 - 2 z_s + z_s (z_s - 3) \log z_s - 2 (z_s - 1) (z_s - 4) \Phi(z_s) \Big) ~, \label{eq:als} \\[8pt]
\Delta a_\ell^a &=& -\frac{g_a^2}{16\pi^2} \Big( 1 + 2 z_a - z_a (z_a - 1) \log z_a + 2 z_a (z_a - 3) \Phi(z_a) \Big) ~, \label{eq:ala} \\[8pt]
\Delta a_\ell^V &=& \frac{g_V^2}{8\pi^2} \Big( 1 - 2 z_V + z_V (z_V - 2) \log z_V - 2 ( 2 + (z_V - 4) z_V) \Phi(z_V) \Big) \nonumber \\
&& \qquad\qquad - \frac{g_A^2}{8\pi^2} \bigg( \frac{2}{z_V} - 5 + 2 z_V - ( 2 + z_V (z_V - 4) ) \log z_V \nonumber \\
&& \qquad\qquad\qquad\qquad\qquad\qquad\qquad + 2 ( z_V - 2)(z_V - 4) \Phi(z_V) \bigg) ~,  \label{eq:alV} 
\end{eqnarray}
with the function
\begin{equation}
 \Phi(z) = \begin{cases} \frac{1}{\sqrt{1 - 4/z}} ~\log \left(\frac{\sqrt{z}+ \sqrt{z-4}}{2} \right) ~, \quad & \text{for} ~~ z > 4 ~, \\ 
 \frac{1}{\sqrt{4/z - 1}} ~\arccos \left(\frac{\sqrt{z}}{2} \right) ~, \quad &  \text{for} ~~z < 4 ~,\end{cases}
\end{equation}
and the mass ratios $z_X = m_X^2/m_\ell^2$ with $X = s,a,V$. The term $\propto 1/z_V$ is unique to the axial-vector scenario.

In the above expression for the vector contributions we have only considered vector and axial vector couplings $g_V, g_A$. If the vector boson instead couples to leptons through the dipole interactions $g_T, g_{T5}$ we find
\begin{eqnarray}
\Delta a_\ell^V &=& \frac{g_T^2}{4\pi^2} \frac{m_\ell^2}{\Lambda^2} \bigg( 6 \log\left(\frac{\Lambda^2}{m_\ell^2}\right) + z_V (9 + 2 z_V) + z_V (8 - z_V (z_V+3)) \log z_V \nonumber \\
&& \qquad\qquad\qquad\qquad\qquad\qquad\qquad  - 2 z_V ( 16 - (z_V +1) z_V) \Phi(z_V) \bigg)  \nonumber\\
&& \qquad\qquad -\frac{g_{T5}^2}{4\pi^2} \frac{m_\ell^2}{\Lambda^2} \bigg( 2 \log\left(\frac{\Lambda^2}{m_\ell^2}\right) - z_V(3 - 2z_V) + z_V^2( 3- z_V ) \log z_V \nonumber \\
&& \qquad\qquad\qquad\qquad\qquad\qquad\qquad + 2 z_V ( z_V - 1)(z_V - 4) \Phi(z_V) \bigg) ~.  \label{eq:alT}
\end{eqnarray}
To the best of our knowledge, this is a new result. The contributions are logarithmically divergent and we have set the renormalization scale equal to the new physics scale $\Lambda$ that parameterizes the size of the dipole coupling. The large logarithms could be re-summed using rernormalization group equations, but this is beyond the scope of our work. We note that in the presence of both $g_T$ and $g_{T5}$ the vector contributes to electric dipole moments of leptons which are strongly constrained. We do not consider such a scenario in the following.

For electrons there is a large region of parameter space for which it is a good approximation to expand in $m_e^2 \ll m_X^2$. Doing so, one finds
\begin{eqnarray}
\Delta a_e^s &\simeq& \frac{1}{8\pi^2} \frac{m_e^2}{m_s^2}~g_s^2\left[ \log\left(\frac{m_s^2}{m_e^2}\right) - \frac{7}{6} \right] ~, \label{eq:aes_approx} \\
\Delta a_e^a &\simeq& -\frac{1}{8\pi^2} \frac{m_e^2}{m_a^2} ~g_a^2 \left[ \log\left(\frac{m_a^2}{m_e^2}\right) - \frac{11}{6} \right] ~, \label{eq:aea_approx} \\
\Delta a_e^V &\simeq& \frac{1}{12\pi^2} \frac{m_e^2}{m_V^2} \Big(g_V^2 - 5 g_A^2 \Big) ~, \label{eq:aeV_approx}\\
\Delta a_e^V &\simeq& \frac{1}{2\pi^2} \frac{m_e^2}{\Lambda^2} \bigg[ \Big(3g_T^2 - g_{T5}^2 \Big) \log\left(\frac{\Lambda^2}{m_V^2}\right) + \frac{1}{6} \Big( g_T^2 - 7 g_{T5}^2 \Big) \bigg]~. \label{eq:aeT_approx}
\end{eqnarray}

\subsubsection{Mono-photon searches}  \label{sec:mono_photon}

Mono-photon searches at the BaBar experiment~\cite{BaBar:2017tiz}, $e^+ e^- \to \gamma \, X$, place relevant constraints on each of the considered models that feature couplings to electrons.  
Analogous searches at the Belle~II experiment are expected to further improve the reach~\cite{Belle-II:2022cgf}.

The BaBar analysis~\cite{BaBar:2017tiz} targets specifically invisibly decaying dark photons and sets limits on the kinetic mixing parameter $\epsilon$ as a function of the dark photon mass. In the following, we describe how we recast the results for our scenarios. 
We first determine the differential cross sections of $e^+ e^- \to \gamma X$, with $X = s, a, V$ in the models we consider. We find
\begin{eqnarray}
\frac{d\sigma(e^+e^-\to \gamma s)}{d\cos{\theta}} &=&
   \frac{e^2g_s^2}{8\pi s} \frac{1}{\sin^2 \theta} \left(1 + \frac{m_s^4}{s^2}\right)
   \left(1-\frac{m_s^2}{s} \right)^{-2} ~, \\
\frac{d\sigma(e^+e^-\to \gamma a)}{d\cos{\theta}} &=&
   \frac{e^2 g_a^2}{8\pi s} \frac{1}{\sin^2 \theta} \left(1 + \frac{m_a^4}{s^2}\right)
   \left(1-\frac{m_a^2}{s} \right)^{-2} ~,
 \\ \label{eq:dsigma_V}
\frac{d\sigma(e^+e^-\to \gamma V)}{d\cos{\theta}} &=&
   \frac{e^2 \left(g_V^2 + g_A^2\right)}{8\pi s} \left[ \frac{2}{\sin^2 \theta} \left(1 + \frac{m_V^4}{s^2}\right)
   \left(1-\frac{m_V^2}{s} \right)^{-2} - 1 \right] ~, \\
\frac{d\sigma(e^+e^-\to \gamma V)}{d\cos{\theta}} &=&
   \frac{e^2 \left(g_T^2 + g_{T5}^2\right)}{2\pi \Lambda^2}  \nonumber \\
   &&  \times \left[ \frac{1}{\sin^2 \theta} \frac{m_V^2}{s} \left(1 + \frac{m_V^4}{s^2}\right)
   \left(1-\frac{m_V^2}{s} \right)^{-2} +1 - \frac{m_V^2}{s} \right] ~.
\end{eqnarray}
In the above expressions, $\sqrt{s} \simeq 10.58 $\,GeV is the center of mass energy at BaBar and $\theta$ is the angle between the emitted photon and the beamline in the center of mass frame. We have taken the limit $m_e\to 0$ which is an excellent approximation.

As the BaBar analysis imposes the cut $-0.4 < \cos\theta < 0.6$, we require that the cross sections of our models in the same region of $\cos\theta$ does not exceed the corresponding cross section of a dark photon
\begin{equation}
 \int_{-0.4}^{0.6} d\cos\theta~ \frac{d\sigma(e^+e^-\to \gamma X)}{d\cos{\theta}} <  \int_{-0.4}^{0.6} d\cos\theta~ \frac{d\sigma(e^+e^-\to \gamma A^\prime)}{d\cos{\theta}} ~,
\end{equation} 
with the differential dark photon cross section given by
\begin{equation}
\frac{d\sigma(e^+e^-\to \gamma A^\prime)}{d\cos{\theta}} =
   \frac{\epsilon^2 e^4}{8\pi s} \left[ \frac{2}{\sin^2 \theta} \left(1 + \frac{m_{A^\prime}^4}{s^2}\right)
   \left(1-\frac{m_{A^\prime}^2}{s} \right)^{-2} - 1 \right] ~.
\end{equation}

\subsubsection{Beam dump constraints} \label{sec:beam_dump}

The NA64 experiment at CERN places constraints on the production of invisible particles coupled to electrons \cite{Gninenko:2016kpg}. More specifically, NA64 is an electron/positron beam dump experiment utilizing 100 GeV beams scattering with an active target and looking for missing energy. 
In their latest analysis based on $9.37\times 10^{11}$ electrons on target \cite{NA64:2023wbi}, NA64 set the most stringent constraint on the parameter space of an invisible dark photon in the $(10^{-3}-0.35)$ GeV mass range\footnote{The NA64 collaboration also sets interesting bounds on the dark photon parameter space using a positron beam run~\cite{NA64:2023ehh}, which exploits dark photon production via resonant annihilation of positrons with atomic electrons in the target nuclei. However, these bounds are not relevant to our analysis, as they are only competitive at higher dark particle masses, $m\gtrsim 0.1$ GeV.}. Analogous searches at the future LDMX missing-momentum experiment are expected to further improve these bounds \cite{LDMX:2025bog}.

Signal efficiencies and yields are expected to change by an $\mathcal O(1)$ amount for other light dark sector particles. Instead of recasting the NA64 dark photon search, we use the NA64 results from~\cite{NA64:2021xzo} which provide directly limits on the couplings of scalars, pseudo-scalars, vectors, and axial-vectors in the mass range from 1\,MeV to 1\,GeV. 

Note that the results in~\cite{NA64:2021xzo} are based on approximately one third of the statistics used in~\cite{NA64:2023wbi}. A recast of the full current data set might improve the limits on the couplings by a factor of approximately $\sqrt{3} \sim 1.7$.

\subsubsection{Kaon decays} \label{sec:kaon}

Similar to the $\pi^+ \to \ell^+ \nu_\ell X$ decays, also the analogous kaon decays $K^+ \to \ell^+ \nu_\ell X$ are sensitive probes of light dark sectors. 
The NA62 experiment has searched for the decay $K^+ \to \mu^+ \nu_\mu X$, with an invisible $X$~\cite{NA62:2021bji}. In the mass range that overlaps with the pion decays, $m_X \lesssim m_\pi - m_\mu$, NA62 has set limits on the $K^+ \to \mu^+ \nu_\mu X$ branching ratio in the range between $2\times 10^{-6}$ and $2 \times 10^{-5}$ for a scalar and a vector model. 

Also flavor-changing neutral current decays of kaons, the $K \to \pi X$ decay in particular, potentially give relevant constraints in the models we are considering.
On the experimental side, the strongest constraints on the $K^+ \to \pi^+ X$ rate have been obtained at BNL-E949 and NA62~\cite{BNL-E949:2009dza, NA62:2020pwi, NA62:2020xlg, NA62:2025upx} (see also \cite{Guadagnoli:2025xnt} for a recast of the NA62 measurement of the $K^+ \to \pi^+ \nu \bar\nu$ branching ratio as a constraint on ALPs). Even if our dark sector particles couple at tree level only to charged leptons, couplings to other SM particles are generated at the loop level. 

This has been extensively explored in the context of axion-like particles, see e.g.~\cite{Bauer:2020jbp, Bauer:2021mvw}.
In particular, ALP couplings to left-handed (right-handed) charged leptons induce flavor changing couplings to strange and down quarks at the 2-loop (3-loop) level, which are strongly constrained by rare kaon decays. However, some of these loop effects depend logarithmically on an unknown UV cut-off scale. Moreover, additional contributions to flavor changing couplings can arise from loops involving couplings to quarks or gauge bosons. Models might even contain tree-level flavor changing couplings. Without making specific assumptions about many couplings, it is thus not possible to model independently assess the constraints from $K \to \pi X$.

As an example, we note that in the leptophilic ALP models considered in~\cite{Altmannshofer:2022ckw}, the searches for $K^+ \to \pi^+ X$ constrain the couplings of the ALP to electrons at the level of $g_a \sim 10^{-4}$ to $10^{-6}$.

\subsection{Sensitivity to dark sector model parameters} \label{sec:3body_results}

\begin{figure}[tb]
\centering
\includegraphics[width=0.48\textwidth]{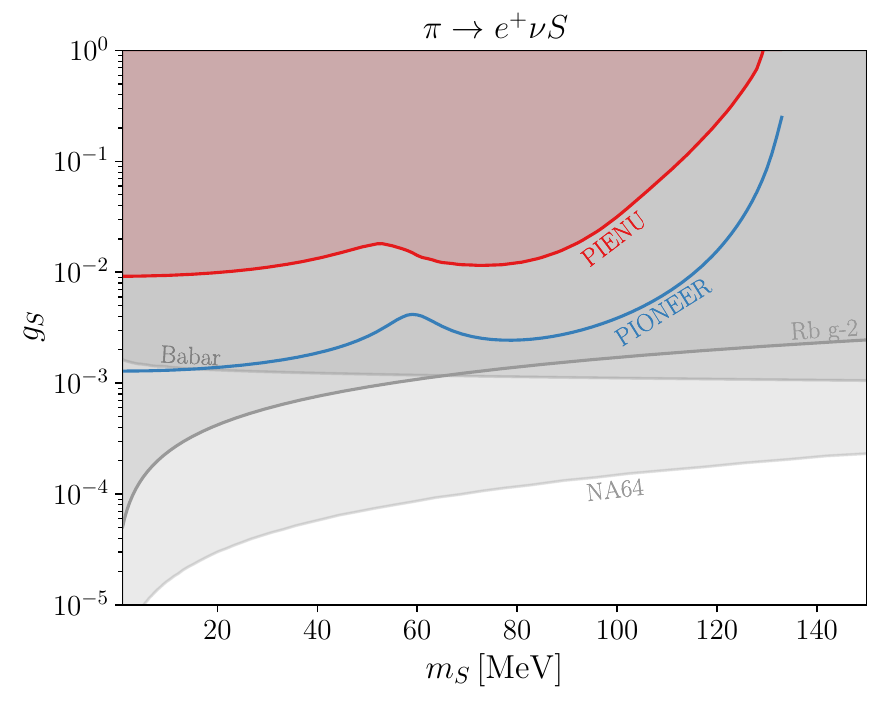} \quad 
\includegraphics[width=0.48\textwidth]{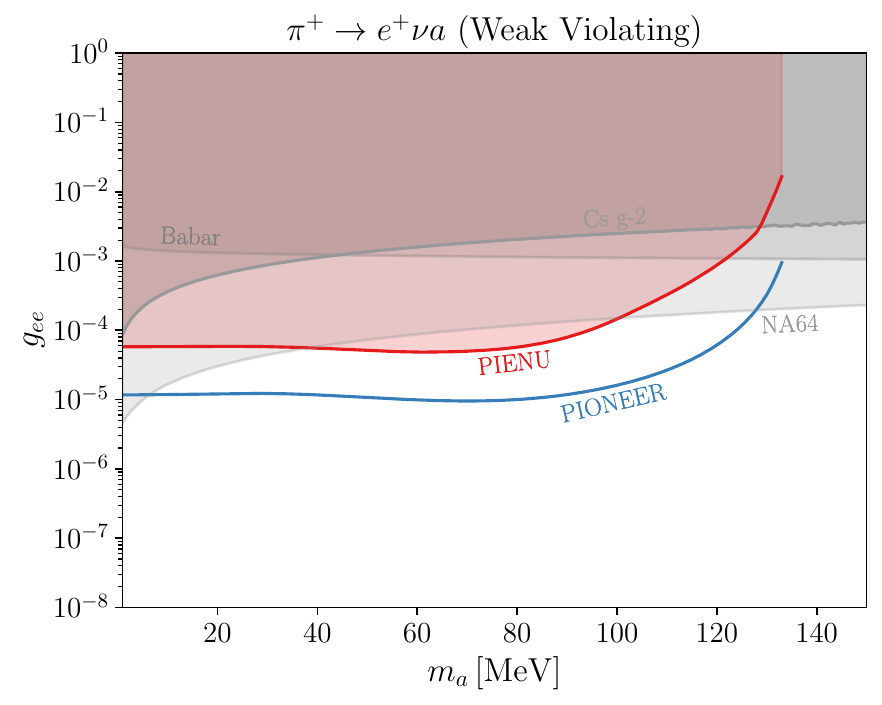} \\[16pt]
\includegraphics[width=0.48\textwidth]{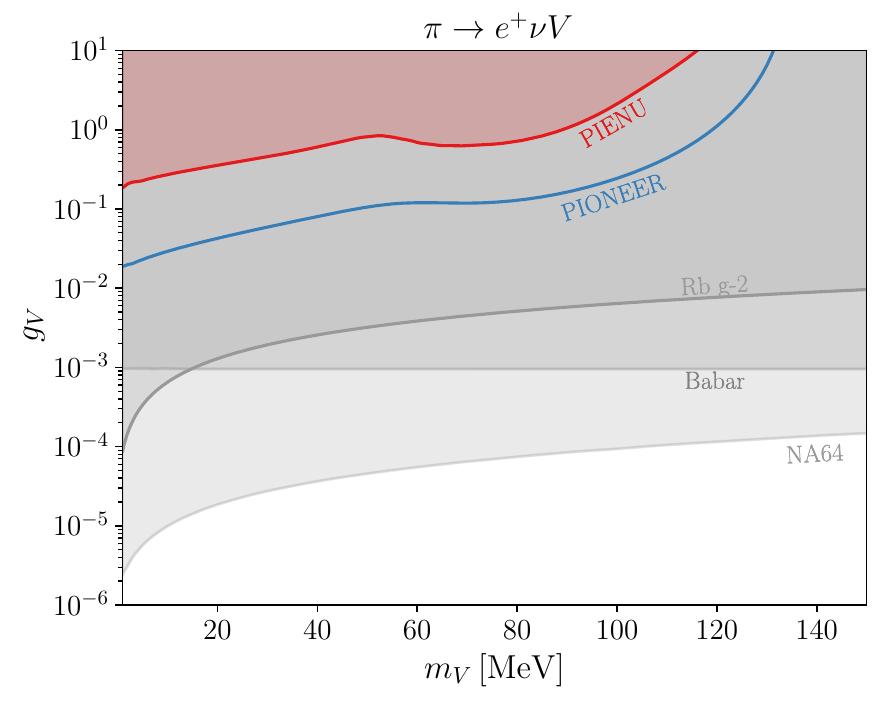} \quad
\includegraphics[width=0.48\textwidth]{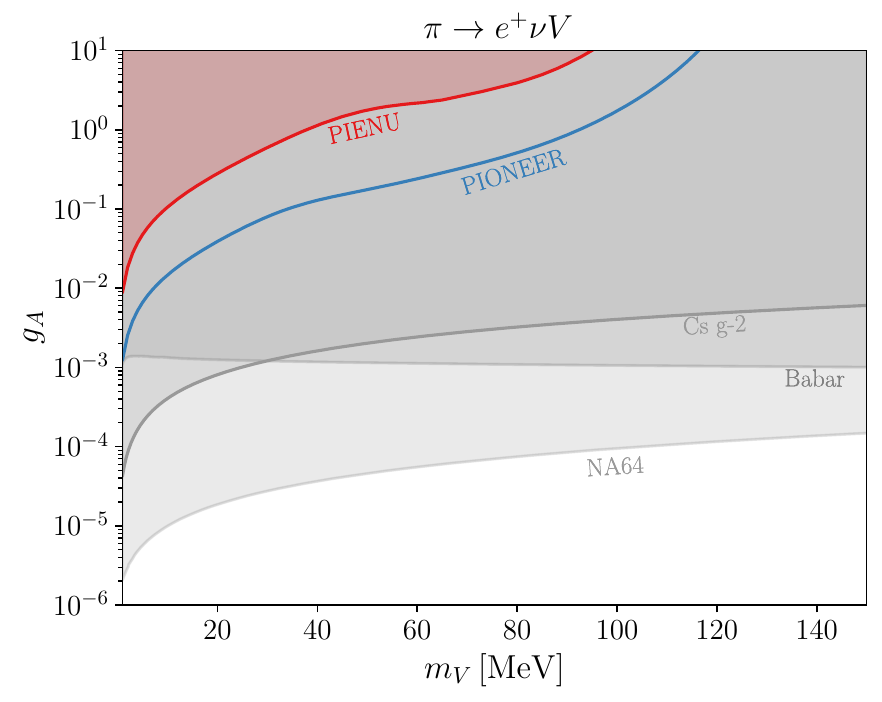}
\caption{Limits on the couplings of dark sector particles to electrons as a function of their mass. {\bf Top left}: scalar; {\bf top right}: axion-like particle with weak violating couplings; {\bf bottom left}: spin-1 with vector coupling; {\bf bottom right}: spin-1 with axial-vector coupling. Existing constraints from the anomalous magnetic moment of the electron, from mono-photon searches at BaBar and from beam dump searches at NA64 are shown in gray. The pion decay constraints from PIENU are shown in red and the PIONEER sensitivity (assuming $0.5 \times 10^8$ recorded $\pi^+ \to e^+ \nu_e$ events) in blue.} 
\label{fig:3body_coupling_e}
\end{figure}
\begin{figure}[tb]
\centering
\includegraphics[width=0.48\textwidth]{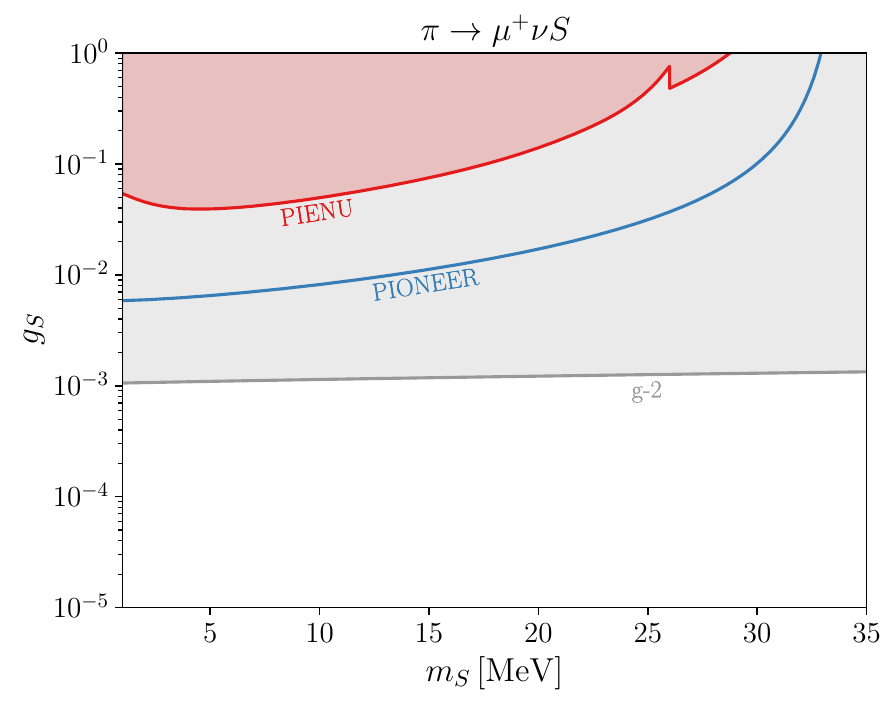} \quad 
\includegraphics[width=0.48\textwidth]{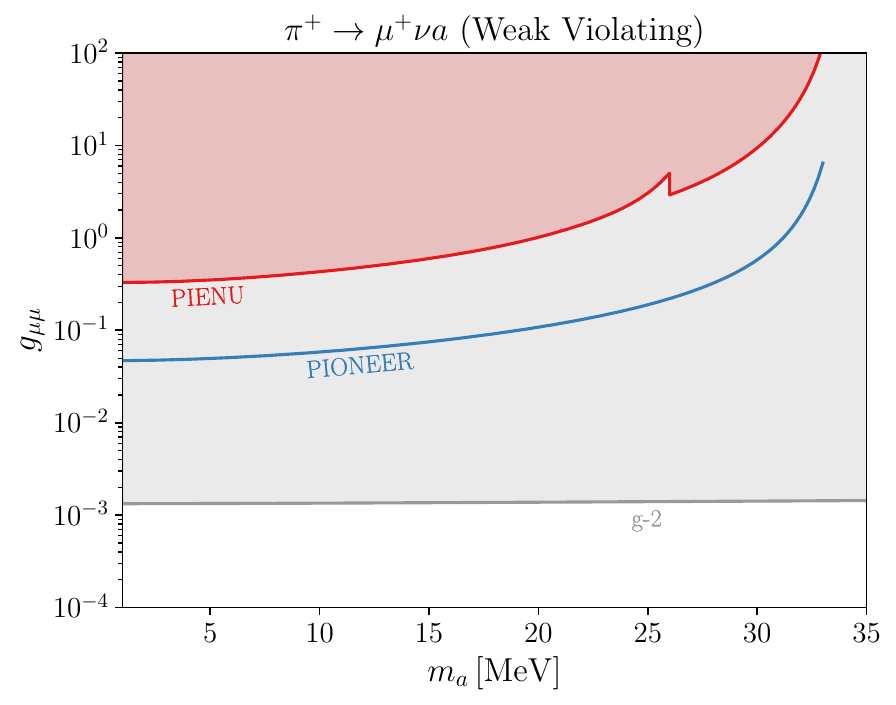} \\[16pt]
\includegraphics[width=0.48\textwidth]{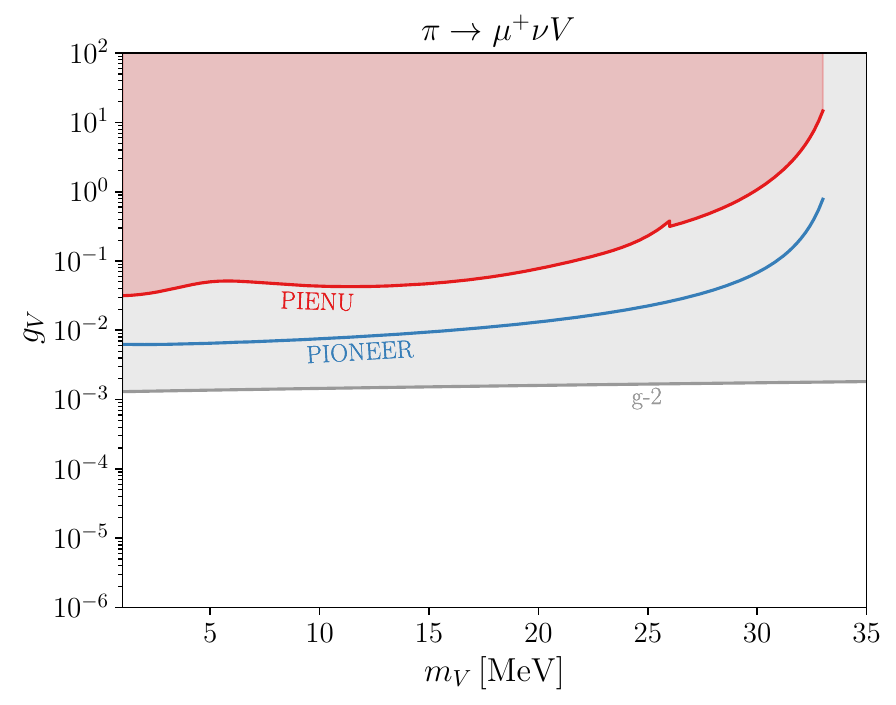} \quad
\includegraphics[width=0.48\textwidth]{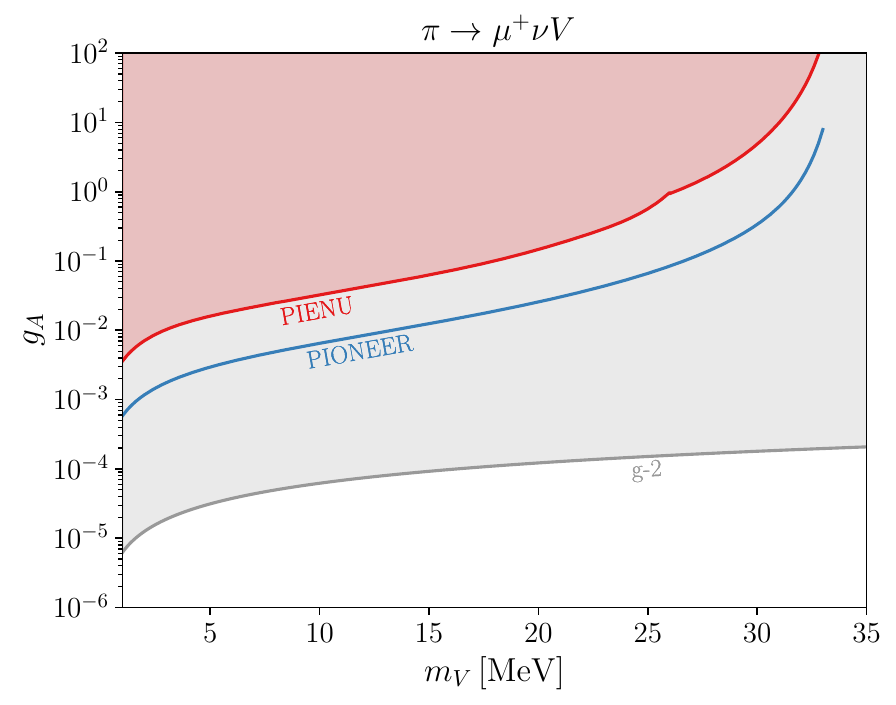}
\caption{Limits on the couplings of dark sector particles to muons as a function of their mass. {\bf Top left}: scalar; {\bf top right}: axion-like particle with weak violating couplings; {\bf bottom left}: spin-1 with vector coupling; {\bf bottom right}: spin-1 with axial-vector coupling. Existing constraints from the anomalous magnetic moment of the muon are shown in gray. The pion decay constraints from PIENU are shown in red and the PIONEER sensitivity (assuming $10^{10}$ recorded $\pi^+ \to \mu^+ \nu_\mu$ events) in blue.} 
\label{fig:3body_coupling_mu}
\end{figure}

In figures~\ref{fig:3body_coupling_e} and~\ref{fig:3body_coupling_mu} we compare the pion decay sensitivity to the sensitivities from the other dark sector searches, the anomalous magnetic moments, mono-photon searches, and beam dump searches in particular.

We show the results for a representative selection of models.
In both figures, the top left panels correspond to a scalar with coupling $g_s$ as defined in equation~\eqref{eq:scalar_couplings}. The top right panels show axion-like particles with weak violating couplings. In the notation of equation~\eqref{eq:axion_couplings} this corresponds to $g_a = - \bar g_a = g_{ee}$ or $g_{\mu\mu}$,~ $g_{a,\nu} = 0$. The bottom left and bottom right panels correspond to spin-1 bosons with vector couplings $g_V$ or axial-vector couplings $g_A$ as defined in equation~\eqref{eq:vector_couplings}. We do not show a weak preserving ALP, as the sensitivities from pion decays turn out to be very similar to the scalar case. We also do not show spin-1 bosons with dipole interactions. Setting the scale $\Lambda$ that enters the couplings in~\eqref{eq:vector_couplings} to $1$\,TeV, we find constraints from pion decays are in the highly non-perturbative regime $g_T, g_{T5} \gg 4\pi$. 

Across all the models we consider, the anomalous magnetic moment of the electron and the BaBar mono-photon searches put limits on the coupling of the dark sector particles to electrons at the level of approximately $10^{-3}$ to $10^{-4}$. The beam dump constraints from NA64 tend to be consistently stronger and reach down to $\sim 10^{-5}$ for low masses $m_X \ll m_\pi$. The anomalous magnetic moment of the muon constrains the scalar, axion-like particle, and vector couplings to muons at around $10^{-3}$. In the case of the axialvector, this constraint is stronger by at least an order of magnitude.
We do not show constraint from the kaon decays $K \to \pi X$ due to their strong model dependence, see the discussion in section~\ref{sec:kaon}.

The existing PIENU searches for $\pi^+ \to e^+ \nu_e X$ cover otherwise unexplored parameter space of the weak violating axion-like particle model. We find that PIONEER can improve over PIENU by approximately one order of magnitude in coupling sensitivity. However, in all but the weak violating axion-like particle model, the other dark sector probes turn out to be more constraining.
In the case of dark sector particles coupled to muons, we find that the anomalous magnetic moment of the muon gives stronger constraints for all models we consider.
In principle, the constraint from $(g-2)_\mu$ can be avoided by invoking a tuned cancellation with other unrelated loop contributions. We do not consider such a scenario.

The above discussion highlights the weak violating axion-like particle coupled to electrons as a useful benchmark model for dark sector searches at PIONEER. 

Another class of models that is very weakly constrained by other probes are light dark sector particles that couple exclusively to neutrinos. So-called Majorons are a prominent example of such particles~\cite{Chikashige:1980ui, Gelmini:1980re, Barger:1981vd, Schechter:1981cv}. A study of the PIONEER sensitivity to light neutrinophilic scalars can be found in~\cite{deLima2026}.

\section{Conclusions} \label{sec:conclusions} 

In this work we have presented a systematic study of light dark sector searches in exotic pion decays at the PIENU experiment and the forthcoming PIONEER experiment. Exploiting the strong helicity or phase-space suppression of Standard Model pion decays, we showed that even very weakly coupled new particles can lead to observable signatures in both two-body and three-body decay channels.
We considered a number of dark sector scenarios, including sterile neutrinos produced in two-body decays $\pi^+ \to \ell^+ N$, as well as invisible scalars, axion-like particles, or dark vectors contributing to the three-body decays $\pi^+ \to \ell^+ \nu_\ell X$.

The sterile neutrinos yield monochromatic positrons from $\pi^+ \to e^+ N$ or muons from $\pi^+ \to \mu^+ N$ and can be searched for using bump hunts. We validated our sterile neutrino analysis framework by successfully reproducing existing PIENU limits. Building on this, we have estimated the sensitivity of PIONEER using detailed simulations based on its current detector design. The main sterile neutrino results are summarized in figures~\ref{fig:sterile_bounds} and~\ref{fig:sterile_bounds_mu}. Already in phase 1, PIONEER will be able improve over PIENU by at least one order of magnitude and will start to reach into see-saw motivated parameter space.

Light, invisible dark sector particles can be produced in $\pi^+ \to e^+ \nu_e X$ and $\pi^+ \to \mu^+ \nu_\mu X$ decays and result in continuous distortions in the low-energy tail of the positron and muon energy spectra. We have successfully reproduced the existing PIENU limits on the $\pi^+ \to \ell^+ \nu_\ell X$ branching ratios in the benchmark scenario used by the experiment. We have then recast the PIENU limits to dark sector scenarios with invisible scalars, axion-like particles, and dark vectors. Due to the different shape of the charged lepton energy spectrum in each model, we find that the pion decay constraints vary by $\mathcal O(1)$ between the different scenarios. Using simulated data we found that PIONEER will extend sensitivity to significantly smaller branching ratios, improving the PIENU bounds by one to two orders of magnitude. These results are summarized in figures~\ref{fig:3body_BR_limits_e} and~\ref{fig:3body_BR_limits_mu}.

By comparing with complementary probes, including lepton anomalous magnetic moments, BaBar mono-photon searches, and NA64 beam-dump searches, we determined the role of pion decays in testing regions of model parameter space. We have found that pion decays into weak-violating axion-like particles emerge as a well-motivated benchmark, with PIENU already probing novel parameter space and PIONEER expected to improve sensitivity by roughly an additional order of magnitude.

Our analysis has focused on fully invisible dark sector particles, where missing-energy signatures dominate. An important extension of this program will be the exploration of visible decay signatures from short-lived dark sector states, which could further enhance the discovery potential of PIONEER. We will study these visible signatures in a follow up paper. Overall, our results stress that exotic pion decays form an important component of PIONEER's program and  provide a relevant avenue for probing light dark sectors.

\section*{Acknowledgements} 

We thank Shintaro Ito for invaluable correspondence about the sterile neutrino searches at PIENU, as well as Douglas Bryman, Quentin Buat, David Hertzog, Peter Kammel, Simone Mazza, Adam Molnar, Jennifer Ott, Bruce Schumm, and Robert Shrock for many useful discussions and comments. We thank the authors of~\cite{deLima2026} for making their work available to us prior to publication and for coordinating submission to the arXiv.

The research of WA, PG, and SG is supported by the U.S. Department of Energy grant number DE-SC0010107. The research of JD is supported in part by the U.S. Department of Energy grant number DESC0025569. Part of this work was performed at the Aspen Center for Physics, which is supported by National Science Foundation grant PHY-2210452. This research was also supported in part by grant NSF PHY-2309135 to the Kavli Institute for Theoretical Physics (KITP). WA acknowledges support by the Munich Institute for Astro-, Particle and BioPhysics (MIAPbP) which is funded by the Deutsche Forschungsgemeinschaft (DFG, German Research Foundation) under Germany's Excellence Strategy - EXC-2094 – 390783311.
The work of PS was supported by the United States Department of Energy grant DE-FG02-97ER41020.

\begin{appendix}
\section{Details on the PIENU Sterile Neutrino Search Recast} \label{app:recast} 

The PIENU experiment performed a bump search for sterile neutrinos in $\pi^+ \to e^+ N$ decay for sterile neutrino masses in the range of $(62 - 134)$~MeV~\cite{PIENU:2017wbj}.
Here we provide a detailed description of our reproduction of the search.

\begin{figure}[tb]
\centering
\includegraphics[width=1.0\textwidth]{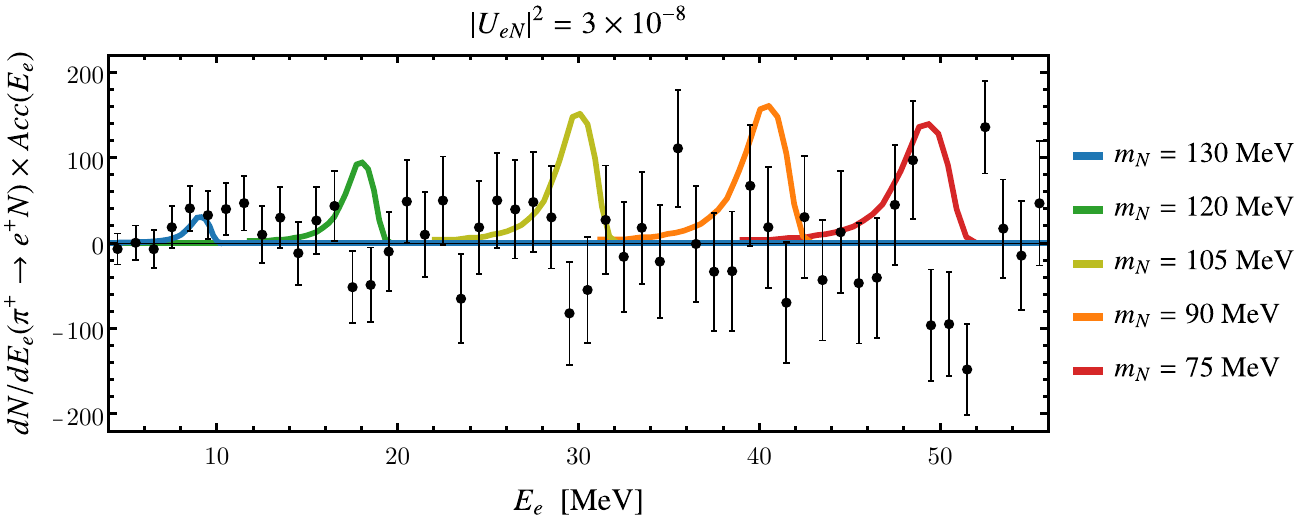} 
\caption{Expected number of detected $\pi^+ \to e^+ N$ decays at the PIENU experiment as a function of the positron energy for several values of the sterile neutrino mass. The sterile neutrino mixing angle is set to $|U_{e N}|^2 = 3 \times 10^{-8}$. For comparison, the residuals extracted from~\cite{PIENU:2017wbj} are shown in black.} 
\label{fig:sterile_spectrum}
\end{figure}

In~\cite{PIENU:2017wbj}, the PIENU collaboration shows the observed positron energy spectrum, the fitted model of the spectrum without a sterile neutrino, and the residuals of the fit for positron energies between 4~MeV and 56~MeV. This range of positron energies corresponds to sterile neutrino masses between approximately 62~MeV and 134~MeV (cf. equation~\eqref{eq:E_peak}). The residuals shown by PIENU are rebinned and need to be interpreted with care. We re-extracted the residuals ourselves from the shown data and the fitted model that consists of the $\pi^+ \to e^+ \nu_e$ decay, as well as backgrounds from muon decay at rest and in flight. Our residuals, including statistical uncertainties only, are shown in figure~\ref{fig:sterile_spectrum}.

The PIENU paper~\cite{PIENU:2017wbj} also shows the positron energy spectrum of the $\pi^+ \to e^+ N$ signal with a sterile neutrino mass $m_N \simeq 90$\,MeV. For this mass, the spectrum has a peak at $\bar E_e \simeq 40$\,MeV and is smeared out, for example, due to the calorimeter response. To obtain the signal shape for different sterile neutrino masses, we assume that the width of the shown spectrum scales with the square root of the positron energy. Such a scaling of the energy resolution with the square root of the energy was used in parts of the PIENU $\pi^+ \to \mu^+ N$ analysis~\cite{PIENU:2019usb}.\footnote{Since this scaling assumption is approximate, we performed several cross-checks. In correspondence with the PIENU collaboration~\cite{Ito:2025}, we obtained an approximate expression for the PIENU energy resolution as a function of the positron energy. As a first check, we modeled the signal as a Gaussian with a width corresponding to this provided resolution. As a second check, we retained the shape of the sterile neutrino signal shown by PIENU, but rescaled its width using the relative energy resolution from the same expression instead of the $\sqrt{E_e}$ scaling. All methods yielded very similar results, confirming the robustness of our approach.}

To obtain the number of $\pi^+ \to e^+ N$ events as a function of the sterile neutrino model parameters, we use
\begin{equation}
\frac{N(\pi^+ \to e^+ N)}{N(\pi^+ \to e^+ \nu_e)} = \frac{\text{BR}(\pi^+ \to e^+ N)}{\text{BR}(\pi^+ \to e^+ \nu_e)} ~,
\end{equation}
with the ratio of branching ratios given in equations~\eqref{eq:sterile_exact1}-\eqref{eq:sterile_exact2}.
From the fitted $\pi^+ \to e^+ \nu_e$ component of the positron energy spectrum shown in~\cite{PIENU:2017wbj}, we extract 
$N(\pi^+ \to e^+ \nu_e) \simeq 1.4 \times 10^6$. 

In figure~\ref{fig:sterile_spectrum}, we show the expected spectra of a sterile neutrino signal for several choices of the sterile neutrino mass. In the plot, we include the detector acceptance correction $Acc(E_e)$ given in~\cite{PIENU:2017wbj} as a function of the positron energy. For illustration, we set the mixing angle to $|U_{e N}|^2 = 3 \times 10^{-8}$ and compare the spectra to our extracted residuals.

To obtain bounds on the mixing angle $|U_{e N}|^2$, we fit the energy spectrum of the sterile neutrino signal to our extracted residuals. Emulating the approach of the PIENU analysis, we include in the fit also components from the SM decay $\pi^+ \to e^+ \nu_e$ and backgrounds from muon decay at rest and muon decay in flight. We keep the shape of these components fixed to the ones shown in~\cite{PIENU:2017wbj}, but let their normalizations float. The uncertainties of the event numbers across the energy bins are taken to be uncorrelated. 

Imposing the physical prior $\text{BR}(\pi^+ \to e^+ N) \geq 0$, we obtain the 90\% C.L. bound on the sterile neutrino mixing angle $|U_{e N}|^2$ shown in figure~\ref{fig:sterile_bounds}.
We also attempted an alternative approach to constrain $|U_{e N}|^2$ without making use of any background shape information. Demanding that the predicted number of sterile neutrino events in a given energy bin does not exceed the observed number of events gives, in principle, a very robust bound. However, we find that it is approximately 1 to 2 orders of magnitude weaker than the constraint from the bump hunt.

\begin{figure}[tb]
\centering
\includegraphics[width=1.0\textwidth]{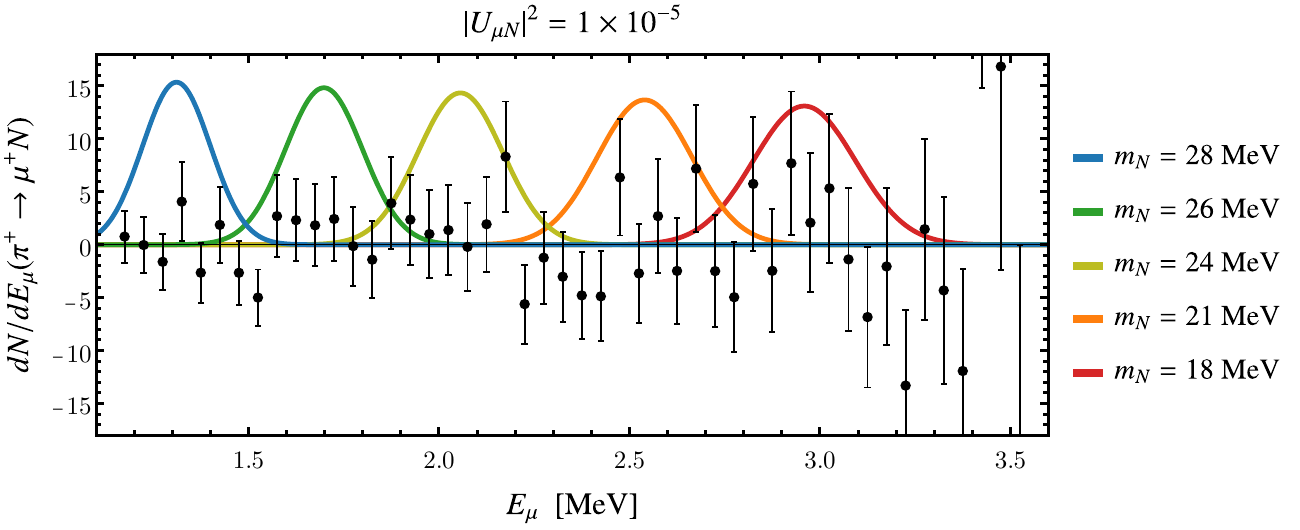} \\[16pt]
\includegraphics[width=1.0\textwidth]{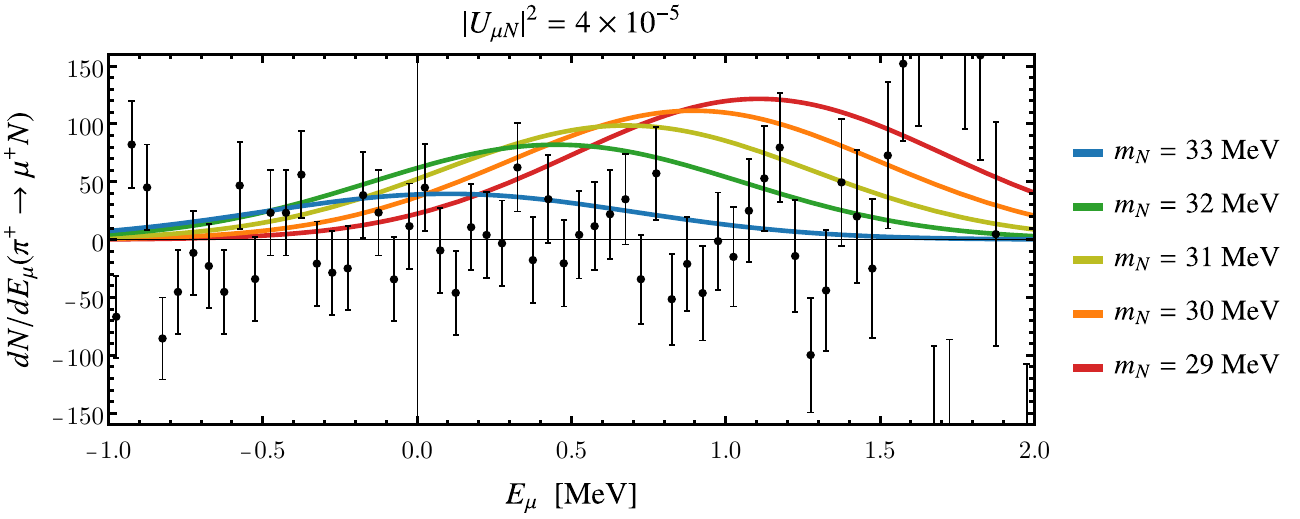} 
\caption{Expected number of detected $\pi^+ \to \mu^+ N$ decays at the PIENU experiment as a function of the muon kinetic energy for several values of the sterile neutrino mass. The top shows the high muon energy analysis, the bottom the low muon energy analysis. The sterile neutrino mixing angles are set to $|U_{\mu N}|^2 = 1 \times 10^{-5}$ and $|U_{\mu N}|^2 = 4 \times 10^{-5}$, respectively. For comparison, the residuals extracted from~\cite{PIENU:2019usb} are shown in black.} 
\label{fig:sterile_spectrum_mu}
\end{figure}

\bigskip
We follow a similar procedure to reproduce the PIENU searches for sterile neutrinos in $\pi^+ \to \mu^+ N$ decays. The PIENU analysis~\cite{PIENU:2019usb} considered separate signal selections with low muon energy and high muon energy, corresponding to sterile neutrino masses of 28\,MeV to 34\,MeV and 16\,MeV to 29\,MeV, respectively.  
We extract residuals between the provided event data and the fitted model that consists of the $\pi^+ \to \mu^+ \nu_\mu$ decay, as well as backgrounds from radiative muon decay (high energy analysis) and pion decay in flight (low energy analysis). Our residuals are shown in figure~\ref{fig:sterile_spectrum_mu}.

The number of expected sterile neutrino signal events is determined by 
\begin{equation}
\frac{N(\pi^+ \to \mu^+ N)}{N(\pi^+ \to \mu^+ \nu_\mu)} = \frac{\text{BR}(\pi^+ \to \mu^+ N)}{\text{BR}(\pi^+ \to \mu^+ \nu_\mu)} ~,
\end{equation}
with $N(\pi^+ \to \mu^+ \nu_\mu) = 9.1 \times 10^6$ in the high energy analysis and $N(\pi^+ \to \mu^+ \nu_\mu) = 1.3 \times 10^8$ in the low energy analysis~\cite{PIENU:2021clt}. The signal shape was chosen to be a Gaussian. In the low energy analysis, we fix the width of the Gaussian to 0.6\,MeV, which corresponds to the width of the main $\pi^+ \to \mu^+ \nu_\mu$ peak. In the high energy analysis, we use an energy dependent width as detailed in~\cite{PIENU:2019usb}: at the main $\pi^+ \to \mu^+ \nu_\mu$ peak ($E_\mu \simeq 4.1$ MeV) the width corresponds to the width of the peak which is approximately 0.16\,MeV. At lower energies, the width scales with the square root of the energy. The resulting signal shapes are shown in the plots of figure~\ref{fig:sterile_spectrum_mu} for a few example choices of the sterile neutrino mass for the high energy (top) and low energy (bottom) analysis. 

To obtain bounds on the sterile neutrino mixing angle with muon neutrinos, $|U_{\mu N}|^2$, we perform a very similar analysis as for the electron events discussed above, including the additional components from $\pi^+ \to \mu^+ \nu_\mu$ and radiative decay backgrounds. 
The pion decay in flight background in the low energy analysis is parameterized by a second order polynomial with three free parameters that are allowed to vary in the fit. 
The 90\% C.L. bound on the mixing angle $|U_{\mu N}|^2$ is
shown in figure~\ref{fig:sterile_bounds_mu}, after imposing the physical prior $\text{BR}(\pi^+ \to \mu^+ N) \geq 0$.

\section{Details on the PIENU Three-Body Pion Decay Recast} \label{app:recast2} 

In this appendix, we describe our reproduction of the PIENU search~\cite{PIENU:2021clt} for a dark sector state $X$ in three-body pion decays, focusing first on the $\pi^+ \to e^+ \nu_e X$ channel, and subsequently addressing the muon mode $\pi^+ \to \mu^+ \nu_\mu X$.

We extracted the experimental data from the two data sets used for the $\pi^+ \to e^+ \nu_e X$ search and presented in the plots of~\cite{PIENU:2021clt}, comprising approximately $5 \times 10^5$ and $8 \times 10^5$ $\pi^+ \to e^+ \nu_e$ events, respectively. To model the data, PIENU included background components from the low-energy tail of the Standard Model $\pi^+ \to e^+ \nu_e$ decay, as well as from muon decays at rest and muon decays in flight.

We convoluted the signal spectrum from~\cite{Batell:2017cmf} with a signal spread function to account for detector resolution effects. Specifically, we assumed that the signal spread at the positron energy of $E_e = 69.8$\,MeV (corresponding to the peak of the SM $\pi^+ \to e^+ \nu_e$ decay) is described by the observed $\pi^+ \to e^+ \nu_e$ line shape. For other positron energies, the spread was scaled proportionally to the square root of the positron energy, similar to our approach in reproducing the PIENU sterile neutrino search described in appendix~\ref{app:recast}. We find that this convolution procedure slightly broadens the signal spectrum and induces a modest downward shift of the peak position by a few MeV. As noted in~\cite{PIENU:2021clt}, no energy-dependent acceptance corrections were applied in the analysis, and we follow the same approach in our reproduction.

\begin{figure}[tb]
\centering
\includegraphics[width=1.0\textwidth]{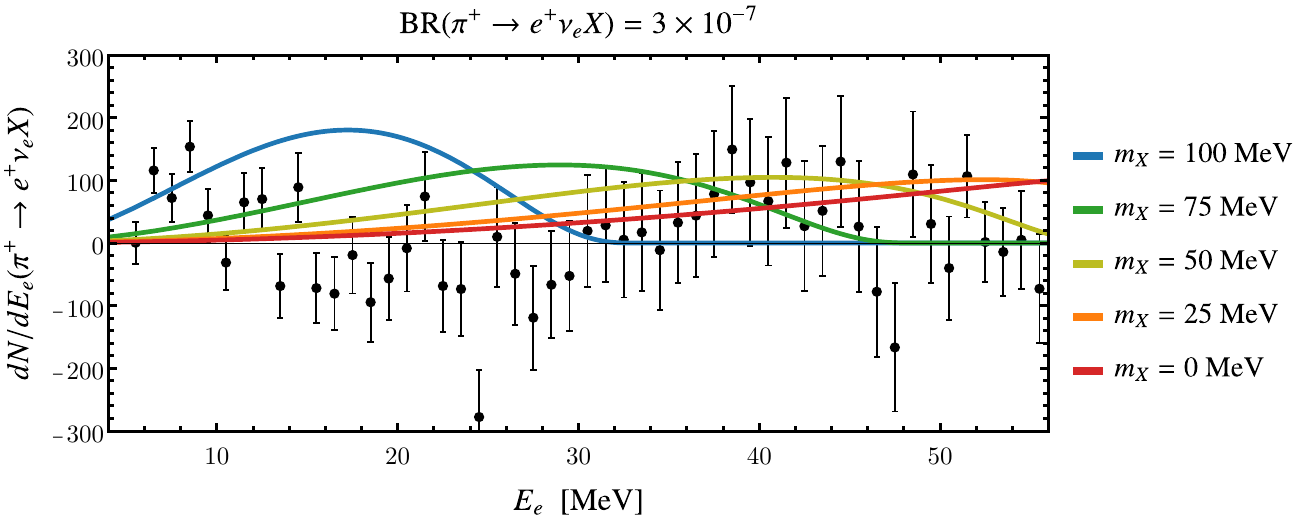} 
\caption{Expected number of detected $\pi^+ \to e^+ \nu_e X$ decays at the PIENU experiment as a function of the positron energy for several values of the $X$ mass. The signal model is taken from~\cite{Batell:2017cmf}. The $\pi^+ \to e^+ \nu_e X$ branching ratio is set to $3 \times 10^{-7}$. For comparison, the residuals extracted from~\cite{PIENU:2021clt} are shown in black.} 
\label{fig:3body_positron_spectrum}
\end{figure}

We find excellent agreement between our reconstructed signal spectra and those presented in~\cite{PIENU:2021clt} for a dark sector mass of $m_X = 80$\,MeV. As an illustration, figure~\ref{fig:3body_positron_spectrum} shows our signal spectra for various $X$ masses, assuming a fixed branching ratio $\text{BR}(\pi^+ \to e^+ \nu_e X) = 3 \times 10^{-7}$. The event numbers and shown residuals correspond to the sum of the two data sets from~\cite{PIENU:2021clt}.

\begin{figure}[tb]
\centering
\includegraphics[width=0.7\textwidth]{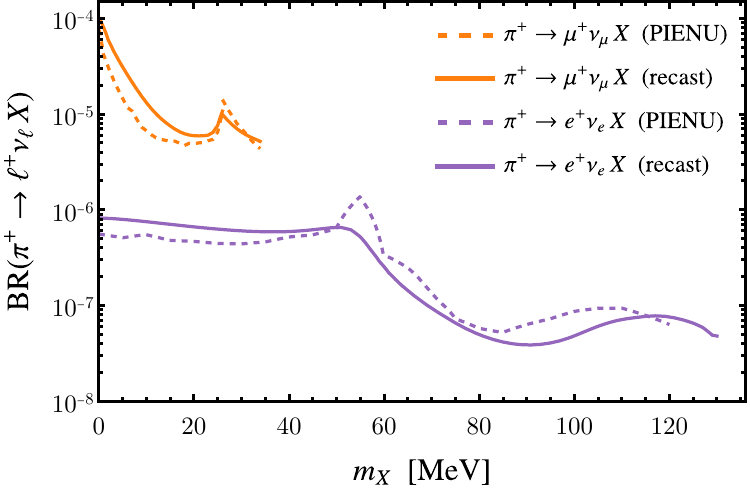} 
\caption{Upper limits on the $\pi^+ \to e^+ \nu_e X$ (purple) and $\pi^+ \to \mu^+ \nu_\mu X$ (orange) branching ratios as a function of the $X$ mass for the signal model from~\cite{Batell:2017cmf}. The PIENU result from~\cite{PIENU:2021clt} is shown in dashed. Our recast is shown by the solid lines.} 
\label{fig:3body_check}
\end{figure}

To extract the branching ratio limits, we performed fits of the signal spectrum to the residuals in the positron energy range $4.5\,\text{MeV} < E_e < 56$\,MeV, including the three background components with floating normalizations. The 90\% C.L. limits on the branching ratio for the $\pi^+ \to e^+ \nu_e X$ decay are shown in figure~\ref{fig:3body_check} by the solid purple curve. They are in good agreement with those reported by the PIENU collaboration shown by the dashed curve, validating our reproduction of the analysis. 

\bigskip
Our reproduction of the $\pi^+ \to \mu^+ \nu_\mu X$ search by PIENU~\cite{PIENU:2021clt} follows a similar procedure. Analogous to the search for sterile neutrinos in $\pi^+ \to \mu^+ N$ decays (described in appendix~\ref{app:recast}), the $\pi^+ \to \mu^+ \nu_\mu X$ search is split into two muon energy regions. At low energies, the data contain approximately $1.3\times 10^8$ muon events and it is well described by the SM $\pi^+ \to \mu^+ \nu_\mu$ peak and a second order polynomial in $E_\mu$ that models contributions from pion decay in flight. The data in the high energy region is comprised of approximately $9.1 \times 10^6$ muon events, and it is well described by the SM peak and a tail from radiative decays $\pi^+ \to \mu^+ \nu_\mu \gamma$ with a shape obtained from Monte Carlo simulations.

\begin{figure}[tb]
\centering
\includegraphics[width=1.0\textwidth]{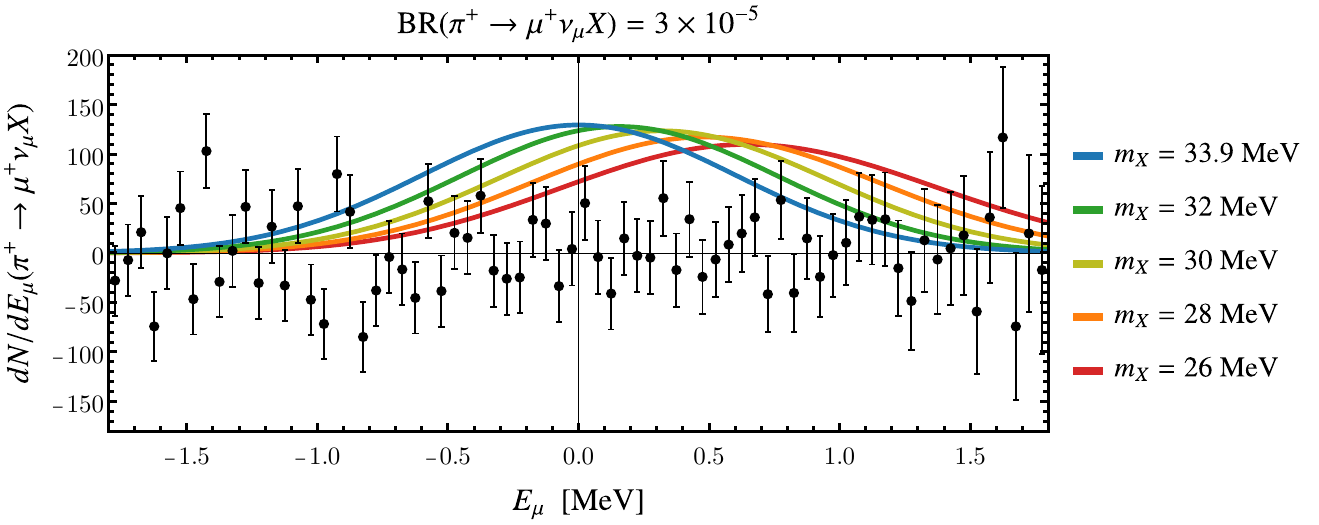} \\[16pt]
\includegraphics[width=1.0\textwidth]{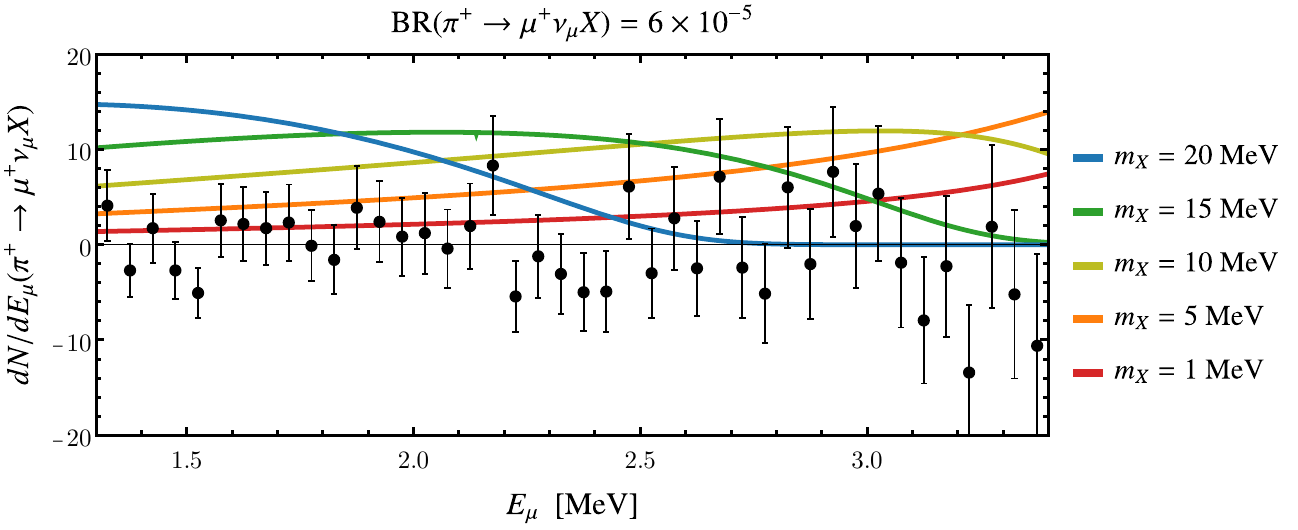} 
\caption{Expected number of detected $\pi^+ \to \mu^+ \nu_\mu X$ decays at the PIENU experiment as a function of the muon kinetic energy for several values of the $X$ mass. The signal model is taken from~\cite{Batell:2017cmf}. The top and bottom plots show the low and high energy analysis with the $\pi^+ \to \mu^+ \nu_\nu X$ branching ratio set to $3 \times 10^{-5}$ and $6 \times 10^{-5}$, respectively. For comparison, the residuals extracted from~\cite{PIENU:2021clt} are shown in black.} 
\label{fig:3body_muon_spectrum}
\end{figure}

To model the signal, we convolute the $\pi^+ \to \mu^+ \nu_\mu X$ spectrum with a Gaussian spread function. Analogous to the sterile neutrino search, in the low energy analysis, we fix the width of the Gaussian to $0.6$\,MeV, while in the high energy analysis, we use a width that scales with the square root of the muon kinetic energy and is $0.16$\,MeV for $E_\mu = 4.1$\,MeV. We find good agreement between our final signal shapes and the shapes shown in~\cite{PIENU:2021clt} for $X$ masses of $15$\,MeV and $33.9$\,MeV. For illustration we show in figure~\ref{fig:3body_muon_spectrum} our signal spectra for several choices of the $X$ mass and compare them to the residuals of the PIENU low energy and high energy analysis~\cite{PIENU:2021clt}.

We then fit the signal spectra to the residuals.
In the low energy analysis, we also vary all three parameters of the polynomial that models the pion decay in flight, while for the high energy analysis we vary the normalization of the radiative decay background, keeping its shape fixed. We show our 90\% C.L. limits on the $\pi^+ \to \mu^+ \nu_\mu X$ branching ratio in figure~\ref{fig:3body_check} by the orange lines. Our recast (solid) agrees very well with the official limit reported by PIENU (dashed).

\end{appendix}

\bibliographystyle{JHEP}
\bibliography{biblio.bib}

\providecommand{\href}[2]{#2}\begingroup\raggedright\begin{thebibliography}{100}

\bibitem{Bryman:2025pet}
D.~Bryman and R.~Shrock, \emph{{Pion Decay}},
  \href{https://arxiv.org/abs/2502.18384}{{\ttfamily 2502.18384}}.

\bibitem{PIONEER:2022yag}
{\scshape PIONEER} collaboration, \emph{{PIONEER: Studies of Rare Pion
  Decays}},  \href{https://arxiv.org/abs/2203.01981}{{\ttfamily 2203.01981}}.

\bibitem{PIONEER:2022alm}
{\scshape PIONEER} collaboration, \emph{{Testing Lepton Flavor Universality and
  CKM Unitarity with Rare Pion Decays in the PIONEER experiment}},  in
  \emph{{Snowmass 2021}}, 3, 2022
  [\href{https://arxiv.org/abs/2203.05505}{{\ttfamily 2203.05505}}].

\bibitem{Aditya:2012ay}
Y.G.~Aditya, K.J.~Healey and A.A.~Petrov, \emph{{Searching for super-WIMPs in
  leptonic heavy meson decays}},
  \href{https://doi.org/10.1016/j.physletb.2012.02.042}{\emph{Phys. Lett. B}
  {\bfseries 710} (2012) 118}
  [\href{https://arxiv.org/abs/1201.1007}{{\ttfamily 1201.1007}}].

\bibitem{Batell:2017cmf}
B.~Batell, T.~Han, D.~McKeen and B.~Shams Es~Haghi, \emph{{Thermal Dark Matter
  Through the Dirac Neutrino Portal}},
  \href{https://doi.org/10.1103/PhysRevD.97.075016}{\emph{Phys. Rev. D}
  {\bfseries 97} (2018) 075016}
  [\href{https://arxiv.org/abs/1709.07001}{{\ttfamily 1709.07001}}].

\bibitem{Alves:2017avw}
D.S.M.~Alves and N.~Weiner, \emph{{A viable QCD axion in the MeV mass range}},
  \href{https://doi.org/10.1007/JHEP07(2018)092}{\emph{JHEP} {\bfseries 07}
  (2018) 092} [\href{https://arxiv.org/abs/1710.03764}{{\ttfamily
  1710.03764}}].

\bibitem{Altmannshofer:2019yji}
W.~Altmannshofer, S.~Gori and D.J.~Robinson, \emph{{Constraining axionlike
  particles from rare pion decays}},
  \href{https://doi.org/10.1103/PhysRevD.101.075002}{\emph{Phys. Rev. D}
  {\bfseries 101} (2020) 075002}
  [\href{https://arxiv.org/abs/1909.00005}{{\ttfamily 1909.00005}}].

\bibitem{Dror:2020fbh}
J.A.~Dror, \emph{{Discovering leptonic forces using nonconserved currents}},
  \href{https://doi.org/10.1103/PhysRevD.101.095013}{\emph{Phys. Rev. D}
  {\bfseries 101} (2020) 095013}
  [\href{https://arxiv.org/abs/2004.04750}{{\ttfamily 2004.04750}}].

\bibitem{Hostert:2020xku}
M.~Hostert and M.~Pospelov, \emph{{Novel multilepton signatures of dark sectors
  in light meson decays}},
  \href{https://doi.org/10.1103/PhysRevD.105.015017}{\emph{Phys. Rev. D}
  {\bfseries 105} (2022) 015017}
  [\href{https://arxiv.org/abs/2012.02142}{{\ttfamily 2012.02142}}].

\bibitem{Bauer:2020jbp}
M.~Bauer, M.~Neubert, S.~Renner, M.~Schnubel and A.~Thamm, \emph{{The
  Low-Energy Effective Theory of Axions and ALPs}},
  \href{https://doi.org/10.1007/JHEP04(2021)063}{\emph{JHEP} {\bfseries 04}
  (2021) 063} [\href{https://arxiv.org/abs/2012.12272}{{\ttfamily
  2012.12272}}].

\bibitem{Bauer:2021wjo}
M.~Bauer, M.~Neubert, S.~Renner, M.~Schnubel and A.~Thamm, \emph{{Consistent
  Treatment of Axions in the Weak Chiral Lagrangian}},
  \href{https://doi.org/10.1103/PhysRevLett.127.081803}{\emph{Phys. Rev. Lett.}
  {\bfseries 127} (2021) 081803}
  [\href{https://arxiv.org/abs/2102.13112}{{\ttfamily 2102.13112}}].

\bibitem{Bauer:2021mvw}
M.~Bauer, M.~Neubert, S.~Renner, M.~Schnubel and A.~Thamm, \emph{{Flavor probes
  of axion-like particles}},
  \href{https://doi.org/10.1007/JHEP09(2022)056}{\emph{JHEP} {\bfseries 09}
  (2022) 056} [\href{https://arxiv.org/abs/2110.10698}{{\ttfamily
  2110.10698}}].

\bibitem{Gallo:2021ame}
J.A.~Gallo, A.W.M.~Guerrera, S.~Pe\~naranda and S.~Rigolin, \emph{{Leptonic
  meson decays into invisible ALP}},
  \href{https://doi.org/10.1016/j.nuclphysb.2022.115791}{\emph{Nucl. Phys. B}
  {\bfseries 979} (2022) 115791}
  [\href{https://arxiv.org/abs/2111.02536}{{\ttfamily 2111.02536}}].

\bibitem{Dutta:2021cip}
B.~Dutta, D.~Kim, A.~Thompson, R.T.~Thornton and R.G.~Van~de Water,
  \emph{{Solutions to the MiniBooNE Anomaly from New Physics in Charged Meson
  Decays}}, \href{https://doi.org/10.1103/PhysRevLett.129.111803}{\emph{Phys.
  Rev. Lett.} {\bfseries 129} (2022) 111803}
  [\href{https://arxiv.org/abs/2110.11944}{{\ttfamily 2110.11944}}].

\bibitem{Bickendorf:2022buy}
G.~Bickendorf and M.~Drees, \emph{{Constraints on light leptophilic dark matter
  mediators from decay experiments}},
  \href{https://doi.org/10.1140/epjc/s10052-022-11128-9}{\emph{Eur. Phys. J. C}
  {\bfseries 82} (2022) 1163}
  [\href{https://arxiv.org/abs/2206.05038}{{\ttfamily 2206.05038}}].

\bibitem{Altmannshofer:2022ckw}
W.~Altmannshofer, J.A.~Dror and S.~Gori, \emph{{New Opportunities for Detecting
  Axion-Lepton Interactions}},
  \href{https://doi.org/10.1103/PhysRevLett.130.241801}{\emph{Phys. Rev. Lett.}
  {\bfseries 130} (2023) 241801}
  [\href{https://arxiv.org/abs/2209.00665}{{\ttfamily 2209.00665}}].

\bibitem{Guerrera:2022ykl}
A.W.M.~Guerrera and S.~Rigolin, \emph{{ALP Production in Weak Mesonic Decays}},
  \href{https://doi.org/10.1002/prop.202200192}{\emph{Fortsch. Phys.}
  {\bfseries 71} (2023) 2200192}
  [\href{https://arxiv.org/abs/2211.08343}{{\ttfamily 2211.08343}}].

\bibitem{DiLuzio:2023ndz}
L.~Di~Luzio, A.W.M.~Guerrera, X.P.~D{\'\i}az and S.~Rigolin, \emph{{On the
  IR/UV flavour connection in non-universal axion models}},
  \href{https://doi.org/10.1007/JHEP06(2023)046}{\emph{JHEP} {\bfseries 06}
  (2023) 046} [\href{https://arxiv.org/abs/2304.04643}{{\ttfamily
  2304.04643}}].

\bibitem{Hostert:2023tkg}
M.~Hostert and M.~Pospelov, \emph{{Pion decay constraints on exotic 17~MeV
  vector bosons}},
  \href{https://doi.org/10.1103/PhysRevD.108.055011}{\emph{Phys. Rev. D}
  {\bfseries 108} (2023) 055011}
  [\href{https://arxiv.org/abs/2306.15077}{{\ttfamily 2306.15077}}].

\bibitem{Eberhart:2025lyu}
A.~Eberhart, M.~Fedele, F.~Kahlhoefer, E.~Ravensburg and R.~Ziegler,
  \emph{{Leptophilic ALPs in Laboratory Experiments}},
  \href{https://arxiv.org/abs/2504.05873}{{\ttfamily 2504.05873}}.

\bibitem{DiLuzio:2025ojt}
L.~Di~Luzio, P.~Paradisi and N.~Selimovic, \emph{{Hunting for a 17 MeV particle
  coupled to electrons}},
  \href{https://doi.org/10.1016/j.nuclphysb.2025.117177}{\emph{Nucl. Phys. B}
  {\bfseries 1021} (2025) 117177}
  [\href{https://arxiv.org/abs/2504.14014}{{\ttfamily 2504.14014}}].

\bibitem{Okawa:2025sak}
S.~Okawa and Y.~Omura, \emph{{Novel bounds on neutrino portal dark matter from
  leptonic meson decays}},  \href{https://arxiv.org/abs/2511.08102}{{\ttfamily
  2511.08102}}.

\bibitem{PIENU:2011aa}
{\scshape PIENU} collaboration, \emph{{Search for Massive Neutrinos in the
  Decay $\pi \to e \nu$}},
  \href{https://doi.org/10.1103/PhysRevD.84.052002}{\emph{Phys. Rev. D}
  {\bfseries 84} (2011) 052002}
  [\href{https://arxiv.org/abs/1106.4055}{{\ttfamily 1106.4055}}].

\bibitem{PiENu:2015seu}
{\scshape PiENu} collaboration, \emph{{Improved Measurement of the $\pi \to
  \textrm{e} \nu$ Branching Ratio}},
  \href{https://doi.org/10.1103/PhysRevLett.115.071801}{\emph{Phys. Rev. Lett.}
  {\bfseries 115} (2015) 071801}
  [\href{https://arxiv.org/abs/1506.05845}{{\ttfamily 1506.05845}}].

\bibitem{PIENU:2017wbj}
{\scshape PIENU} collaboration, \emph{{Improved search for heavy neutrinos in
  the decay $\pi\rightarrow e\nu$}},
  \href{https://doi.org/10.1103/PhysRevD.97.072012}{\emph{Phys. Rev. D}
  {\bfseries 97} (2018) 072012}
  [\href{https://arxiv.org/abs/1712.03275}{{\ttfamily 1712.03275}}].

\bibitem{PIENU:2019usb}
{\scshape PIENU} collaboration, \emph{{Search for heavy neutrinos in $\pi \to
  \mu\nu$ decay}},
  \href{https://doi.org/10.1016/j.physletb.2019.134980}{\emph{Phys. Lett. B}
  {\bfseries 798} (2019) 134980}
  [\href{https://arxiv.org/abs/1904.03269}{{\ttfamily 1904.03269}}].

\bibitem{PIENU:2020loi}
{\scshape PIENU} collaboration, \emph{{Improved search for two body muon decay
  ${\mu}^+{\rightarrow}e^+X_H$}},
  \href{https://doi.org/10.1103/PhysRevD.101.052014}{\emph{Phys. Rev. D}
  {\bfseries 101} (2020) 052014}
  [\href{https://arxiv.org/abs/2002.09170}{{\ttfamily 2002.09170}}].

\bibitem{PIENU:2021clt}
{\scshape PIENU} collaboration, \emph{{Search for three body pion decays
  ${\pi}^+{\to} \ell^+{\nu}X$}},
  \href{https://doi.org/10.1103/PhysRevD.103.052006}{\emph{Phys. Rev. D}
  {\bfseries 103} (2021) 052006}
  [\href{https://arxiv.org/abs/2101.07381}{{\ttfamily 2101.07381}}].

\bibitem{NA62:2020mcv}
{\scshape NA62} collaboration, \emph{{Search for heavy neutral lepton
  production in $K^+$ decays to positrons}},
  \href{https://doi.org/10.1016/j.physletb.2020.135599}{\emph{Phys. Lett. B}
  {\bfseries 807} (2020) 135599}
  [\href{https://arxiv.org/abs/2005.09575}{{\ttfamily 2005.09575}}].

\bibitem{NA62:2021bji}
{\scshape NA62} collaboration, \emph{{Search for $K^+$ decays to a muon and
  invisible particles}},
  \href{https://doi.org/10.1016/j.physletb.2021.136259}{\emph{Phys. Lett. B}
  {\bfseries 816} (2021) 136259}
  [\href{https://arxiv.org/abs/2101.12304}{{\ttfamily 2101.12304}}].

\bibitem{NA62:2025csa}
{\scshape NA62} collaboration, \emph{{Search for heavy neutral leptons in
  $\pi^+$ decays to positrons}},
  \href{https://arxiv.org/abs/2507.07345}{{\ttfamily 2507.07345}}.

\bibitem{Marciano:1993sh}
W.J.~Marciano and A.~Sirlin, \emph{{Radiative corrections to $pi_{\ell 2}$
  decays}}, \href{https://doi.org/10.1103/PhysRevLett.71.3629}{\emph{Phys. Rev.
  Lett.} {\bfseries 71} (1993) 3629}.

\bibitem{Knecht:1999ag}
M.~Knecht, H.~Neufeld, H.~Rupertsberger and P.~Talavera, \emph{{Chiral
  perturbation theory with virtual photons and leptons}},
  \href{https://doi.org/10.1007/s100529900265}{\emph{Eur. Phys. J. C}
  {\bfseries 12} (2000) 469}
  [\href{https://arxiv.org/abs/hep-ph/9909284}{{\ttfamily hep-ph/9909284}}].

\bibitem{Cirigliano:2007xi}
V.~Cirigliano and I.~Rosell, \emph{{Two-loop effective theory analysis of $\pi
  (K) \to e \bar \nu_e [\gamma]$ branching ratios}},
  \href{https://doi.org/10.1103/PhysRevLett.99.231801}{\emph{Phys. Rev. Lett.}
  {\bfseries 99} (2007) 231801}
  [\href{https://arxiv.org/abs/0707.3439}{{\ttfamily 0707.3439}}].

\bibitem{Cirigliano:2007ga}
V.~Cirigliano and I.~Rosell, \emph{{$\pi/K \to e \bar \nu_e$ branching ratios
  to $O(e^2 p^4)$ in Chiral Perturbation Theory}},
  \href{https://doi.org/10.1088/1126-6708/2007/10/005}{\emph{JHEP} {\bfseries
  10} (2007) 005} [\href{https://arxiv.org/abs/0707.4464}{{\ttfamily
  0707.4464}}].

\bibitem{Bryman:1985bv}
D.A.~Bryman, M.S.~Dixit, R.~Dubois, J.A.~Macdonald, T.~Numao, B.~Olaniyi
  et~al., \emph{{Measurement of the $\pi \to e \nu_e$ Branching Ratio}},
  \href{https://doi.org/10.1103/PhysRevD.33.1211}{\emph{Phys. Rev. D}
  {\bfseries 33} (1986) 1211}.

\bibitem{Britton:1992pg}
D.I.~Britton et~al., \emph{{Measurement of the $pi^+ \to e^+ \nu_e$ branching
  ratio}}, \href{https://doi.org/10.1103/PhysRevLett.68.3000}{\emph{Phys. Rev.
  Lett.} {\bfseries 68} (1992) 3000}.

\bibitem{Czapek:1993kc}
G.~Czapek et~al., \emph{{Branching ratio for the rare pion decay into positron
  and neutrino}}, \href{https://doi.org/10.1103/PhysRevLett.70.17}{\emph{Phys.
  Rev. Lett.} {\bfseries 70} (1993) 17}.

\bibitem{ParticleDataGroup:2024cfk}
{\scshape Particle Data Group} collaboration, \emph{{Review of particle
  physics}}, \href{https://doi.org/10.1103/PhysRevD.110.030001}{\emph{Phys.
  Rev. D} {\bfseries 110} (2024) 030001}.

\bibitem{Pocanic:2003pf}
D.~Pocanic et~al., \emph{{Precise measurement of the $\pi^+ \to \pi^0 e^+ \nu$
  branching ratio}},
  \href{https://doi.org/10.1103/PhysRevLett.93.181803}{\emph{Phys. Rev. Lett.}
  {\bfseries 93} (2004) 181803}
  [\href{https://arxiv.org/abs/hep-ex/0312030}{{\ttfamily hep-ex/0312030}}].

\bibitem{Jodidio:1986mz}
A.~Jodidio et~al., \emph{{Search for Right-Handed Currents in Muon Decay}},
  \href{https://doi.org/10.1103/PhysRevD.34.1967}{\emph{Phys. Rev. D}
  {\bfseries 34} (1986) 1967}.

\bibitem{TWIST:2014ymv}
{\scshape TWIST} collaboration, \emph{{Search for two body muon decay
  signals}}, \href{https://doi.org/10.1103/PhysRevD.91.052020}{\emph{Phys. Rev.
  D} {\bfseries 91} (2015) 052020}
  [\href{https://arxiv.org/abs/1409.0638}{{\ttfamily 1409.0638}}].

\bibitem{Calibbi:2020jvd}
L.~Calibbi, D.~Redigolo, R.~Ziegler and J.~Zupan, \emph{{Looking forward to
  lepton-flavor-violating ALPs}},
  \href{https://doi.org/10.1007/JHEP09(2021)173}{\emph{JHEP} {\bfseries 09}
  (2021) 173} [\href{https://arxiv.org/abs/2006.04795}{{\ttfamily
  2006.04795}}].

\bibitem{Jho:2022snj}
Y.~Jho, S.~Knapen and D.~Redigolo, \emph{{Lepton-flavor violating axions at MEG
  II}}, \href{https://doi.org/10.1007/JHEP10(2022)029}{\emph{JHEP} {\bfseries
  10} (2022) 029} [\href{https://arxiv.org/abs/2203.11222}{{\ttfamily
  2203.11222}}].

\bibitem{Hill:2023dym}
R.J.~Hill, R.~Plestid and J.~Zupan, \emph{{Searching for new physics at
  {\ensuremath{\mu}}{\textrightarrow}e facilities with $\mu^+$ and $\pi^+$
  decays at rest}},
  \href{https://doi.org/10.1103/PhysRevD.109.035025}{\emph{Phys. Rev. D}
  {\bfseries 109} (2024) 035025}
  [\href{https://arxiv.org/abs/2310.00043}{{\ttfamily 2310.00043}}].

\bibitem{Knapen:2023zgi}
S.~Knapen, K.~Langhoff, T.~Opferkuch and D.~Redigolo, \emph{{A robust search
  for lepton flavour violating axions at Mu3e}},
  \href{https://doi.org/10.1007/JHEP07(2025)243}{\emph{JHEP} {\bfseries 07}
  (2025) 243} [\href{https://arxiv.org/abs/2311.17915}{{\ttfamily
  2311.17915}}].

\bibitem{Altmannshofer_inprep}
W.~Altmannshofer et~al., ``{in preparation}.''

\bibitem{Shrock:1980vy}
R.E.~Shrock, \emph{{New Tests For, and Bounds On, Neutrino Masses and Lepton
  Mixing}}, \href{https://doi.org/10.1016/0370-2693(80)90235-X}{\emph{Phys.
  Lett. B} {\bfseries 96} (1980) 159}.

\bibitem{Shrock:1980ct}
R.E.~Shrock, \emph{{General Theory of Weak Leptonic and Semileptonic Decays. 1.
  Leptonic Pseudoscalar Meson Decays, with Associated Tests For, and Bounds on,
  Neutrino Masses and Lepton Mixing}},
  \href{https://doi.org/10.1103/PhysRevD.24.1232}{\emph{Phys. Rev. D}
  {\bfseries 24} (1981) 1232}.

\bibitem{deGouvea:2015euy}
A.~de~Gouv{\^e}a and A.~Kobach, \emph{{Global Constraints on a Heavy
  Neutrino}}, \href{https://doi.org/10.1103/PhysRevD.93.033005}{\emph{Phys.
  Rev. D} {\bfseries 93} (2016) 033005}
  [\href{https://arxiv.org/abs/1511.00683}{{\ttfamily 1511.00683}}].

\bibitem{Bryman:2019bjg}
D.A.~Bryman and R.~Shrock, \emph{{Constraints on Sterile Neutrinos in the MeV
  to GeV Mass Range}},
  \href{https://doi.org/10.1103/PhysRevD.100.073011}{\emph{Phys. Rev. D}
  {\bfseries 100} (2019) 073011}
  [\href{https://arxiv.org/abs/1909.11198}{{\ttfamily 1909.11198}}].

\bibitem{Abdullahi:2022jlv}
A.M.~Abdullahi et~al., \emph{{The present and future status of heavy neutral
  leptons}}, \href{https://doi.org/10.1088/1361-6471/ac98f9}{\emph{J. Phys. G}
  {\bfseries 50} (2023) 020501}
  [\href{https://arxiv.org/abs/2203.08039}{{\ttfamily 2203.08039}}].

\bibitem{Deppisch:2024izn}
F.F.~Deppisch, T.E.~Gonzalo, C.~Majumdar and Z.~Zhang, \emph{{Relaxing limits
  from Big Bang Nucleosynthesis on Heavy Neutral Leptons with axion-like
  particles}}, \href{https://doi.org/10.1088/1475-7516/2025/02/054}{\emph{JCAP}
  {\bfseries 02} (2025) 054}
  [\href{https://arxiv.org/abs/2410.06970}{{\ttfamily 2410.06970}}].

\bibitem{Esteban:2024eli}
I.~Esteban, M.C.~Gonzalez-Garcia, M.~Maltoni, I.~Martinez-Soler, J.P.~Pinheiro
  and T.~Schwetz, \emph{{NuFit-6.0: updated global analysis of three-flavor
  neutrino oscillations}},
  \href{https://doi.org/10.1007/JHEP12(2024)216}{\emph{JHEP} {\bfseries 12}
  (2024) 216} [\href{https://arxiv.org/abs/2410.05380}{{\ttfamily
  2410.05380}}].

\bibitem{Planck:2019nip}
{\scshape Planck} collaboration, \emph{{Planck 2018 results. V. CMB power
  spectra and likelihoods}},
  \href{https://doi.org/10.1051/0004-6361/201936386}{\emph{Astron. Astrophys.}
  {\bfseries 641} (2020) A5}
  [\href{https://arxiv.org/abs/1907.12875}{{\ttfamily 1907.12875}}].

\bibitem{ACT:2023kun}
{\scshape ACT} collaboration, \emph{{The Atacama Cosmology Telescope: DR6
  Gravitational Lensing Map and Cosmological Parameters}},
  \href{https://doi.org/10.3847/1538-4357/acff5f}{\emph{Astrophys. J.}
  {\bfseries 962} (2024) 113}
  [\href{https://arxiv.org/abs/2304.05203}{{\ttfamily 2304.05203}}].

\bibitem{DESI:2024mwx}
{\scshape DESI} collaboration, \emph{{DESI 2024 VI: cosmological constraints
  from the measurements of baryon acoustic oscillations}},
  \href{https://doi.org/10.1088/1475-7516/2025/02/021}{\emph{JCAP} {\bfseries
  02} (2025) 021} [\href{https://arxiv.org/abs/2404.03002}{{\ttfamily
  2404.03002}}].

\bibitem{Bryman:1983cja}
D.A.~Bryman, R.~Dubois, T.~Numao, B.~Olaniyi, A.~Olin, M.S.~Dixit et~al.,
  \emph{{Search for Massive Neutrinos in $\pi \to e \nu_e$ Decay}},
  \href{https://doi.org/10.1103/PhysRevLett.50.1546}{\emph{Phys. Rev. Lett.}
  {\bfseries 50} (1983) 1546}.

\bibitem{Azuelos:1986eg}
G.~Azuelos et~al., \emph{{Constraints on Massive Neutrinos in $\pi \to e \nu_e$
  Decay}}, \href{https://doi.org/10.1103/PhysRevLett.56.2241}{\emph{Phys. Rev.
  Lett.} {\bfseries 56} (1986) 2241}.

\bibitem{DeLeener-Rosier:1991luz}
N.~De~Leener-Rosier et~al., \emph{{Search for massive neutrinos in $\pi \to e
  \nu$ decay}}, \href{https://doi.org/10.1103/PhysRevD.43.3611}{\emph{Phys.
  Rev. D} {\bfseries 43} (1991) 3611}.

\bibitem{Britton:1992xv}
D.I.~Britton et~al., \emph{{Improved search for massive neutrinos in $\pi^+ \to
  e^+ \nu$ decay}}, \href{https://doi.org/10.1103/PhysRevD.46.R885}{\emph{Phys.
  Rev. D} {\bfseries 46} (1992) R885}.

\bibitem{Abela:1981nf}
R.~Abela, M.~Daum, G.H.~Eaton, R.~Frosch, B.~Jost, P.R.~Kettle et~al.,
  \emph{{Search for an Admixture of Heavy Neutrino in Pion Decay}},
  \href{https://doi.org/10.1016/0370-2693(81)90884-4}{\emph{Phys. Lett. B}
  {\bfseries 105} (1981) 263}.

\bibitem{Minehart:1981fv}
R.C.~Minehart, K.O.H.~Ziock, R.~Marshall, W.A.~Stephens, M.~Daum, B.~Jost
  et~al., \emph{{A Search for Admixture of Massive Neutrinos in the Decay $\pi
  \to \mu \nu$}}, \href{https://doi.org/10.1103/PhysRevLett.52.804}{\emph{Phys.
  Rev. Lett.} {\bfseries 52} (1984) 804}.

\bibitem{Daum:1987bg}
M.~Daum, B.~Jost, R.M.~Marshall, R.C.~Minehart, W.A.~Stephens and K.O.H.~Ziock,
  \emph{{Search for Admixtures of Massive Neutrinos in the Decay $\pi^+ \to
  \mu^+ \nu$}}, \href{https://doi.org/10.1103/PhysRevD.36.2624}{\emph{Phys.
  Rev. D} {\bfseries 36} (1987) 2624}.

\bibitem{Daum:1995hs}
M.~Daum, R.~Frosch, W.~Hajdas, M.~Janousch, P.R.~Kettle, S.~Ritt et~al.,
  \emph{{Search for a neutral particle of mass 33.9 MeV in pion decay}},
  \href{https://doi.org/10.1016/0370-2693(95)01165-M}{\emph{Phys. Lett. B}
  {\bfseries 361} (1995) 179}.

\bibitem{Bryman:1996xd}
D.A.~Bryman and T.~Numao, \emph{{Search for massive neutrinos in $\pi^+ \to
  \mu^+ \nu$ decay}},
  \href{https://doi.org/10.1103/PhysRevD.53.558}{\emph{Phys. Rev. D} {\bfseries
  53} (1996) 558}.

\bibitem{Assamagan:1998vy}
K.~Assamagan, C.~Bronnimann, M.~Daum, R.~Frosch, P.R.~Kettle and C.~Wigger,
  \emph{{Search for a heavy neutrino state in the decay $\pi^+ \to \mu^+
  \nu_\mu$}}, \href{https://doi.org/10.1016/S0370-2693(98)00727-8}{\emph{Phys.
  Lett. B} {\bfseries 434} (1998) 158}.

\bibitem{Daum:2000ac}
M.~Daum, M.~Janousch, P.R.~Kettle, J.~Koglin, D.~Pocanic, J.~Schottmuller
  et~al., \emph{{The KARMEN time anomaly: Search for a neutral particle of mass
  33.9 MeV in pion decay}},
  \href{https://doi.org/10.1103/PhysRevLett.85.1815}{\emph{Phys. Rev. Lett.}
  {\bfseries 85} (2000) 1815}
  [\href{https://arxiv.org/abs/hep-ex/0008014}{{\ttfamily hep-ex/0008014}}].

\bibitem{Asano:1981he}
Y.~Asano et~al., \emph{{Search for a Heavy Neutrino Emitted in $K^+ \to \mu^+$
  Neutrino Decay}},
  \href{https://doi.org/10.1016/0370-2693(81)90860-1}{\emph{Phys. Lett. B}
  {\bfseries 104} (1981) 84}.

\bibitem{Hayano:1982wu}
R.S.~Hayano et~al., \emph{{HEAVY NEUTRINO SEARCH USING K(mu2) DECAY}},
  \href{https://doi.org/10.1103/PhysRevLett.49.1305}{\emph{Phys. Rev. Lett.}
  {\bfseries 49} (1982) 1305}.

\bibitem{E949:2014gsn}
{\scshape E949} collaboration, \emph{{Search for heavy neutrinos in
  $K^+\to\mu^+\nu_H$ decays}},
  \href{https://doi.org/10.1103/PhysRevD.91.052001}{\emph{Phys. Rev. D}
  {\bfseries 91} (2015) 052001}
  [\href{https://arxiv.org/abs/1411.3963}{{\ttfamily 1411.3963}}].

\bibitem{NA62:2017qcd}
{\scshape NA62} collaboration, \emph{{Search for heavy neutral lepton
  production in $K^+$ decays}},
  \href{https://doi.org/10.1016/j.physletb.2018.01.031}{\emph{Phys. Lett. B}
  {\bfseries 778} (2018) 137}
  [\href{https://arxiv.org/abs/1712.00297}{{\ttfamily 1712.00297}}].

\bibitem{ParticleDataGroup:2022pth}
{\scshape Particle Data Group} collaboration, \emph{{Review of Particle
  Physics}}, \href{https://doi.org/10.1093/ptep/ptac097}{\emph{PTEP} {\bfseries
  2022} (2022) 083C01}.

\bibitem{patrick1}
P.~Schwendimann, \emph{Pioneer: A next generation rare pion decay experiment
  relying on 5d tracking},  Workshop on ACTS Tracking for Nuclear Physics 2025,
  LBNL May 12-15, 2025.

\bibitem{patrick2}
P.~Schwendimann, \emph{Overview and current status of the simulation, reco and
  analysis},  PIONEER collaboration meeting, TRIUMF Jan 7-10, 2025.

\bibitem{Beesley:2024mts}
O.~Beesley et~al., \emph{{Measurements of a LYSO crystal array from threshold
  to 100MeV}}, \href{https://doi.org/10.1016/j.nima.2025.170320}{\emph{Nucl.
  Instrum. Meth. A} {\bfseries 1075} (2025) 170320}
  [\href{https://arxiv.org/abs/2409.14691}{{\ttfamily 2409.14691}}].

\bibitem{Boyarsky:2020dzc}
A.~Boyarsky, M.~Ovchynnikov, O.~Ruchayskiy and V.~Syvolap, \emph{{Improved big
  bang nucleosynthesis constraints on heavy neutral leptons}},
  \href{https://doi.org/10.1103/PhysRevD.104.023517}{\emph{Phys. Rev. D}
  {\bfseries 104} (2021) 023517}
  [\href{https://arxiv.org/abs/2008.00749}{{\ttfamily 2008.00749}}].

\bibitem{Bondarenko:2021cpc}
K.~Bondarenko, A.~Boyarsky, J.~Klaric, O.~Mikulenko, O.~Ruchayskiy, V.~Syvolap
  et~al., \emph{{An allowed window for heavy neutral leptons below the kaon
  mass}}, \href{https://doi.org/10.1007/JHEP07(2021)193}{\emph{JHEP} {\bfseries
  07} (2021) 193} [\href{https://arxiv.org/abs/2101.09255}{{\ttfamily
  2101.09255}}].

\bibitem{Dev:2025pru}
P.S.B.~Dev, Q.-f.~Wu and X.-J.~Xu, \emph{{No Hiding in the Dark: Cosmological
  Bounds on Heavy Neutral Leptons with Dark Decay Channels}},
  \href{https://arxiv.org/abs/2507.12270}{{\ttfamily 2507.12270}}.

\bibitem{Bolton:2019pcu}
P.D.~Bolton, F.F.~Deppisch and P.S.~Bhupal~Dev, \emph{{Neutrinoless double beta
  decay versus other probes of heavy sterile neutrinos}},
  \href{https://doi.org/10.1007/JHEP03(2020)170}{\emph{JHEP} {\bfseries 03}
  (2020) 170} [\href{https://arxiv.org/abs/1912.03058}{{\ttfamily
  1912.03058}}].

\bibitem{Dekens:2020ttz}
W.~Dekens, J.~de~Vries, K.~Fuyuto, E.~Mereghetti and G.~Zhou, \emph{{Sterile
  neutrinos and neutrinoless double beta decay in effective field theory}},
  \href{https://doi.org/10.1007/JHEP06(2020)097}{\emph{JHEP} {\bfseries 06}
  (2020) 097} [\href{https://arxiv.org/abs/2002.07182}{{\ttfamily
  2002.07182}}].

\bibitem{Deppisch:2020ztt}
F.F.~Deppisch, L.~Graf, F.~Iachello and J.~Kotila, \emph{{Analysis of light
  neutrino exchange and short-range mechanisms in $0\nu\beta\beta$ decay}},
  \href{https://doi.org/10.1103/PhysRevD.102.095016}{\emph{Phys. Rev. D}
  {\bfseries 102} (2020) 095016}
  [\href{https://arxiv.org/abs/2009.10119}{{\ttfamily 2009.10119}}].

\bibitem{deVries:2024mla}
J.~de~Vries, H.K.~Dreiner, J.~Groot, J.Y.~G{\"u}nther and Z.S.~Wang,
  \emph{{Probing light sterile neutrinos in left-right symmetric models with
  displaced vertices and neutrinoless double beta decay}},
  \href{https://doi.org/10.1007/JHEP04(2025)007}{\emph{JHEP} {\bfseries 04}
  (2025) 007} [\href{https://arxiv.org/abs/2406.15091}{{\ttfamily
  2406.15091}}].

\bibitem{Ballett:2019bgd}
P.~Ballett, T.~Boschi and S.~Pascoli, \emph{{Heavy Neutral Leptons from
  low-scale seesaws at the DUNE Near Detector}},
  \href{https://doi.org/10.1007/JHEP03(2020)111}{\emph{JHEP} {\bfseries 03}
  (2020) 111} [\href{https://arxiv.org/abs/1905.00284}{{\ttfamily
  1905.00284}}].

\bibitem{Berryman:2019dme}
J.M.~Berryman, A.~de~Gouvea, P.J.~Fox, B.J.~Kayser, K.J.~Kelly and J.L.~Raaf,
  \emph{{Searches for Decays of New Particles in the DUNE Multi-Purpose Near
  Detector}}, \href{https://doi.org/10.1007/JHEP02(2020)174}{\emph{JHEP}
  {\bfseries 02} (2020) 174}
  [\href{https://arxiv.org/abs/1912.07622}{{\ttfamily 1912.07622}}].

\bibitem{Coloma:2020lgy}
P.~Coloma, E.~Fern{\'a}ndez-Mart{\'\i}nez, M.~Gonz{\'a}lez-L{\'o}pez,
  J.~Hern{\'a}ndez-Garc{\'\i}a and Z.~Pavlovic, \emph{{GeV-scale neutrinos:
  interactions with mesons and DUNE sensitivity}},
  \href{https://doi.org/10.1140/epjc/s10052-021-08861-y}{\emph{Eur. Phys. J. C}
  {\bfseries 81} (2021) 78} [\href{https://arxiv.org/abs/2007.03701}{{\ttfamily
  2007.03701}}].

\bibitem{Breitbach:2021gvv}
M.~Breitbach, L.~Buonocore, C.~Frugiuele, J.~Kopp and L.~Mittnacht,
  \emph{{Searching for physics beyond the Standard Model in an off-axis DUNE
  near detector}}, \href{https://doi.org/10.1007/JHEP01(2022)048}{\emph{JHEP}
  {\bfseries 01} (2022) 048}
  [\href{https://arxiv.org/abs/2102.03383}{{\ttfamily 2102.03383}}].

\bibitem{Carenza:2021pcm}
P.~Carenza and G.~Lucente, \emph{{Supernova bound on axionlike particles
  coupled with electrons}},
  \href{https://doi.org/10.1103/PhysRevD.104.103007}{\emph{Phys. Rev. D}
  {\bfseries 104} (2021) 103007}
  [\href{https://arxiv.org/abs/2107.12393}{{\ttfamily 2107.12393}}].

\bibitem{Fiorillo:2025sln}
D.F.G.~Fiorillo, T.~Pitik and E.~Vitagliano, \emph{{Supernova production of
  axionlike particles coupling to electrons, reloaded}},
  \href{https://doi.org/10.1103/y1r2-gtb5}{\emph{Phys. Rev. D} {\bfseries 112}
  (2025) 083008} [\href{https://arxiv.org/abs/2503.15630}{{\ttfamily
  2503.15630}}].

\bibitem{Fox:2011qd}
P.J.~Fox, J.~Liu, D.~Tucker-Smith and N.~Weiner, \emph{{An Effective Z'}},
  \href{https://doi.org/10.1103/PhysRevD.84.115006}{\emph{Phys. Rev. D}
  {\bfseries 84} (2011) 115006}
  [\href{https://arxiv.org/abs/1104.4127}{{\ttfamily 1104.4127}}].

\bibitem{Fan:2022eto}
X.~Fan, T.G.~Myers, B.A.D.~Sukra and G.~Gabrielse, \emph{{Measurement of the
  Electron Magnetic Moment}},
  \href{https://doi.org/10.1103/PhysRevLett.130.071801}{\emph{Phys. Rev. Lett.}
  {\bfseries 130} (2023) 071801}
  [\href{https://arxiv.org/abs/2209.13084}{{\ttfamily 2209.13084}}].

\bibitem{Muong-2:2025xyk}
{\scshape Muon g-2} collaboration, \emph{{Measurement of the Positive Muon
  Anomalous Magnetic Moment to 127~ppb}},
  \href{https://doi.org/10.1103/7clf-sm2v}{\emph{Phys. Rev. Lett.} {\bfseries
  135} (2025) 101802} [\href{https://arxiv.org/abs/2506.03069}{{\ttfamily
  2506.03069}}].

\bibitem{Aliberti:2025beg}
R.~Aliberti et~al., \emph{{The anomalous magnetic moment of the muon in the
  Standard Model: an update}},
  \href{https://doi.org/10.1016/j.physrep.2025.08.002}{\emph{Phys. Rept.}
  {\bfseries 1143} (2025) 1}
  [\href{https://arxiv.org/abs/2505.21476}{{\ttfamily 2505.21476}}].

\bibitem{Parker:2018vye}
R.H.~Parker, C.~Yu, W.~Zhong, B.~Estey and H.~M\"uller, \emph{{Measurement of
  the fine-structure constant as a test of the Standard Model}},
  \href{https://doi.org/10.1126/science.aap7706}{\emph{Science} {\bfseries 360}
  (2018) 191} [\href{https://arxiv.org/abs/1812.04130}{{\ttfamily
  1812.04130}}].

\bibitem{Morel:2020dww}
L.~Morel, Z.~Yao, P.~Clad\'e and S.~Guellati-Kh\'elifa, \emph{{Determination of
  the fine-structure constant with an accuracy of 81 parts per trillion}},
  \href{https://doi.org/10.1038/s41586-020-2964-7}{\emph{Nature} {\bfseries
  588} (2020) 61}.

\bibitem{Aoyama:2019ryr}
T.~Aoyama, T.~Kinoshita and M.~Nio, \emph{{Theory of the Anomalous Magnetic
  Moment of the Electron}},
  \href{https://doi.org/10.3390/atoms7010028}{\emph{Atoms} {\bfseries 7} (2019)
  28}.

\bibitem{Volkov:2024yzc}
S.~Volkov, \emph{{Calculation of the total 10th order QED contribution to the
  electron magnetic moment}},
  \href{https://doi.org/10.1103/PhysRevD.110.036001}{\emph{Phys. Rev. D}
  {\bfseries 110} (2024) 036001}
  [\href{https://arxiv.org/abs/2404.00649}{{\ttfamily 2404.00649}}].

\bibitem{Aoyama:2024aly}
T.~Aoyama, M.~Hayakawa, A.~Hirayama and M.~Nio, \emph{{Verification of the
  tenth-order QED contribution to the anomalous magnetic moment of the electron
  from diagrams without fermion loops}},
  \href{https://doi.org/10.1103/PhysRevD.111.L031902}{\emph{Phys. Rev. D}
  {\bfseries 111} (2025) L031902}
  [\href{https://arxiv.org/abs/2412.06473}{{\ttfamily 2412.06473}}].

\bibitem{Aoyama:2020ynm}
T.~Aoyama et~al., \emph{{The anomalous magnetic moment of the muon in the
  Standard Model}},
  \href{https://doi.org/10.1016/j.physrep.2020.07.006}{\emph{Phys. Rept.}
  {\bfseries 887} (2020) 1} [\href{https://arxiv.org/abs/2006.04822}{{\ttfamily
  2006.04822}}].

\bibitem{BaBar:2017tiz}
{\scshape BaBar} collaboration, \emph{{Search for Invisible Decays of a Dark
  Photon Produced in ${e}^{+}{e}^{-}$ Collisions at BaBar}},
  \href{https://doi.org/10.1103/PhysRevLett.119.131804}{\emph{Phys. Rev. Lett.}
  {\bfseries 119} (2017) 131804}
  [\href{https://arxiv.org/abs/1702.03327}{{\ttfamily 1702.03327}}].

\bibitem{Belle-II:2022cgf}
{\scshape Belle-II} collaboration, \emph{{Snowmass White Paper: Belle II
  physics reach and plans for the next decade and beyond}},
  \href{https://arxiv.org/abs/2207.06307}{{\ttfamily 2207.06307}}.

\bibitem{Gninenko:2016kpg}
S.N.~Gninenko, N.V.~Krasnikov, M.M.~Kirsanov and D.V.~Kirpichnikov,
  \emph{{Missing energy signature from invisible decays of dark photons at the
  CERN SPS}}, \href{https://doi.org/10.1103/PhysRevD.94.095025}{\emph{Phys.
  Rev. D} {\bfseries 94} (2016) 095025}
  [\href{https://arxiv.org/abs/1604.08432}{{\ttfamily 1604.08432}}].

\bibitem{NA64:2023wbi}
{\scshape NA64} collaboration, \emph{{Search for Light Dark Matter with NA64 at
  CERN}}, \href{https://doi.org/10.1103/PhysRevLett.131.161801}{\emph{Phys.
  Rev. Lett.} {\bfseries 131} (2023) 161801}
  [\href{https://arxiv.org/abs/2307.02404}{{\ttfamily 2307.02404}}].

\bibitem{NA64:2023ehh}
{\scshape NA64} collaboration, \emph{{Probing light dark matter with positron
  beams at NA64}},
  \href{https://doi.org/10.1103/PhysRevD.109.L031103}{\emph{Phys. Rev. D}
  {\bfseries 109} (2024) L031103}
  [\href{https://arxiv.org/abs/2308.15612}{{\ttfamily 2308.15612}}].

\bibitem{LDMX:2025bog}
{\scshape LDMX} collaboration, \emph{{LDMX - The Light Dark Matter
  eXperiment}},  \href{https://arxiv.org/abs/2508.11833}{{\ttfamily
  2508.11833}}.

\bibitem{NA64:2021xzo}
{\scshape NA64} collaboration, \emph{{Constraints on New Physics in Electron
  $g-2$ from a Search for Invisible Decays of a Scalar, Pseudoscalar, Vector,
  and Axial Vector}},
  \href{https://doi.org/10.1103/PhysRevLett.126.211802}{\emph{Phys. Rev. Lett.}
  {\bfseries 126} (2021) 211802}
  [\href{https://arxiv.org/abs/2102.01885}{{\ttfamily 2102.01885}}].

\bibitem{BNL-E949:2009dza}
{\scshape BNL-E949} collaboration, \emph{{Study of the decay $K^+\to\pi^+\nu
  \bar\nu$ in the momentum region $140 < P_\pi < 199$ MeV/c}},
  \href{https://doi.org/10.1103/PhysRevD.79.092004}{\emph{Phys. Rev. D}
  {\bfseries 79} (2009) 092004}
  [\href{https://arxiv.org/abs/0903.0030}{{\ttfamily 0903.0030}}].

\bibitem{NA62:2020pwi}
{\scshape NA62} collaboration, \emph{{Search for $\pi^0$ decays to invisible
  particles}}, \href{https://doi.org/10.1007/JHEP02(2021)201}{\emph{JHEP}
  {\bfseries 02} (2021) 201}
  [\href{https://arxiv.org/abs/2010.07644}{{\ttfamily 2010.07644}}].

\bibitem{NA62:2020xlg}
{\scshape NA62} collaboration, \emph{{Search for a feebly interacting particle
  $X$ in the decay $K^{+}\rightarrow\pi^{+}X$}},
  \href{https://doi.org/10.1007/JHEP03(2021)058}{\emph{JHEP} {\bfseries 03}
  (2021) 058} [\href{https://arxiv.org/abs/2011.11329}{{\ttfamily
  2011.11329}}].

\bibitem{NA62:2025upx}
{\scshape NA62} collaboration, \emph{{Searches for hidden sectors using
  K$^{+}${\textrightarrow} {\ensuremath{\pi}}$^{+}$X decays}},
  \href{https://doi.org/10.1007/JHEP11(2025)143}{\emph{JHEP} {\bfseries 11}
  (2025) 143} [\href{https://arxiv.org/abs/2507.17286}{{\ttfamily
  2507.17286}}].

\bibitem{Guadagnoli:2025xnt}
D.~Guadagnoli, A.~Iohner, C.~Lazzeroni, D.~Martinez~Santos, J.C.~Swallow and
  C.~Toni, \emph{{New bound on the vectorial axion-down-strange coupling from
  $K^+ \to \pi^+ \nu \bar \nu$ data}},
  \href{https://arxiv.org/abs/2503.05865}{{\ttfamily 2503.05865}}.

\bibitem{Chikashige:1980ui}
Y.~Chikashige, R.N.~Mohapatra and R.D.~Peccei, \emph{{Are There Real Goldstone
  Bosons Associated with Broken Lepton Number?}},
  \href{https://doi.org/10.1016/0370-2693(81)90011-3}{\emph{Phys. Lett. B}
  {\bfseries 98} (1981) 265}.

\bibitem{Gelmini:1980re}
G.B.~Gelmini and M.~Roncadelli, \emph{{Left-Handed Neutrino Mass Scale and
  Spontaneously Broken Lepton Number}},
  \href{https://doi.org/10.1016/0370-2693(81)90559-1}{\emph{Phys. Lett. B}
  {\bfseries 99} (1981) 411}.

\bibitem{Barger:1981vd}
V.D.~Barger, W.-Y.~Keung and S.~Pakvasa, \emph{{Majoron Emission by
  Neutrinos}}, \href{https://doi.org/10.1103/PhysRevD.25.907}{\emph{Phys. Rev.
  D} {\bfseries 25} (1982) 907}.

\bibitem{Schechter:1981cv}
J.~Schechter and J.W.F.~Valle, \emph{{Neutrino Decay and Spontaneous Violation
  of Lepton Number}},
  \href{https://doi.org/10.1103/PhysRevD.25.774}{\emph{Phys. Rev. D} {\bfseries
  25} (1982) 774}.

\bibitem{deLima2026}
C.H.~de~Lima, D.~McKeen, J.~Ng and D.~Tuckler, \emph{Hadronic probes of
  non-standard neutrino interactions},  2026.

\bibitem{Ito:2025}
S.~Ito. private communication.

\end{thebibliography}\endgroup

\end{document}